\documentclass[%
preprint,
amsmath,amssymb,
aps,
prb,
floatfix,
]{revtex4-2}

\PassOptionsToPackage{backend=biber}{biblatex}
\usepackage[T1]{fontenc}
\usepackage[utf8]{inputenc}

\usepackage{siunitx}
\usepackage{amsmath} 
\usepackage{bm}

\usepackage{xcolor}
\usepackage{hyperref}
\hypersetup{colorlinks=true, citecolor=blue, urlcolor=blue, linkcolor=black}

\usepackage{xr}
\makeatletter
\newcommand*{\addFileDependency}[1]{
  \typeout{(#1)}
  \@addtofilelist{#1}
  \IfFileExists{#1}{}{\typeout{No file #1.}}
}
\makeatother

\newcommand*{\myexternaldocument}[2][]{%
    \externaldocument[#1]{#2}%
    \addFileDependency{#2.tex}%
    \addFileDependency{#2.aux}%
}
\myexternaldocument[supp-]{supplement}

\usepackage{graphicx} 
\graphicspath{ {DGraphics/} }
\usepackage{caption}
\usepackage{subfig}
\usepackage{wrapfig}
\usepackage[export]{adjustbox}
\usepackage{gensymb} 

\usepackage{relsize}
\usepackage{multirow}
\usepackage{dcolumn}
\usepackage{array}
\usepackage{booktabs}

\newcommand{\Figref}[2][Fig.~]{#1\ref{#2}}
\newcommand{\figref}[2][Fig.~]{#1\ref{#2}}

\newcommand{\tabref}[2][Table~]{#1\ref{#2}}

\renewcommand{\eqref}[2][Equation~]{#1(\ref{#2})}

\DeclareSIUnit\angstrom{\protect \text {Å}}

\definecolor{amethyst}{rgb}{0.6, 0.4, 0.8}

\newcolumntype{P}[1]{>{\centering\arraybackslash}p{#1}}

\newcommand{\etal}{\emph{et al.}}

\renewcommand{\o}{O$_2$}
\renewcommand{\ao}{$A$O}
\newcommand{\bo}{$B$O$_2$}
\newcommand{\abo}{$AB$O$_3$}

\newcommand{\ato}{$A$TiO$_3$}
\newcommand{\azo}{$A$ZrO$_3$}
\newcommand{\aso}{$A$SnO$_3$}

\newcommand{\cbo}{Ca$B$O$_3$}
\newcommand{\sbo}{Sr$B$O$_3$}
\newcommand{\bbo}{Ba$B$O$_3$}

\newcommand{\cto}{CaTiO$_3$}
\newcommand{\sto}{SrTiO$_3$}
\newcommand{\bto}{BaTiO$_3$}
\newcommand{\czo}{CaZrO$_3$}
\newcommand{\szo}{SrZrO$_3$}
\newcommand{\bzo}{BaZrO$_3$}
\newcommand{\cso}{CaSnO$_3$}
\newcommand{\sso}{SrSnO$_3$}
\newcommand{\bso}{BaSnO$_3$}

\begin{document}
\title{Instability of oxide perovskite surfaces induced by vacancy formation}
\author{Ned Thaddeus Taylor}
\email{n.t.taylor@exeter.ac.uk}
\affiliation{Department of Physics and Astronomy, University of Exeter, Stocker Road, Exeter, EX4 4QL, United Kingdom}
\author{Michael Thomas Morgan}
\affiliation{Catalan Institute of Nanoscience and Nanotechnology (ICN2), CSIC and BIST, Campus UAB, Bellaterra, 08193 Barcelona, Spain}
\affiliation{Physics Department, Autonomous University of Barcelona (UAB),  Campus UAB, Bellaterra, 08193 Barcelona, Spain}
\affiliation{Royal Melbourne Institute of Technology, School of Engineering, GPO Box 2476, Melbourne VIC 3001, Australia}
\author{Steven Paul Hepplestone}%
\email{s.p.hepplestone@exeter.ac.uk}
\affiliation{Department of Physics and Astronomy, University of Exeter, Stocker Road, Exeter, EX4 4QL, United Kingdom}

\begin{abstract}

This work presents a first principles study of the (001) surface energetics of nine oxide perovskites, with a focus on the role of surface vacancies in determining termination stability.
Additionally, investigation into the behaviour of vacancies as a function of depth from the surface in these perovskites, \abo{} ($A$=Ca, Sr, Ba; $B$=Ti, Zr, Sn), is carried out, and results are compared to formation of the vacancies in bulk.
Combining results from these investigations reveals a general trend for all nine perovskites -- the undefected \ao{} surface is more energetically favourable to form than the \bo{} surface.
This dominance of the \ao{} over the \bo{} surface is further enforced by the phase diagrams of perovskite surfaces.
However, $A$-site vacancies at the \ao{} surface are far more favourable ($1$--$2$~\si{\electronvolt} lower in energy) than $B$-site vacancies at the \bo{} surface.
Charged vacancies only drive this further under oxygen-rich conditions, showing a smaller range of stability for the \ao{} than the \bo{} surface.
These results indicates that, whilst the \ao{} surface is easier to form, the \bo{} surface will display better long term stability, making it more suitable for use in potential applications.
This study furthers the understanding of oxide perovskite (001)-terminated surface stability, which will aid in surface growth and manufacturing.

\end{abstract}


\maketitle

\section{Introduction}

The basic chemical formula of oxide perovskites, \abo{}, can be easily customised by changing the cation sites ($A$ and $B$), which results in materials with different properties. Because of this, the material grouping has received great attention for many years in a wide range of fields, including metal-insulator transition~\cite{GonzalezRosillo2020}, two-dimensional electron gas formation~\cite{Yang2016,Eom2022,Mahatara2024EnhancedCarrierDensities}, catalysis~\cite{Zhu2014,Zhang2021}, photocatalysis~\cite{Taylor2020}, and many more~\cite{Kubicek2017,He2020a,Zuzic2022,Yoon2024PerovskiteStannateHeterojunctions}. Much of the desirability of oxide perovskites stems from application of their surfaces and interfaces, so understanding their surface properties is key to determining their potential. With different surfaces exhibiting a variety of properties, finding a way to reliably manufacture the desired surface -- in addition to assessing the stability of said surface -- is essential.

\abo{} structures tend to involve covalent bonds forming between $B$ and O atoms, which in turn form a cage around each ionically-bonded $A$-site atom~\cite{Tkaczmiech2000}.
Recent attempts to manufacture (001) perovskite oxide surfaces have highlighted the difficulties involved, with multiple experimental studies producing different pristine (001) perovskite oxide surfaces~\cite{Woo2015,Setvin2018,Sokolovi2021,Florencio2014,Wang2018,Koo2018,Bachelet2009,Druce2014}. Many attempts to form pristine surfaces have focused on using one or a combination of annealing, grinding, polishing, and cleaving. Samples terminated on the (001) plane that are produced through cleaving and post annealing contain both \ao{} and \bo{} terminations~\cite{Sokolovi2021}, whilst, for ground and polished processes, the surface area is found to be dominated by \bo{} termination ($75$--$95$~\% for \sto{} single crystalline substrates)~\cite{Florencio2014}. Further investigations have shown that, whilst \bo{} surfaces are more dominant for short surface treatment timescales and low temperatures, the \ao{} surface becomes dominant the longer it is treated and higher temperatures employed~\cite{Wang2018,Koo2018} (for example, \sto{} shows total \ao{} coverage after $72$ hours of annealing at $1300$~\si{\celsius} in air~\cite{Bachelet2009}). Additionally, it has been shown that, by employing a combination of polishing and post annealing~\cite{Druce2014}, there is an overwhelming tendency towards \ao{}-termination. Surface reconstructions of (001) terminated perovskites are another clear and present obstacle to the production of pristine surfaces. A well documented reconstruction on \sto{} is composed of TiO$_2$ structures on a TiO$_2$ bulk termination~\cite{Deak2006,Mochizuki2023}.

Surface formation in perovskites is often more nuanced than a simple comparison of surface energies.
For the (001) surfaces of oxide perovskites, both \ao{}- and \bo{}-terminations are typically considered~\cite{Eglitis2018,Li2023StabilityElectronicProperties}, as either can form depending on the chemical conditions employed~\cite{Heifets2007}.
Once exposed, the surface can evolve due to changes in chemical potential, leading to compositional shifts through processes such as degradation, adsorption, and vacancy formation~\cite{Guo1998,May2012,Han2016,Fabbri2017,Cheng2019,Lopes2021}.
Over time, and under favourable conditions, elemental diffusion tends to occur from regions of high to low concentration~\cite{Schmidt2013,Koo2018,Shin2018,Heelweg2021}, resulting in the formation of atomic vacancies as species precipitate or vaporise to form more stable compounds~\cite{Duan2018,Yang2018,Mehmood2019}.
The stability of a surface is therefore not only determined by its formation energy, but also by its resistance to vacancy formation, which ideally requires a high vacancy formation energy per unit cell relative to the formation energy of the bulk.

Surfaces may also reconstruct to lower their energy following cleavage~\cite{Erdman2003,Deak2006,Riva2018,Yue2019,Mochizuki2023}, yet predicting such reconstructions remains computationally expensive~\cite{Pickard2011,Mochizuki2023,Kempen2025InverseCatalystsTuning,Pickard2025BeyondTheoryDriven}, and
first principles searches are typically limited to small supercells.
These difficulties are compounded by the sensitivity of surface stability to the chemical potentials of constituent elements, which vary with synthesis or cleavage conditions~\cite{Koo2018}, and by the possibility of multiple reconstructions coexisting on a single surface~\cite{Ohtake2008,Sin2020}, further complicating the energetic landscape.
While global optimisation methods have been developed for surface systems~\cite{Christiansen2022AtomisticGlobalOptimization}, they often focus on structural diversity rather than thermodynamic stability.
In light of these challenges, examining the role of vacancies in \ao{}- and \bo{}-terminated slabs offers an alternative perspective, providing insight into potential degradation pathways, chemical evolution, and the mechanisms that may underpin surface reconstruction.

Recent machine-learned interatomic potentials have accelerated bulk structure searches~\cite{batatia2023FoundationModelAtomistic,Deng2023CHGNETPretrainedUniversalNeural,Pickard2025BeyondTheoryDriven}, but their transferability to surfaces remains limited, often misidentifying energetic rankings even near the ground state~\cite{Pitfield2025AugmentationUniversalPotentials}.
Fine-tuning these models for surface-specific applications is ongoing~\cite{Pitfield2025AugmentationUniversalPotentials}, though it typically requires large, tailored datasets.

In this work, we explore, via first principles, the role of vacancies on surface stability of a set of nine alkaline-earth oxide perovskites (\abo{}, where $A$=Ca, Sr, Ba and $B$=Ti, Zr, Sn). We do so by first considering the energy of formation of the two potential (001) surface terminations, \ao{} and \bo{}. We take this discussion further by exploring how the favourability of the two terminations are affected by changes in growth chemical environments. We then study the energetics of $A$-, $B$-, and O-site vacancies both in the bulk perovskites, and in slabs as a function of depth.

\section{Methods}
\label{sec:methods}

\subsection{Computational methods}

In this work, first principles techniques based on density functional theory (DFT) were used to determine the structural and energetic properties of a set of nine oxide perovskites (\cto{}, \sto{}, \bto{}, \czo{}, \szo{}, \bzo{}, \cso{}, \sso{}, and \bso{}, hereafter referred to simply as \textit{the perovskites}). These calculations were performed using the Vienna ab initio simulation package (VASP)~\cite{Kresse1996,Kresse1996a}. The valence electrons for each atomic species were considered as follows: Ca 3p$^6$ 4s$^2$, Sr  4s$^2$ 4p$^2$ 5s$^2$, Ba 5s$^2$ 5p$^6$ 6s$^2$,  Zr 4s$^2$ 4p$^2$ 5s$^2$ 4d$^2$, Ti 3p$^6$ 3d$^4$ 4s$^2$, Sn 5s$^2$ 5p$^2$, O 2s$^2$ 2p$^4$. The projector augmented wave method was used to describe the interaction between core and valence electrons, and a plane-wave basis set was used with an energy cutoff of $700$~\si{\electronvolt}.
In a heterogeneous system, the presence of multiple substructures with distinct periodic characters necessitates  higher plane-wave cutoff to resolve the more complex superposition of basis functions.
Unless otherwise stated, calculations were completed using the generalised gradient approximation Perdew-Burke-Ernzerhof (GGA-PBE) functional~\cite{Perdew1996}.
All calculations were performed with spin-polarised settings.
The PBE functional is chosen for three key reasons:
1) GGA-PBE has been extensively employed previously for oxide perovskite studies~\cite{Armiento2011,Kuklja2011,Jacobs2019,Dar2020,Taylor2020,Xiong2021,Stoch2012,Kuganathan2021}, allowing for comparison with this work, and deviations from experiment are, as such, well documented,
2) it shows reasonable geometric accuracy and self-consistency~\cite{Curnan2014,Medasani2015,Heifets2001}, and
3) in addition to prohibitive computational cost, more accurate functionals, such as HSE06~\cite{Krukau2006}, are still found to deviate from experiment, in some cases even exhibiting greater inaccuracy than PBE~\cite{Jacobs2019}, or require exchange fraction fitting~\cite{Alkauskas2011} and other times showing reliably predictable corrections~\cite{Li2022}.
For geometrical relaxations, all forces were relaxed to less than $0.01$~\si{\electronvolt/\angstrom} per atom, and electronic self-consistency is accurate to within $10^{-7}$~\si{\electronvolt}.

The crystal structures of the perovskites are as follows:  \cto{}, \czo{}, \cso{},  \szo{}, and \sso{} are all orthorhombic (space group $Pnma$),  \bto{}, and  \sto{} are tetragonal (space group $P4mm$),  and \bzo{}, and \bso{} are cubic (space group $Pm\bar{3}m$).
Initial bulk perfect crystals for the perovskites were obtained from the Materials Project database~\cite{Jain2013,mp_all}, with subsequent structural relaxations then performed using VASP (for \sto{}, the structure relaxes to a tetragonal phase from the cubic one found on the Materials Project).
Primitive cells of the orthorhombic perovskites contain four perovskite units ($A_{4}B_{4}$O$_{12}$), whereas tetragonal and cubic perovskite cells contain a single perovskite unit (\abo{}). All considerations of \textit{k}-point grids are performed using a single cubic perovskite unit as the reference, using $10 \times 10 \times 10$ Monkhorst-Pack grid~\cite{Monkhorst1976}, or equivalents.
Both the lattice constants and the atomic positions for the undefected bulk cells are relaxed (cell-free relaxation) to obtain the theoretical ground state.
Bulk vacancies are modelled in $2 \times 2 \times 2$ supercells, giving a vacancy concentration of $0.031$ and $0.125$~vac/unit (average of $4.8\times10^{20}$ and $1.8\times10^{21}$~\si{\per\centi\meter\cubed}) for orthorhombic and tetragonal/cubic crystals, respectively (a representation of the supercells is provided in
Fig.~S2~\cite{supplementary}%
).
Previous studies have shown a reasonable convergence of supercell size on vacancy formation energy for oxide perovskites~\cite{Curnan2014}.
Further details regarding the lattice constants, phases, structural setup, and calculated band gaps of the bulk materials are provided in
Sections~SII~\hspace{-0.15em}A and SIII~\hspace{-0.15em}A and
Table~SIII~\cite{supplementary}%
.
For bulk vacancy calculations, cell-free relaxation is performed (see
Section SIII~\hspace{-0.15em}B~\cite{supplementary}
for further details).

A \textit{slab} is a thin film of a material bordered by vacuum either side and is used to model surfaces in computational studies.
It consists of a sufficiently large number of atomic layers to ensure that the central layers retain bulk-like properties, while still allowing for surface effects to be studied.
In this work, the term \textit{slab} refers to the entire structure, while \textit{surface} specifically denotes the outermost atomic layer (top layer) of the slab.
The layer immediately beneath the top layer is referred to as the \textit{sub-surface}, and all deeper layers are labelled according to their layer number, as shown in \figref{fig:surface}.
For each perovskite, the (001) \ao{}- and \bo{}-terminated slabs were generated using ARTEMIS~\cite{Taylor2020a}. \ao{}-terminated (\bo-terminated) slabs include six layers of complete units of \abo{}, with an additional layer of \ao{} (\bo{}) to result in two symmetrically equivalent surfaces on the slab. A $14$~\si{\angstrom} vacuum gap is introduced along the [001] Miller direction to separate the two surfaces of a slab.
The slab surfaces are then extended to retrieve $2 \times 2$ ($\mathbf{a} \times \mathbf{b}$) extensions (for further details, see
Section~SII~\hspace{-0.15em}A~\cite{supplementary})%
. Vacancies in slabs are modelled on the $2 \times 2$ supercells, with 6 layers of perovskite units, resulting in vacancy concentrations of $0.021$ and $0.042$~vac/unit (average of $7.7\times10^{17}$ and $1.5\times10^{18}$~\si{\per\centi\meter\squared}) for orthorhombic and tetragonal/cubic crystals, respectively (a representation of the slabs is provided in
Fig.~S2~\cite{supplementary}%
).
For undefected and defected slab calculations, the cell is fixed whilst atomic positions are relaxed (fixed-cell relaxation).

In real materials, vacancy defects can adopt different charge states, which are balanced by the surrounding environment.
The influence of charge states on defect formation energies in \sso{} is investigated by explicitly modelling charged vacancies for a range of charge states.
Specifically, the following charge states are considered:
$V_{\mathrm{Sr}}$ $-3$ to $+3$,  
$V_{\mathrm{Sn}}$ $-4$ to $+3$, and
$V_{\mathrm{O}}$ $-2$ to $+2$.
Whilst, in bulk materials, these defects are known to take on particular charge states, it is unknown what effect a nearby surface will have on the available charge states and, thus, a wider range than typical is explored.

Defect calculations are performed using a fixed-cell approach, where the lattice constant is constrained to its bulk value, allowing atomic relaxations while maintaining a consistent dielectric background.
This is necessary to properly account for image charge interactions between periodic images.
The force and energy convergence criteria for charged defect calculations is consistent with those outlined above for non-charged calculations.
The respective charge neutral bulk vacancies are recalculated using fixed-cell relaxation, as required by the correction methods.
The difference between the cell-free and fixed-cell charge neutral vacancies for \sso{} is $0.03$~\si{\electronvolt}/vac.

A comparison of the formation energies of surfaces and neutral charge defects within oxide perovskites for machine-learned potentials versus DFT is presented in
Section SII~\hspace{-0.15em}G~\cite{supplementary}%
, where it is found that these models, whilst able to accurately capture the energetic ordering of their (001) surfaces, is wholly unable to capture energetic trends of bulk and near-surface vacancies.
As such, without further training, these models are not applicable to the space of vacancies in oxide perovskites.

Corrections to the formation energies of bulk charged defects are applied using \texttt{doped}~\cite{Kavanagh2024DopedPythonToolkit}, which implements the extended Freysoldt, Neugebauer, and Van de Walle (eFNV) method~\cite{Kumagai2014ElectrostaticBasedFinite} to mitigate finite-size effects arising from periodic boundary conditions.
For charged defects in slab geometries, the \texttt{qdef2d}~\cite{Tan2019ChargedDefectsFramework} and \texttt{sxdefectalign2d}~\cite{Freysoldt2009,Freysoldt2018FirstPrinciplesCalculations} packages are used to automate the application of appropriate correction schemes for systems that include vacuum regions.
The total relative dielectric values used for \sso{} are $\varepsilon_{r,xx} = 18.239$, $\varepsilon_{r,yy} = 18.850$, and $\varepsilon_{r,zz} = 16.329$, with all other values set to 0 (obtained from the Materials Project~\cite{Jain2013}.
For slab calculations, the $c$-axis is parallel to the surface normal vector, meaning that the dielectric constant along the slab is $\varepsilon_{r,zz}$.
Whilst a permittivity of a slab should be calculated directly, this has a high associated computational cost; as it is known that permittivity reduces for thin films~\cite{Zaccone2024ExplainingThicknessDependent}, the choice of the bulk dielectric values can be taken as the upper limit.
The method employed for studying image charges and the associated correction methods are discussed exhaustively in Refs.~\cite{Kumagai2014ElectrostaticBasedFinite,Freysoldt2018FirstPrinciplesCalculations,Freysoldt2022LimitationsEmpiricalSupercell}.

To address inaccuracies in band gap and band alignment inherent to the GGA-PBE functional, additional calculations are performed using the HSE06 hybrid functional~\cite{Krukau2006} in a \textit{one-shot} approach.
For the undefected bulk and \ao{}- and \bo{}-terminated slabs of \sso{}, a single self-consistent calculation is carried out on GGA-PBE-relaxed structures to obtain more accurate electronic and energetic properties.
These refined values are then incorporated into the defect calculations using the averaged electrostatic potential-referenced band edge shift (BES) method~\cite{Broberg2023HighThroughputCalculations} to further improve accuracy.
Whilst HSE06 does not accurately capture the experimental band gap of \sso{}, it is an improvement from GGA-PBE, thus providing more qualitatively meaningful results.

\subsection{Surface formation energy}

In order to compare the energetics of different surface terminations, we require a surface formation energy for each. Formation energies are used to compare the energetic favourability of elements in different compounds or phases. For this work, we consider two approaches for defining energetic surface stability. The first is for simple comparisons based upon taking the relevant binary oxides (\ao{} and \bo{}) as references. The second approach uses the generally more in-depth evaluation involving chemical potentials. For the first, we use the \abo{} perovskite, and the \ao{} and \bo{} binary oxides as comparison compounds, due to their stability and favourability in oxygen-rich environments~\cite{Diebold2010,Pujari2014} (see
Table~SIV~\cite{supplementary}%
). Energy values, $E_{\textrm{X}}$, denoted here correspond to a DFT calculated total energy of one chemical unit of the X compound. Finally, for surfaces, formation energy is given per unit surface area, which makes the value independent of surface supercell employed.

The equation used to calculate surface formation energy for a particular slab is

\begin{equation}
        E_{\mathrm{form}}^{i}=\frac{\left[ E_\textrm{slab}^{i} - N_\textrm{\abo{}}^{i} E_\textrm{\abo{}}^\textrm{bulk} - N_{i\textrm{-ox}} E_{i\textrm{-ox}}^\textrm{bulk}\right]}{2A},
\label{eq:form:surf}
\end{equation}

\noindent
where $E_\textrm{\abo{}}^\textrm{bulk}$ is the energy of the bulk perovskite, $E_\textrm{slab}^{i}$ is the energy of the perovskite slab with surface termination $i$, $N_\textrm{\abo{}}^{i}$ is the number of complete units of \abo{} in the slab, and $E_{i\textrm{-ox}}^\textrm{bulk}$ and $N_{i\textrm{-ox}}$ are the energy and number of additional units of the $i$-corresponding binary oxide (\ao{} or \bo{}). $A$ is the area of the surface present in the cell (an additional $2$ is included to account for there being two symmetrically equivalent surfaces in the cell).

\subsection{Surface Gibbs free energy}

Whilst \eqref{eq:form:surf} provides an easy comparison of the data, it does not take into consideration the chemical environment in which the surface is formed. To account for this surface Gibbs free energy must be used. The method used to assess surface Gibbs free energy follows the method outlined by Heifets \etal{}~\cite{Heifets2007} and is provided in further detailed in
Section~SII~\hspace{-0.15em}C~\cite{supplementary}%
.

We can identify the most energetically favourable surface at different chemical conditions (defined by the chemical potentials, $\mu$) as the one with the lowest surface Gibbs free energy. The surface Gibbs free energy per unit cell for an $i$-terminated slab, $\Omega^{i}$, is defined as

\begin{equation}
    \Omega^{i}=\frac{1}{2}\left[G_\textrm{slab}^{i}-N_{A}^{i} \mu_{A}-N_{B}^{i} \mu_{B}-N_\textrm{O}^{i} \mu_\textrm{O}\right],
\label{eq:gibbs:surface}
\end{equation}

\noindent
where $G_\textrm{slab}^{i}$ is the Gibbs free energy of the slab, $N_{A}^{i}$, $N_{B}^{i}$, $N_\textrm{O}^{i}$ are the number of $A$, $B$, and O atoms in the slab respectively, and $\mu_{A}$, $\mu_{B}$, $\mu_\textrm{O}$ are the chemical potentials of the $A$, $B$, and O species respectively. Halving the first term accounts for the two symmetrically equivalent surfaces present in the slab.

As the system is in equilibrium with the bulk \abo{}, we can show that the chemical potential of the perovskite is interdependent on the chemical potential of its three constituent elements

\begin{equation}
    \mu_\textrm{\abo{}} = \mu_{A} + \mu_{B} + 3 \mu_\textrm{O}.
    \label{eq:gibbs:chempot}
\end{equation}

\noindent
We approximate the chemical potential of the perovskite to its Gibbs free energy ($G_\textrm{\abo{}}^\textrm{bulk}$), which we define as the DFT energy relating to one chemical unit of \abo{}.

\begin{equation}
    \mu_\textrm{\abo{}} \approx G_\textrm{\abo{}}^\textrm{bulk} \approx E_\textrm{\abo{}}^\textrm{bulk},
    \label{eq:chempot:approx}
\end{equation}

\noindent
As $\mu_\textrm{\abo{}}$ is considered a fixed value, and the chemical potentials (e.g relative concentrations, temperatures) of $A$, $B$, and O are interdependent, there exist only two truly independent variables -- we choose these to be those for $A$ and O. We choose to redefine the number of $B$ atoms in terms of a difference in its number compared to $A$ and O atoms (denoted using X),

\begin{equation}
    \Gamma_{{B}, X}^{i}=\frac{1}{2}\left(N_{X}^{i}-N_{B}^{i} \frac{N_{X}^\textrm{bulk}}{N_{B}^\textrm{bulk}}\right).
\label{eq:gibbs:gamma}
\end{equation}

\noindent
$\Gamma_{{B},\textrm{X}}^{i}$ is the number of excess atoms of species X, in $i$-terminated slab, with respect to the number of atoms of species $B$. These $\Gamma$-factors allow us remove the $\Delta\mu_{B}$ dependency from surface Gibbs free energy $\Omega^i$, allowing us to more clearly consider the effects of different growth and environmental conditions.

Next, we shift the zero of the chemical potentials of species $A$ and O ($\mu_\textrm{X}$) to their bulk energies ($\Delta\mu_\textrm{X}$) which are $A$ bulk and \o{} molecule, respectively. Now we can re-postulate the surface Gibbs free energy from \eqref{eq:gibbs:surface} in terms of merely the chemical potential of the two species $A$ and O, with that of species $B$ being wholly dependent on $A$ and O.

\begin{equation}
    \Omega^{i}=\phi^{i}-\Gamma_{{B}, {A}}^{i} \Delta \mu_{A}-\Gamma_{{B}, \textrm{O}}^{i} \Delta \mu_\textrm{O},
    \label{eq:gibbs:surface:reduced:3}
\end{equation}

\noindent
where

\begin{equation}
\phi^{i} \approx \frac{1}{2}\left[E_\textrm{slab}^{i}-N_{B}^{i} E_\textrm{\abo{}}^\textrm{bulk}\right]-\Gamma_{{B}, {A}}^{i} E_{A}^\textrm{bulk}-\frac{1}{2} \Gamma_{{B}, \textrm{O}}^{i} E_{\textrm{O}_{2}},
\label{eq:gibbs:surface:zero}
\end{equation}

\noindent
and $E_{A}^\textrm{bulk}$ ($E_{\textrm{O}_{2}}$) is the energy of an individual $A$ atom in its bulk (individual O atom in an \o{} molecule). We can now vary the relative chemical potentials, $\Delta\mu_{A}$ and $\Delta\mu_\textrm{O}$, to find the formation energy of each material at different environmental conditions; for a set of chemical potentials ($\Delta\mu_{A}$ and $\Delta\mu_\textrm{O}$) the material with the lowest formation energy, $\Omega^i$, is the most energetically favourable material at that chemical condition.

\subsection{Vacancy formation energy}

To calculate the formation energy of a vacancy in the bulk, we use the binary oxides as the relative compounds for comparison (i.e. determine whether the vacant atom would prefer to reside in the perovskite bulk or in its corresponding binary oxide),

\begin{equation}
    E_{\mathrm{form},\textrm{X}}^\textrm{bulk} = \left( E_\textrm{X-vac}^\textrm{bulk} + E_\textrm{X-ox}^\textrm{bulk} \right) - \left( N_\textrm{cells} E_\textrm{\abo{}}^\textrm{bulk} + \frac{N_\textrm{O}^\textrm{X}}{2} E_\textrm{\o{}} \right).
\label{eq:form:vac:bulk}
\end{equation}

\begin{figure*}[ht]
    \centering
    \subfloat[\ao{} surface]{\fbox{\includegraphics[scale=0.13]{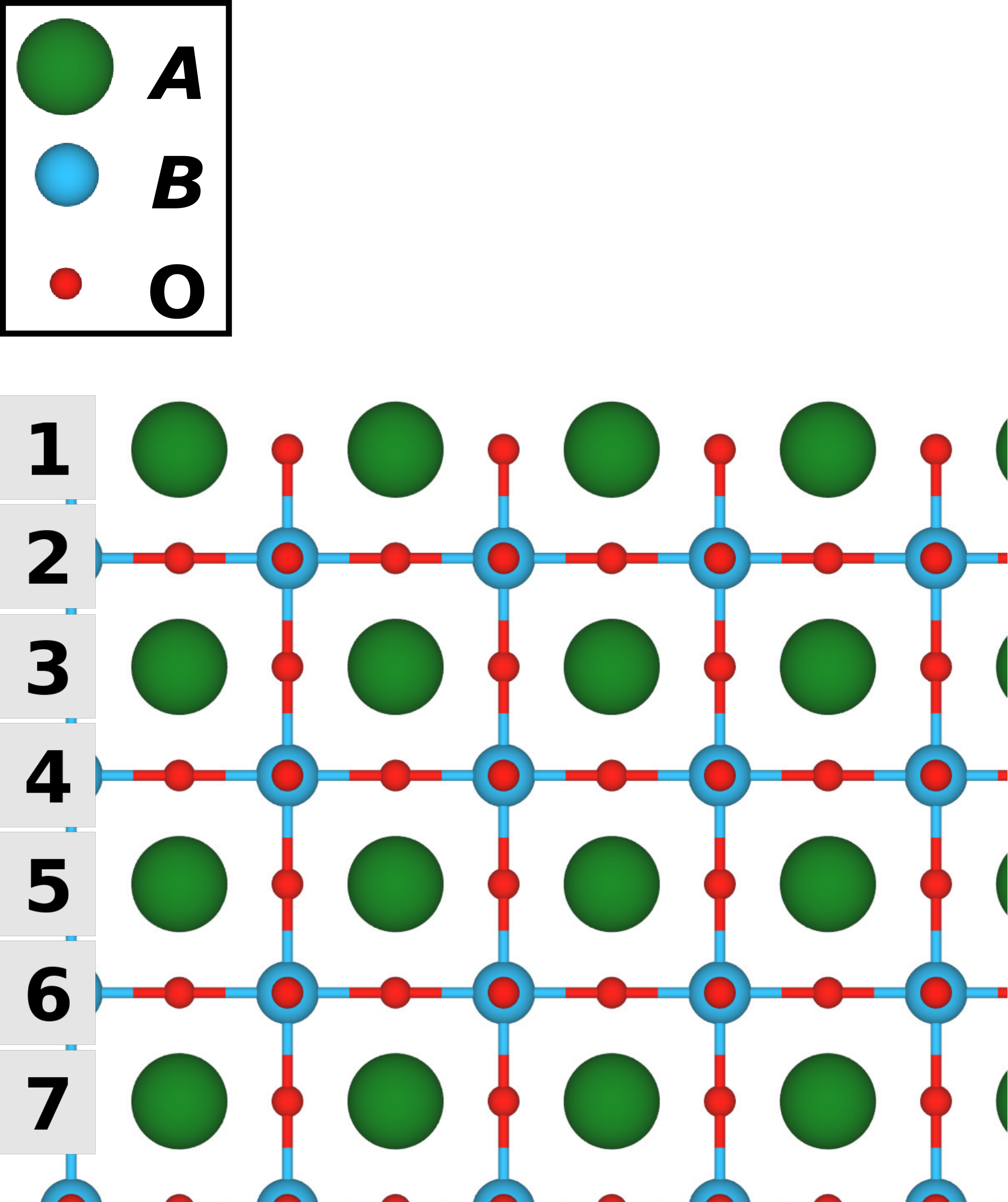}}\label{fig:surface:AO}}\hspace{2em}%
    \subfloat[\bo{} surface]{\fbox{\includegraphics[scale=0.13]{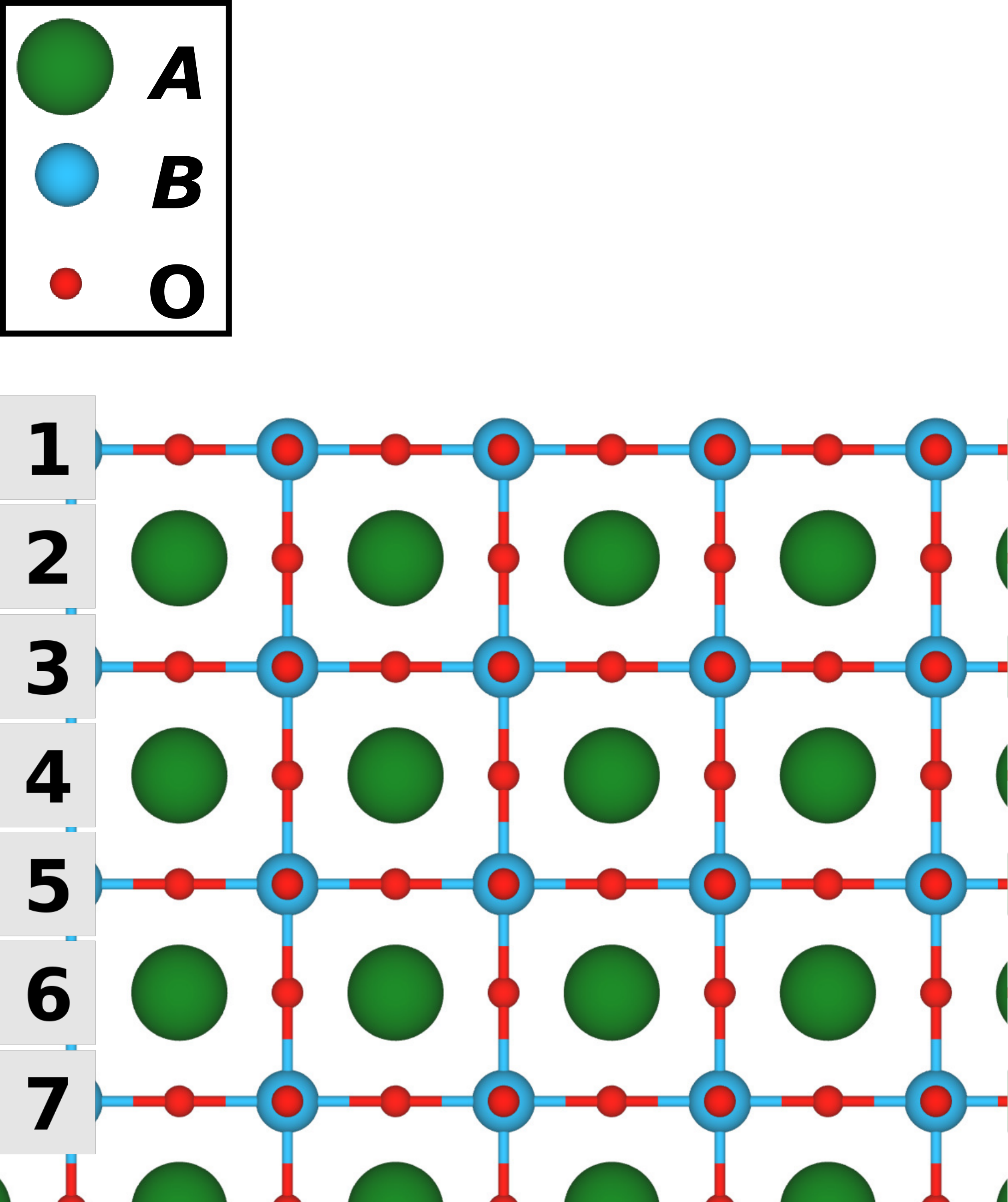}}\label{fig:surface:BO2}}%
    \caption{
        Ball-and-stick representation of the two (001) surface terminations of a tetragonal (and cubic) oxide perovskite: \protect\subref{fig:surface:AO} \ao{} and \protect\subref{fig:surface:BO2} \bo{}.
        Whilst the orthorhombic phase exhibits slight tilting, these two terminations are the only unique (001) surfaces, with minor distortions to the structure.
        Layer 1 is defined as the \textit{surface} layer, layer 2 is the \textit{sub-surface} layer, and all other layers are referred to using their respective numbers as labelled in the figure.
        Figure generated using VESTA~\cite{Momma2011}.        
    }
    \label{fig:surface}
\end{figure*}

\noindent
Here, $E_\textrm{X-vac}^\textrm{bulk}$ is the energy of the vacated perovskite bulk supercell, $E_\textrm{X-ox}^\textrm{bulk}$ is the energy of the $X$-based binary oxide, $N_\textrm{cells}$ is the number of perovskite units in the undefected supercell, $E_\textrm{\o{}}$ is the energy of an isolated \o{} molecule, and $N_\textrm{O}^\textrm{X}$ is the number of oxygen atoms in the chemical unit of the X-based binary oxide.

Formation energies for vacancies in the slab are given with respect to the corresponding bulk vacancy,

\begin{equation}
    E_{\mathrm{form},\textrm{X}}^{i} = \left( E_\textrm{X-vac}^{i} + N_\textrm{cells} E_\textrm{\abo{}}^\textrm{bulk} \right) - \left( E_\textrm{slab}^{i} + E_\textrm{X-vac}^\textrm{bulk} \right),
    \label{eq:form:vac:slab}
\end{equation}

\noindent
where $E_\textrm{X-vac}^{i}$ is the energy of the vacated slab with surface termination $i$. Clearly, \eqref{eq:form:vac:bulk} depends on choice of chemical potentials. However, for vacancies in slabs, we are more interested in the favourability of a vacancy as a function of depth and, as such, \eqref{eq:form:vac:slab} allows us to focus solely on this, avoiding the issue of an arbitrary choice of chemical potential.

For the nine perovskites, there exists one symmetrically unique $A$-site and one $B$-site.
For the tetragonal and cubic perovskites, there exists one symmetrically unique O-site.
However, for the orthorhombic perovskites, there exist two inequivalent O-sites, the $4c$ and $8d$ Wyckoff positions.
Vacancies have been explored in the bulk on both sites, but only the most energetically favourable site is presented in this work (see
Section SIII~\hspace{-0.15em}B~\cite{supplementary}
for further details).
For the slab vacancies, all symmetrically inequivalent sites are explored for all vacancies.

\section{Results and Discussion}

\subsection{Clean (001) surfaces}

\begin{figure*}[ht]
    \centering
    \subfloat[Surface formation energy]{\includegraphics[width=0.44\linewidth]{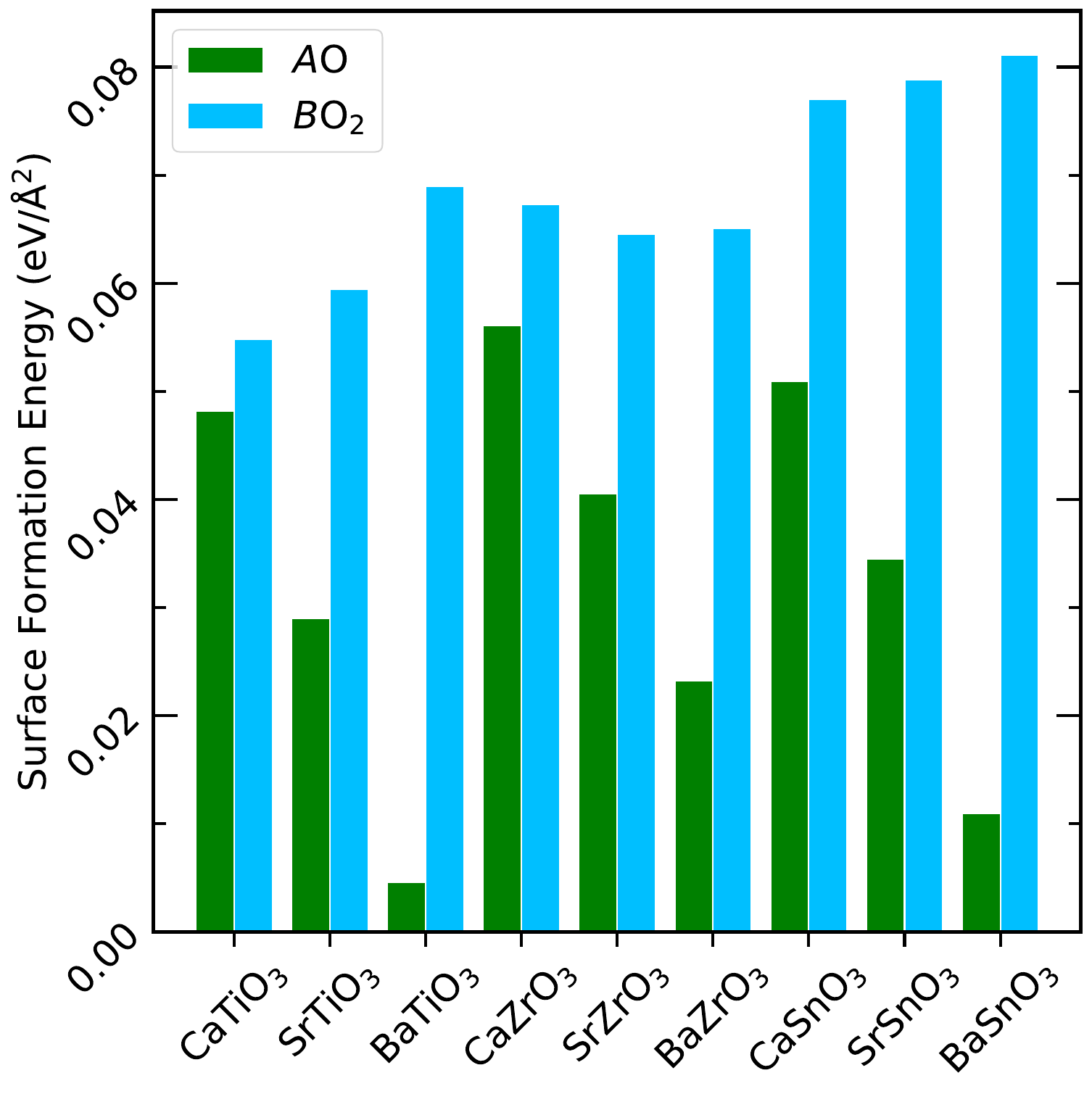}\label{fig:form:slab}}%
    \hspace{1em}%
    \subfloat[Chemical potential plot of surface favourability]{\includegraphics[width=0.515\linewidth]{BaTiO3_chempot}\label{fig:form:chem}}%
    \caption{The surface formation energy for nine oxide perovskites (\abo{}) terminated on the (001) plane with either their \ao{} and \bo{} surface. The formation energies for the \ao{}- and \bo{}-terminated slabs are given with respect to the composite binary oxide, \ao{} and \bo{}, respectively. \protect\subref{fig:form:chem} chemical potential plot for \bto{}, following the method outlined in \cite{Heifets2007}. The chemical potential values of Ba and O are varied, with that of Ti being inherently linked to the others by the chemical formula \bto{}. Note, the surfaces are never more favourable than the undefected bulks, so denote where the surfaces would form if the perfect crystal did not exist using a hashed region. The \bo{}-termination does not appear on the chemical potential plot as it is never more favourable than the other compounds present.}
    \label{fig:form}
\end{figure*}

To understand perovskite surfaces, we first explore clean slabs terminated on the (001) Miller plane (see \figref{fig:surface} for a representation of the two (001) surface terminations, \ao{} and \bo{}). We examine the surface formation energy, which allow us to understand the relative stability of each surface. Higher formation energies correspond to a less energetically favourable surface whilst lower formation energies correspond to a more favourable or more easily formed surface. A negative surface formation energy indicates that spontaneous surface formation occurs and the bulk material is energetically unstable. The formation energies for the two potential (001) surface terminations (\ao{}- and \bo{}-terminated) of the nine oxide perovskites are shown in \figref{fig:form:slab}; here, the values are given relative to composite binary oxides (\ao{} and \bo{}) and the bulk perovskite (\abo{}) are presented in \figref{fig:form:slab}. All the surfaces formation energies are positive (see \figref{fig:form:slab}), which highlights that that spontaneous formation of (001) surfaces does not occur and, as such, the bulk systems can be considered energetically stable.

We find that, in all perovskites considered, the \ao{} surface is more energetically favourable than their corresponding \bo{} surface -- this agrees with experimental works that have shown a preference for \sto{} and 3,3 perovskites to form the \ao{} surface when first grown~\cite{Koo2018,Druce2014}. From considering both
Fig.~S4~\cite{supplementary}
and \figref{fig:form:slab}, it is clear that the phase of the perovskite, is unrelated to which surface is more favourable. For Ca-based oxide perovskites (\cbo{}), \ao{} and \bo{} surfaces are the closest in formation energies. As we change period of the $A$-site ion, the formation energies of the \ao{} and \bo{} surfaces diverge, with Ba-based oxide perovskites (\bbo{}) exhibiting the largest differences. These two surface types show opposite trends for their formation energies. For \ao{} surfaces within a particular $B$-ion (e.g. group IV perovskites, \ato{}), the formation energy decreases as the $A$-ion increases in period. For \bo{} surfaces, the formation energy either remains relatively constant, or increases slightly with increasing period of the $A$-site ion within group II.

We also perform a study of the favourability of the two surfaces for different chemical environments and find that the \ao{}-termination preference holds for almost all environments (i.e. different temperatures, pressures, and elemental concentrations). \Figref{fig:form:chem}  shows that, for \bto{}, the \ao{} surface is the only (001) termination to be more stable than its composite binary oxide compounds for any set of chemical potential values (all nine perovskites are presented in
Fig.~S9~\cite{supplementary}
and generally hold to this trend); this agrees with what has previously been seen for \bto{} and \bzo{}~\cite{Fredrickson2013,Heifets2007}. This is of particular interest for systems such as \sto{}, which have shown both surfaces experimentally~\cite{Sokolovi2021}. For further details regarding the generation of these chemical potential plots, see Section~SIII~\hspace{-0.15em}F~\cite{supplementary}.

The universal energetic favourability of \ao{} surfaces over the corresponding \bo{} surface is ascribed to a more significant change in surface charge in the \ao{} surface than the \bo{}. The charges on the oxygen atoms at the \ao{} surface are found to lie closer to their ideal oxidation number of $-2$ than at the \bo{} surface. The oxidation number of $-2$ represents a full outer most orbital on the oxygen atoms. Therefore oxidation numbers closer to $-2$ are more energetically favourable than numbers closer to $-1$, such as on the \bo{}-terminated surface.

We suggest that the general trend of decreasing surface formation energy with higher $A$-site atomic number can be attributed to the decreasing first and second ionisation energy with increasing period (see Fig.~S1~\cite{supplementary}). As the $A$-site cation period increases, the shielding and distance of the electrons with respect to the nucleus also increases, resulting in a lower ionisation energy than elements with a smaller period in the same group. Although the total amount of charge redistributed from each $A$-site ion is roughly equal, the energetic cost of removing these electrons from an $A$-site decreases with increasing period, and, as such, they are energetically cheaper to donate.

We compare the energies of perovskites with the same $A$-site cation, but with different $B$-site ions. For \ao{}-terminated slabs, we find similar formation energies between them. For example, \cbo{} formulae have an \ao{} surface formation energy of approximately $0.051\pm{}0.003$~\si{\electronvolt/\angstrom\squared}, \sbo{} have an \ao{} surface formation energy of roughly $0.036\pm{}0.007$~\si{\electronvolt/\angstrom\squared}, and \bbo{} have \ao{} surface formation energies of roughly $0.013\pm{}0.009$~\si{\electronvolt/\angstrom\squared}.

Overall, we see little change in formation energy for \ao{}-terminated slabs when the $B$-site cation is changed, and similarly, little change in formation energy for \bo{}-terminated slabs when the $A$-site cation changes period. This is not unexpected as the surface cation should be the dominant contributor to the change in surface formation energy. In addition, we suggest that the reason for the titanate \bo{} surface formation energy increasing with respect to increasing $A$-site period is that the Ti ion has significantly fewer core electrons than Zr and Sn. This in turn means that it has less core screening. As such, changing $A$-site cation more strongly affects the bonding of Ti-O than it would for Zr-O and Sn-O.

\subsection{Neutral vacancies in the bulk}
\label{sec:bulk_vacancies:neutral}

\begin{figure}[ht]
\includegraphics[width=8cm]{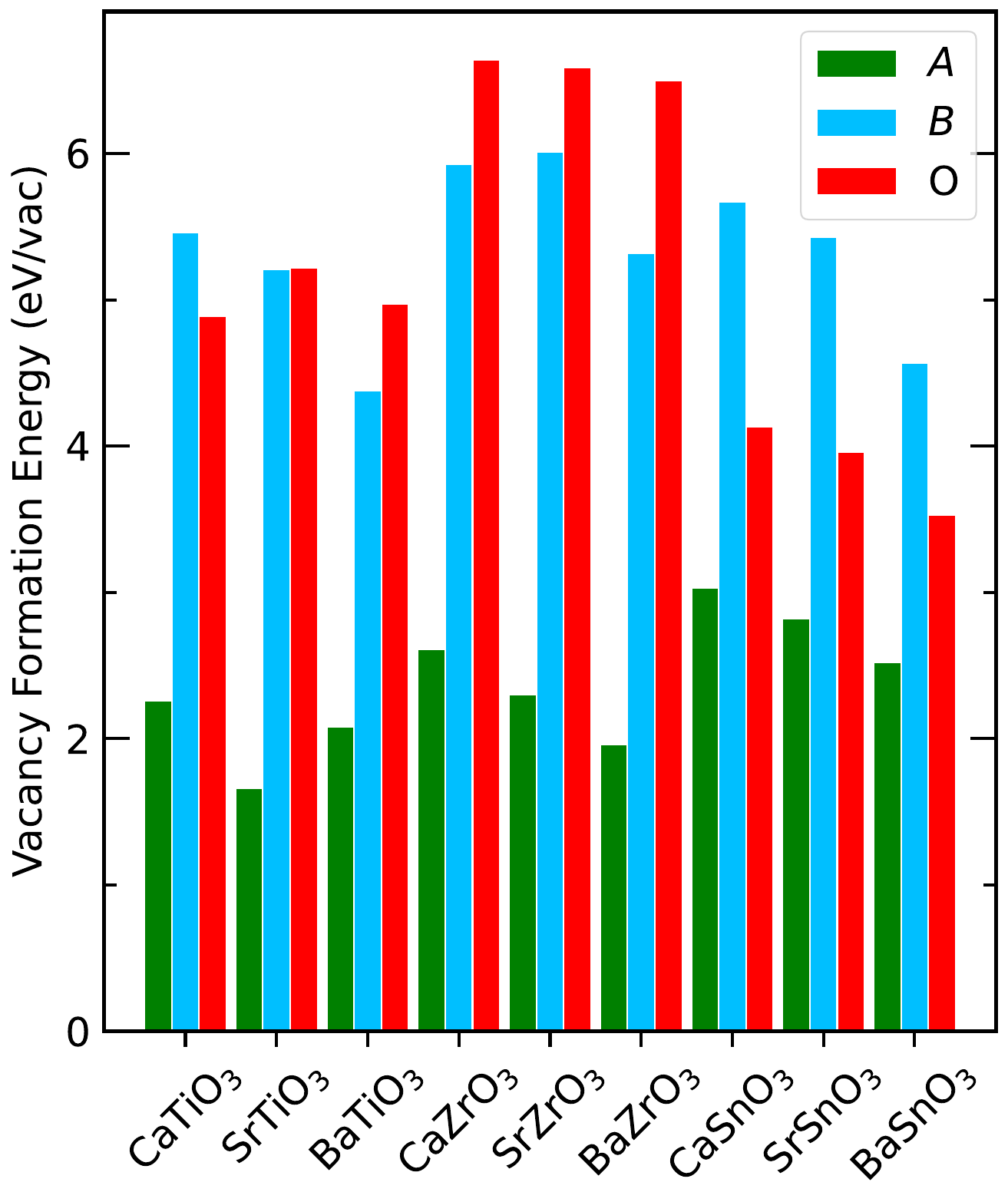}
\caption{
    The formation energy of different $A$-, $B$- and O-site vacancies in various oxide perovskite (\abo{}) bulk crystals (evaluated in the O-rich limit, see \tabref{tab:chem_pots:o-rich_limit}).
    For the orthorhombic-phase perovskites, the O-site corresponding to the lowest energy vacancy is presented; for \cto{}, \szo{}, and \cso{}, this is the $4c$ Wyckoff position, whilst for \czo{} and \sso{}, it is the $8d$ position.
}
\label{fig:vac:bulk}
\end{figure}

\begin{table}[]
    \centering
    \begin{tabular}{cccc}
    \hline
        \multirow{2}{*}{O-rich limit} & \multicolumn{3}{c}{Chemical potential (\si{\electronvolt})} \\\cmidrule(lr){2-4}
         & $A$-site & $B$-site & O-site \\\hline
         \cto{}-TiO$_2$-\o{} & -6.62 & -9.20 & 0.00\\
         \sto{}-TiO$_2$-\o{} & -6.59 & -9.20 & 0.00\\
         \bto{}-TiO$_2$-\o{} & -6.21 & -9.20 & 0.00\\\hline
         
         \czo{}-ZrO$_2$-\o{} & -6.29 & -10.37 & 0.00\\
         \szo{}-ZrO$_2$-\o{} & -6.24 & -10.37 & 0.00\\
         \bzo{}-ZrO$_2$-\o{} & -6.12 & -10.37 & 0.00\\\hline
         
         \cso{}-SnO$_2$-\o{} & -6.33 &  -5.10 & 0.00\\
         \sso{}-SnO$_2$-\o{} & -6.18 &  -5.10 & 0.00\\
         \bso{}-SnO$_2$-\o{} & -5.93 &  -5.10 & 0.00\\\hline
         
    \end{tabular}
    \caption{
    Chemical potential limits for the oxygen-rich environment for each oxide perovskite.
    The chemical potential values are given with respect to the elemental reference phases, i.e. $A$ and $B$ bulk metals and \o{} gas.
    }
    \label{tab:chem_pots:o-rich_limit}
\end{table}

Before considering the role of vacancies at a surface, we first explore the energetics of $A$, $B$, and O vacancies in bulk.

\figref{fig:vac:bulk} shows the energy of formation of $A$-, $B$-, and O-site vacancies in bulk, given in an O-rich environment (i.e. the \abo{}-\bo{}-\o{} limit); this choice is made due to the abundance of oxygen within the atmosphere and, thus, the likely condition to find these perovskites in.
The values for the chemical potentials used are presented in \tabref{tab:chem_pots:o-rich_limit}.
In Fig.~S4~\cite{supplementary}, we also considered these vacancies with respect to the bulk binary oxides and the bulk metals.
Under oxygen-rich conditions, the $A$-site vacancies are always the most favourable vacancy.
$B$-site and O-site vacancies are similar in energies, with O-site vacancies being more favourable in \aso{}, less favourable in \azo{}, and comparable energies in \ato{}.
This ordering can be changed based on the choice of comparison compounds (such as comparing to the $A$ and $B$ bulks or their binary oxides).
Note, when considering these vacancies against forming the bulk metal, the O-site vacancies becomes more favourable, but these energies are substantially higher than with respect to the oxides.
When comparing to the binary oxides, the $A$-site vacancy is always found to be the most favourable vacancy, $B$-site the second, and $O$-site the least favourable vacancy.

Our results compare well with literature values for titanate-based perovskites~\cite{Choi2011,Brown2018,Xia2019} -- our results and literature both show the formation energies of bulk (or deep level) O-site vacancies in titanate-based perovskites to fall within the range of 5--6~\si{\electronvolt}.

In general, the vacancy formation energy results indicate that $A$-site ions are more likely to vacate the bulk under O-rich conditions than the $B$-site ions.

Our results show that, within a certain $B$-based compound (i.e. titanates, zirconates, or stannates), both all vacancy sites exhibit a decreasing formation energy with increasing $A$-site period (with the exception of \sto{} for $A$-sites, \szo{} for $B$-sites, and \cto{} for O-sites).

All vacancies show similar comparability between $B$-site groupings, with the exception of oxygen vacancies in zirconates.
Within materials of the same $A$-site, stannates show the smallest variation in vacancy formation energy of different defects ($\sim3.1$~\si{\electronvolt}).
Titanates show a larger variation in formation energy ($\sim3.8$~\si{\electronvolt}), and zirconates show the largest variation in vacancy formation energy ($\sim4.6$~\si{\electronvolt}).
The oxygen vacancy formation energy of zirconates is about $2$~\si{\electronvolt} higher than in the stannates or titanates, where all other vacancy energies are more consistent.

\subsection{Charged vacancies in bulk \sso{}}
\label{sec:bulk_vacancies:charged}

\begin{figure}[ht]
    \centering
   \includegraphics[width=1.0\linewidth]{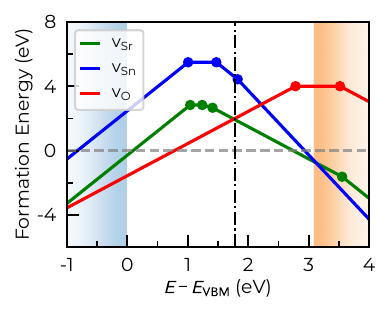}%
    \caption{
        Formation energies of $A$-, $B$-, and O-site charged vacancies in bulk \sso{} (evaluated in the O-rich limit, see \tabref{tab:chem_pots:o-rich_limit}).
        Line gradients correspond to the charge state (in $|$\si{\elementarycharge}$|$) and points of discontinuity identify transition levels.
        The valence (conduction) band region is shaded in blue (orange) and the vertical dash-dotted line is the equilibrium Fermi level.
        Corrections to the charge state energies are calculated using \texttt{doped}~\cite{Kavanagh2024DopedPythonToolkit}, which employs band edge shifting~\cite{Broberg2023HighThroughputCalculations} and image charge corrections~\cite{Kumagai2014ElectrostaticBasedFinite}.
    }
    \label{fig:charge_state:bulk}
\end{figure}

The transition levels of charged vacancies for \sso{} are presented in \figref{fig:charge_state:bulk} for the O-rich limit (\sso{}-\o{}-SrO).
\sso{} is chosen as the example oxide perovskite for charge state studies for three reasons:
1) prior literature studies of charged defects in bulk~\cite{Weston2018,Lin2024TraceImpurityMatters},
2) it shows the least structural distortion brought on by introducing vacancies (see 
Section SIII~\hspace{-0.15em}H~\cite{supplementary}%
), and
3) it shows reasonable convergence of vacancy energies to their bulk counterparts in both surface terminations (see \figref{fig:vac:slab:sn}).
The theoretical band gap for \sso{} bulk is calculated as $3.09$~\si{\electronvolt} using HSE06.

The charge-neutral defect formation energies align with those presented in \figref{fig:vac:bulk}.
Additional plots are presented in 
Fig. S10~\cite{supplementary}
for different chemical potential limits.
Under the same chemical environment (see 
Fig. S6~\cite{supplementary}
, the charged defects agree with those presented in Ref.~\cite{Weston2018}; the band gap is in disagreement due to the choice of mixing employed by the hybrid functional (this study uses the default parameters of HSE06).

The O ($+2$/0) is a shallow defect, along with O (0/$-2$) and Sr ($-2$/$-3$) that fall above the conduction band minimum.
The other Sr and Sn transitions are deep level transitions, sitting near the middle of the band gap.
In this oxygen-rich limit, the pinned Fermi level sits at $1.78$~\si{\electronvolt} above the valence band maximum (VMB); see 
Table SXII~\cite{supplementary}
for Fermi level under different chemical conditions.
The formation energy of the charged defects lie above $0$~\si{\electronvolt} for most of the band gap, i.e. $0.5$--$2.5$~\si{\electronvolt}.

The dopability of \sso{}, in agreement with literature~\cite{Weston2018,Lin2024TraceImpurityMatters} (and other oxide perovskites~\cite{Willis2023PossibilityPType}), is found to be $n$-type dominated with oxygen vacancies under the SnO-\sso{}-Sn limit than $p$-type oxygen vacancies under O-rich conditions (see 
Section SIII~\hspace{-0.15em}I~\cite{supplementary}%
).
This is the case because a higher formation energy for defects at the band edges suggests a system that is more stable to those vacancies, so can support more.
The dopability limit in the O-rich condition lies from $0.50$--$2.47$~\si{\electronvolt}, showing a range for the Fermi level of $1.97$~\si{\electronvolt}.

\subsection{Neutral vacancies at surfaces and in slabs}
\label{sec:slab_vacancies:neutral}

\begin{figure*}
    \subfloat[]{\includegraphics[height=0.6\textheight,keepaspectratio]{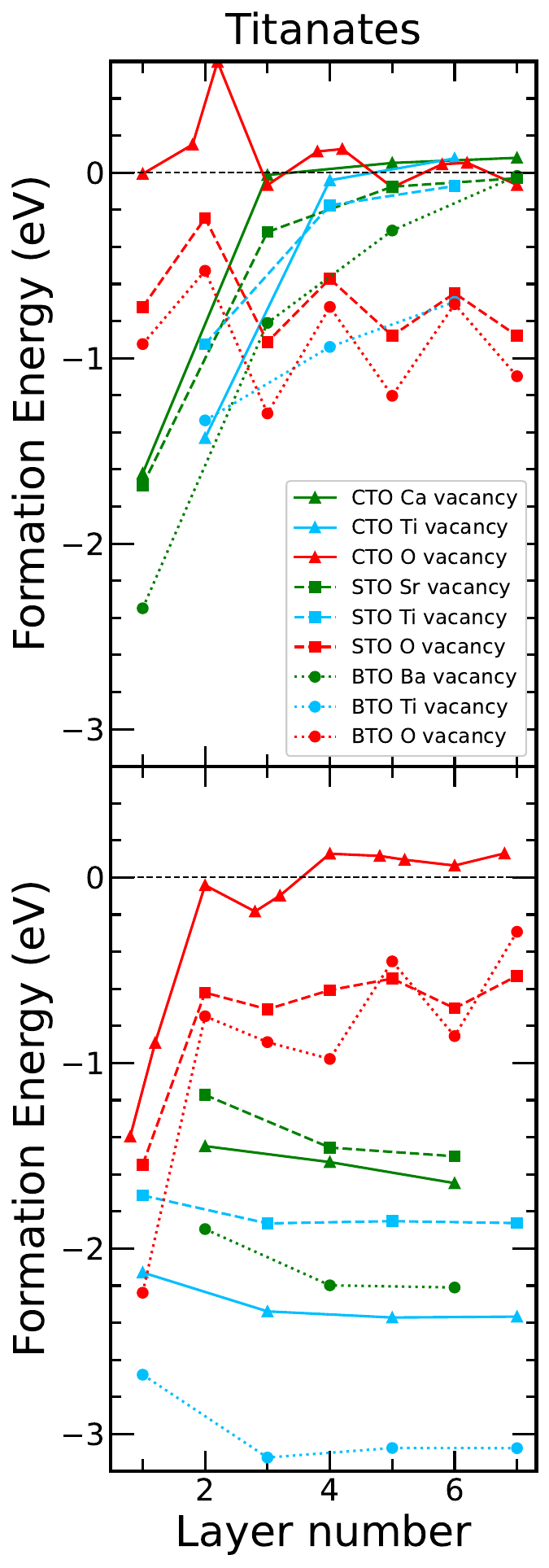}\label{fig:vac:slab:ti}}\hspace{0em}%
    \subfloat[]{\includegraphics[height=0.6\textheight,keepaspectratio]{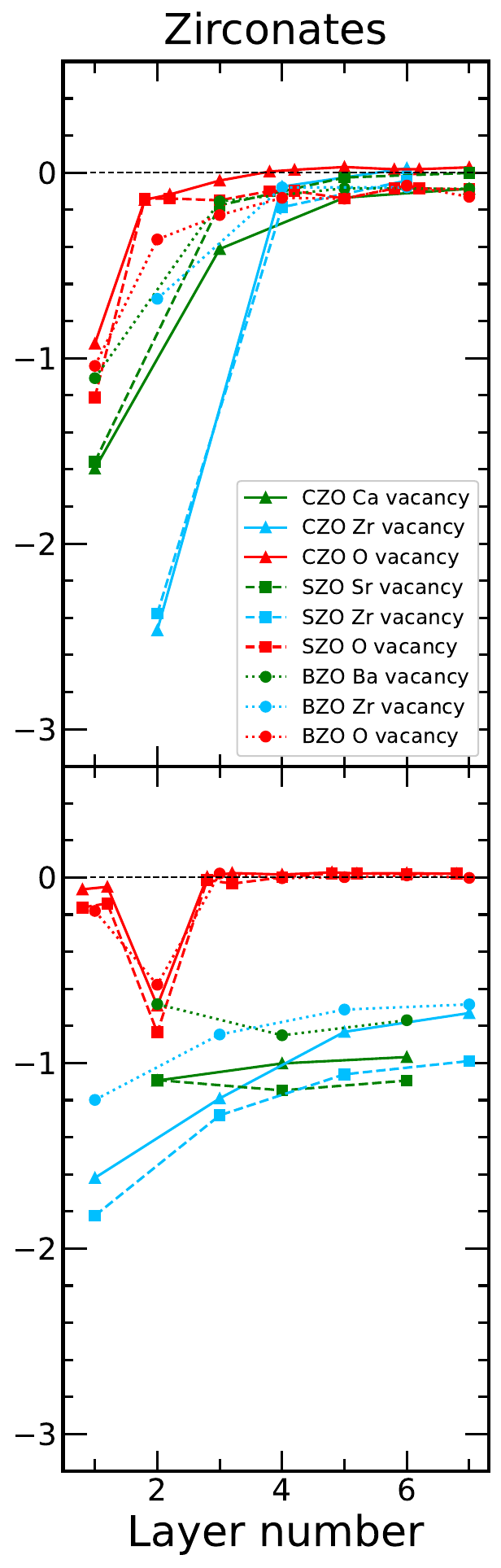}\label{fig:vac:slab:zr}}\hspace{0em}%
    \subfloat[]{\includegraphics[height=0.6\textheight,keepaspectratio]{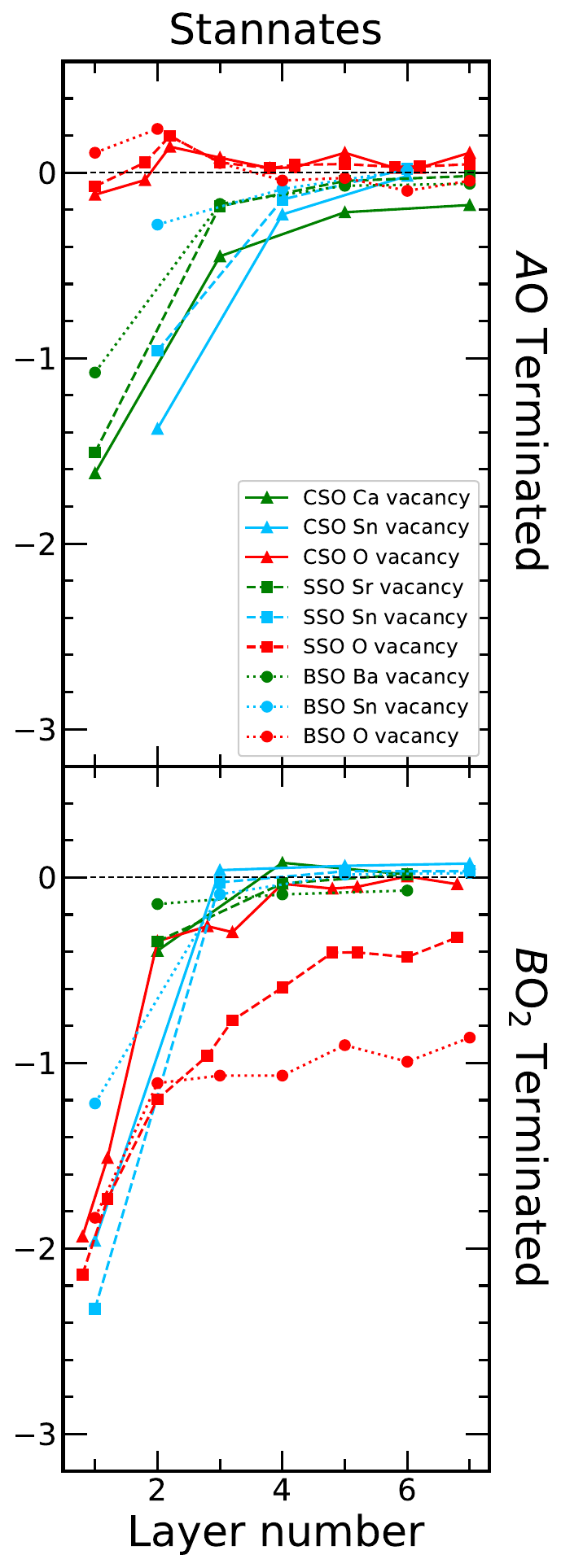}\label{fig:vac:slab:sn}}%
    \caption{Formation energies of $A$-, $B$- and, O-site vacancies within \protect\subref{fig:vac:slab:ti} titanate, \protect\subref{fig:vac:slab:zr} zirconate, and \protect\subref{fig:vac:slab:sn} stannate oxide perovskites, as a function of layer depth from the surface. The upper (lower) panels show vacancies in \ao{}- (\bo{}-) terminated slabs. Vacancy formation energies are given with respect to the vacancy formation energies in bulk, as seen in \figref{fig:vac:bulk}. For reference, we note that the $A$ vacancies in bulk are generally much more energetically favourable than the $B$ vacancies in bulk. For systems that exhibit a tilt (i.e. orthorhombic phases), there exist two unique oxygen sites in the \bo{} layers -- one lying slightly above the $B$-site plane, and one slightly below. To distinguish between these two O-sites, they have been presented at $0.2$ above and below the corresponding $B$ layer number.}
    \label{fig:vac:slab}
\end{figure*}

We now evaluate the role of vacancies in slab structures.
For clean surfaces, we have shown that, based on arguments for the surface formation energies for all the unvacated slabs, the \bo{} surfaces are highly unfavourable and unlikely to occur.

We have investigated, for the nine oxide perovskite considered here, the role of $A$-, $B$- and O-site vacancies as a function of depth from the surface (i.e. layer number).
In \figref{fig:vac:slab}, we show the formation energy of the various vacancies considered.  The vacancy formation energies presented show how much more (or less) energetically favourable these surface vacancies are compared to the bulk. Our results show  that the vacancies in the surface layer are more favourable than in bulk, in agreement with previous first principles studies that have observed the same surface preference of vacancies in \ato{} perovskites and metal~\cite{Zhu2015,Brown2018}.  This trend indicates that for all the perovskites, one can expect higher concentrations of vacancies at the surface than in the bulk.  Our results also agree with the expectation that the vacancy formation energies (with respect to the bulk vacancy) tend towards zero as they approach the deepest point in our simulated slabs (i.e. tend to the bulk vacancy result).  In general, this  trend of vacancy formation energy converging to the bulk value is followed by \ao{} surfaces  more so than \bo{} surfaces which have a number of exceptions.  In many of the structures, we observe an oscillation of oxygen vacancy formation energies between the \ao{} and \bo{} layers. This has previously been reported for \sto{} slabs~\cite{Behtash2018,He2020}, but we show that this can be generalised to further perovskites, as seen in \figref{fig:vac:slab}.

The $A$- and $B$-site surface vacancies are more energetically favourable than in bulk by approximately $1.6$~\si{\electronvolt} and $1.9$~\si{\electronvolt}, respectively. Due to this high energy preference of surface vacancies to those in bulk, defect density at the surface will be substantially higher than in the bulk (by several orders of magnitude). As mentioned earlier in this work, for bulk formation energy, we see that, in all perovskites presented here, the $A$-site bulk vacancies are the most energetically favourable, with a formation energy of around $3.2$~\si{\electronvolt} whereas the average $B$-site bulk vacancies is $4.5$~\si{\electronvolt}.  This results in $A$-site and $B$-site vacancies at the surface with formation energies of around $1.6$~\si{\electronvolt}, and $2.7$~\si{\electronvolt}.  Therefore $A$-site vacancies are more favourable at \ao{}-terminated surfaces than $B$-site vacancies at \bo{}-terminated surfaces. As most distribution functions (such as Boltzmann) have an exponential tendency for these energies, this results in  substantially more $A$ vacancies in a \ao{} surface than $B$ vacancies in a \bo{} surface.  Hence \ao{}-surfaces, whilst more energetically favourable, are more prone to vacancies and, as such, are likely easier to degrade, revealing a \bo{} layer below. We provide an individual breakdown of these analyses for each perovskite in
Section~SIII~\hspace{-0.15em}G~\cite{supplementary}%
, but the trend holds.

Whilst the $A$ and $B$ vacancies trends are clear to understand, the oxygen vacancies show the most variation in their behaviour. We noted earlier that the formation energies of these oscillate between \ao{} and \bo{} layers, but  our results show that this trend is reduced for zirconate systems, and for these zirconates, oxygen vacancies are more favourable in \ao{} surfaces, whereas, in the stannates and titanates, the \bo{} surfaces are more favourable.

There are a few anomalies to note, \cto{} oxygen vacancies in both \ao{} and \bo{} terminated slabs have higher formation energies than their other titanate counterparts (where \cto{} O-site vacancies converge to the bulk value, whereas \sto{} and \bto{} O-site vacancies converge to a value approximately $0.5$~\si{\electronvolt} below their bulk values). This is likely linked to the  \cto{}  phase being orthorhombic, whilst \sto{} and \bto{} are tetragonal. \bso{} O-site vacancies in the \bo{} terminated slab has a similar energy shift, which is the only cubic phase stannate, when compared to \cso{} and \sso{}, which are orthorhombic. When studying the \bso{} \bo{} oxygen vacated slabs, we find that the relaxed structure exhibits an orthorhombic-like phase, with the $B$-O substructure becoming tilted (see
Section~SIII~\hspace{-0.15em}H~\cite{supplementary}%
).

One final anomaly with the results is that $A$- and $B$-site vacancies in \bo{}-terminated titanates and zirconates converge to energies below the respective vacancy formation energy in bulk. In
Table~SVIII~\cite{supplementary}%
, we highlight how the O--$B$--O bonds become distorted in these structures to a polar phase of the perovskite structure, which is more energetically favourable than the non-polar system. Polarisation due to the presence of vacancies has previously been observed in DFT studies of \sto{}~\cite{He2020}.

Aside from the exceptions detailed previously, there is a general trend, in particular for cations, for the vacancy formation energy to be lowest in the first 2 layers of the material. More negative vacancy formation energies compared to bulk implies that the material is susceptible to physical degradation therefore deeming the surface unstable. This corroborates the results of recent experimental findings for \sto{} where all surfaces contained both \bo{}- and \ao{}-terminations~\cite{Florencio2014,Sokolovi2021}, presumably, in part, due to the fact that degradation of one layer will reveal the preceding layer. The phenomenon of Sr segregation on the (001) surface for \sto{} has also been demonstrated through long annealing times at temperatures of around $1300$~\si{\celsius}~\cite{Bachelet2009}. This seems to be in agreement with our results showing that vacancies are more favourable at the surface than in bulk. It is common for vacancies to form during sample manufacturing~\cite{Monama2022}, annealing at high temperature would increase the rate of vacancy diffusion into the material resulting in Sr precipitation at the surface.

\subsection{Charged vacancies at \sso{} surfaces and in slabs}
\label{sec:slab_vacancies:charged}

\begin{figure*}[ht]
    \centering
    \subfloat[Charge defects in \ao{} slab, O-rich limit]{\includegraphics[width=0.45\linewidth]{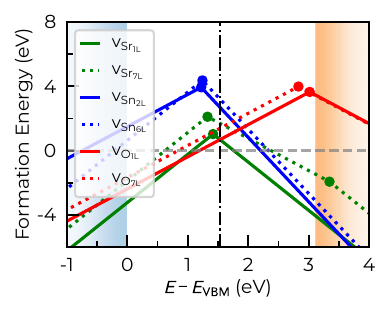}\label{fig:charge_state:AO}}%
    \hspace{1em}%
    \subfloat[Charge defects in \bo{} slab, O-rich limit]{\includegraphics[width=0.45\linewidth]{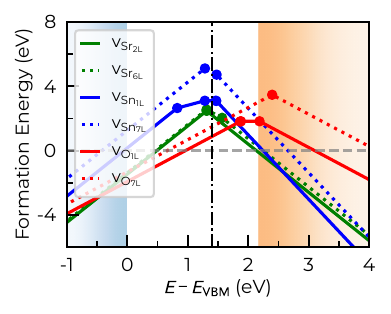}\label{fig:charge_state:BO2}}%
    \caption{
        Formation energies of $A$-, $B$- and, O-site charged vacancies in \sso{} \protect\subref{fig:charge_state:AO} \ao{}- and \protect\subref{fig:charge_state:BO2} \bo{}-terminated slabs under O-rich conditions (see \tabref{tab:chem_pots:o-rich_limit}).
        Line gradients correspond to the charge state (in $|$\si{\elementarycharge}$|$) and points of discontinuity identify transition levels.
        The valence (conduction) band region is shaded in blue (orange) and the vertical dash-dotted line is the equilibrium Fermi level.
        Corrections to the charge state energies are calculated using the following codes: \texttt{doped}~\cite{Kavanagh2024DopedPythonToolkit}, \texttt{qdef2d}~\cite{Tan2019ChargedDefectsFramework}, and \texttt{sxdefectalign2d}~\cite{Freysoldt2009,Freysoldt2018FirstPrinciplesCalculations}, which employ band edge shifting~\cite{Broberg2023HighThroughputCalculations}, image charge~\cite{Kumagai2014ElectrostaticBasedFinite}, and slab charge~\cite{Freysoldt2009,Freysoldt2018FirstPrinciplesCalculations} corrections.
        }
    \label{fig:charge_state:slab}
\end{figure*}

The concentration and stability of vacancies depend on their charge states, governed by the Fermi energy.
While bulk charge states in oxide perovskites are well-studied~\cite{Weston2018}, surface influence remains less explored.
Prior studies on perovskite stannates have focused on fully ionised cation vacancies, such as V$_{\mathrm{Sr}}^{-2}$ and V$_{\mathrm{Sn}}^{-4}$ in \sso{}.
Since no specific dopant environment is assumed, we examine a range of compensating vacancies, including those deviating from expected ionised charge states.
Given the interest in both $p$- and $n$-type doping for stannate perovskites~\cite{Weston2018,Willis2023PossibilityPType}, a comprehensive defect study is necessary.

We analyse how surface termination affects charged vacancy states in \sso{} (see \figref{fig:charge_state:slab}).
We note at the beginning that, overall, charge defects still support $A$-site vacancies being dominant.
Charge states are explored under the oxygen-rich limit (see 
Figs. S11 and S12~\cite{supplementary}
for variations in chemical potential).
In both \ao{} and \bo{}-terminated slabs, charge states at the centre resemble the bulk, but being shifted $0.5$--$1$~\si{\electronvolt} toward the valence band (more significantly for the \bo{}-termianted slab).
Most charge states either remain bulk-like or become more stable at the surface, except for Sn vacancies in the \ao{}-terminated slab due to charge state crossings.
This aligns with \figref{fig:vac:slab}, which shows that the \ao{} surface further reduces vacancy formation energy.
Local potential analysis indicates positive (negative) charges localise around anion (cation) vacancies, while negative (positive) charges delocalise for anion (cation) vacancies.

The \bo{} surface exhibits a reduced theoretical band gap ($2.17$~\si{\electronvolt}) relative to the bulk ($3.09$~\si{\electronvolt})~\cite{Taylor2020}, whereas the \ao{} surface retains a bulk-like gap ($3.11$~\si{\electronvolt}).

In the bulk, most vacancies have positive formation energies, but this is not the case for slabs.
To discuss this, we focus on the oxygen-rich condition.
At both the \ao{} and \bo{}, the  surface Sr ($A$-site) and O vacancies are dominant for the entire band gap (sub-surface Sr for the \bo{} surface).
For the \ao{} surface, Sr vacancies exhibit negative formation energies across most of the band gap and O vacancies nearer to the valence band, making \ao{} surfaces highly susceptible to vacancies.
At the \bo{} surface, it exhbits a much wider range of energies over which no native charged defect has a negative formation energy; oxygen has negative formation energies near the valence bands, comparable to the \ao{} surface, and Sr sub-surface vacancies become favourable just at the conduction band minimum.
Both surface show a clear preference to Sr (sub)surface vacancies, highlight a likely degredation of the \ao{} surface.
For both terminations, V$_{\mathrm{O}}$ is typically the only shallow donor, except for slab-centre V$_{\mathrm{Sr}}$ in \ao{}-terminated slabs, with no native shallow acceptors observed.
Overall, both (001) surfaces are less stable than the bulk due to lower vacancy formation energies at the surface.
The order of vacancy favourability changes with chemical environment, explored further in 
Section SIII~\hspace{-0.15em}J~\cite{supplementary}%
; one key result is that there exists only one limit in which Sn (sub)surface vacancies are more favourable than Sr (sub)surface vacancies, further supporting the preference towards $A$-site vacancies near the surface.

Regarding dopability limits -- the range within the band gap that can support dopants without forming compensating native vacancies -- the \ao{}- and \bo{}-terminated slabs exhibit ranges of $0.57$~\si{\electronvolt} ($1.19$--$1.76$~\si{\electronvolt}) and $1.18$~\si{\electronvolt} ($0.96$--$2.14$~\si{\electronvolt}), respectively.
With a smaller band gap ($2.17$~\si{\electronvolt}), the \bo{} surface's Fermi level dopability range is closer to the band edges, making it more stable under doping.
This supports the argument that $A$-site vacancies dominate at the \ao{} surface, while $B$-site vacancies are less prevalent at the \bo{} surface, indicating greater chemical stability.

\section{Conclusion}

Our results show that for the (001) termination, \ao{} surfaces of the perovskites considered are much more favourable than \bo{} surfaces, despite both being observed experimentally.
We propose that \bo{} surfaces, whilst less energetically favourable, are likely to form from \ao{} surfaces that degrade through vacancy formation, exposing the sub-surface \bo{} layer.
$A$-site vacancies are found to be roughly $1$~\si{\electronvolt} more favourable at \ao{} surfaces than $B$-site vacancies at the \bo{} surface, meaning that \bo{} surfaces are more stable against degradation under vacancy formation than \ao{} surfaces.
In the bulk perovskites examined under O-rich conditions, $A$-site vacancies were found to almost always be the most favourable, with O-site vacancies always being the least favourable.
Surface and shallow vacancies show a dominant trend of increasing formation energy with increasing vacancy depth.
A number of notable exceptions are discussed.
Phase diagrams agree with surface formation energy analysis; for all perovskites studied, under conditions when bulk \abo{} is sustainable, the only stable (001) surface is the \ao{}, with the exceptions of \cto{} and \cso{}.
The former shows both \ao{} and \bo{} to exist, whilst the latter has no stable \ao{} or \bo{} (001) surface, which indicates other surfaces will form.
Charge defects studied in \sso{} show further instability of the \ao{} (due to charged Sr surface vacancies), more so than the \bo{} surface (less susceptible to Sn and O surface vacancies).
With this in  mind, a suitable method for manufacturing \bo{} surfaces of these materials may be to first form an \ao{} surface, then treat the newly formed surface to obtain \bo{} surfaces, which supports the methodology outlined in Ref.~\cite{Kawasaki1994}.
Our main conclusion is that, although the \ao{} surface is more energetically favourable, the formation of $A$-site vacancies will lead to the \bo{} surface being more prevalent than initially expected.
We believe that these results will aid in future manufacturing and modelling of perovskite surfaces and also expand understanding of real long term surface stability of these systems.

\hspace{1em}

The data that support the findings of this study are openly available from the University of Exeter’s Institutional repository~\cite{dataset}.

\section*{Acknowledgments}
The authors thank the Leverhulme for funding this research via Grant RPG-2021-086.
The work provided by N. T. Taylor was supported in part by the Government Office for Science and the Royal Academy of Engineering under the UK Intelligence Community Postdoctoral Research Fellowships scheme.
Via our membership of the UK's HEC Materials Chemistry Consortium, which is funded by EPSRC (EP/R029431), this work used the ARCHER2 UK National Supercomputing Service (http://www.archer2.ac.uk).
The authors would also like to acknowledge the use of the University of Exeter High-Performance Computing (HPC) facility in carrying out this work.
Data retrieved from the Materials Project for 
\cto{} (mp-4019),
\sto{} (mp-5229), 
\bto{} (mp-5986), 
\czo{} (mp-4571),
\szo{} (mp-4387),
\bzo{} (mp-3834),
\cso{} (mp-4438),
\sso{} (mp-2879), and
\bso{} (mp-3163)
from database version v2022.10.28~\cite{Jain2013,mp_all}.

\section*{Author Contributions}
We acknowledge the following individuals for their contributions. 
N.T.T. contributed conceptualisation, validation, formal analysis, investigation, data curation, writing - original draft, writing - review and editing, visualisation, supervision, project administration, and funding acquisition.
M.T.M. contributed validation, formal analysis, investigation, data curation, writing - original draft, writing - review and editing, and visualisation.
S.P.H. contributed conceptualisation, formal analysis, writing - review and editing, supervision, project administration, and funding acquisition.

\bibliography{main}

\end{document}


\title{Supplementary Material\\{}Instability of oxide perovskite surfaces induced by vacancy 
formation}
\author{Ned Thaddeus Taylor}
\email{n.t.taylor@exeter.ac.uk}
\affiliation{Department of Physics and Astronomy, University of Exeter, Stocker Road, Exeter, EX4 4QL, United Kingdom}
\author{Michael Thomas Morgan}
\affiliation{Catalan Institute of Nanoscience and Nanotechnology (ICN2), CSIC and BIST, Campus UAB, Bellaterra, 08193 Barcelona, Spain}
\affiliation{Physics Department, Autonomous University of Barcelona (UAB),  Campus UAB, Bellaterra, 08193 Barcelona, Spain}
\affiliation{Royal Melbourne Institute of Technology, School of Engineering, GPO Box 2476, Melbourne VIC 3001, Australia}
\author{Steven Paul Hepplestone}%
\email{s.p.hepplestone@exeter.ac.uk}
\affiliation{Department of Physics and Astronomy, University of Exeter, Stocker Road, Exeter, EX4 4QL, United Kingdom}

\maketitle

\tableofcontents
\listoffigures
\listoftables

\section{Ionisation Energies}
\label{sec:ionisation}

\begin{figure}[ht!]
    \centering
    \includegraphics[scale=0.4]{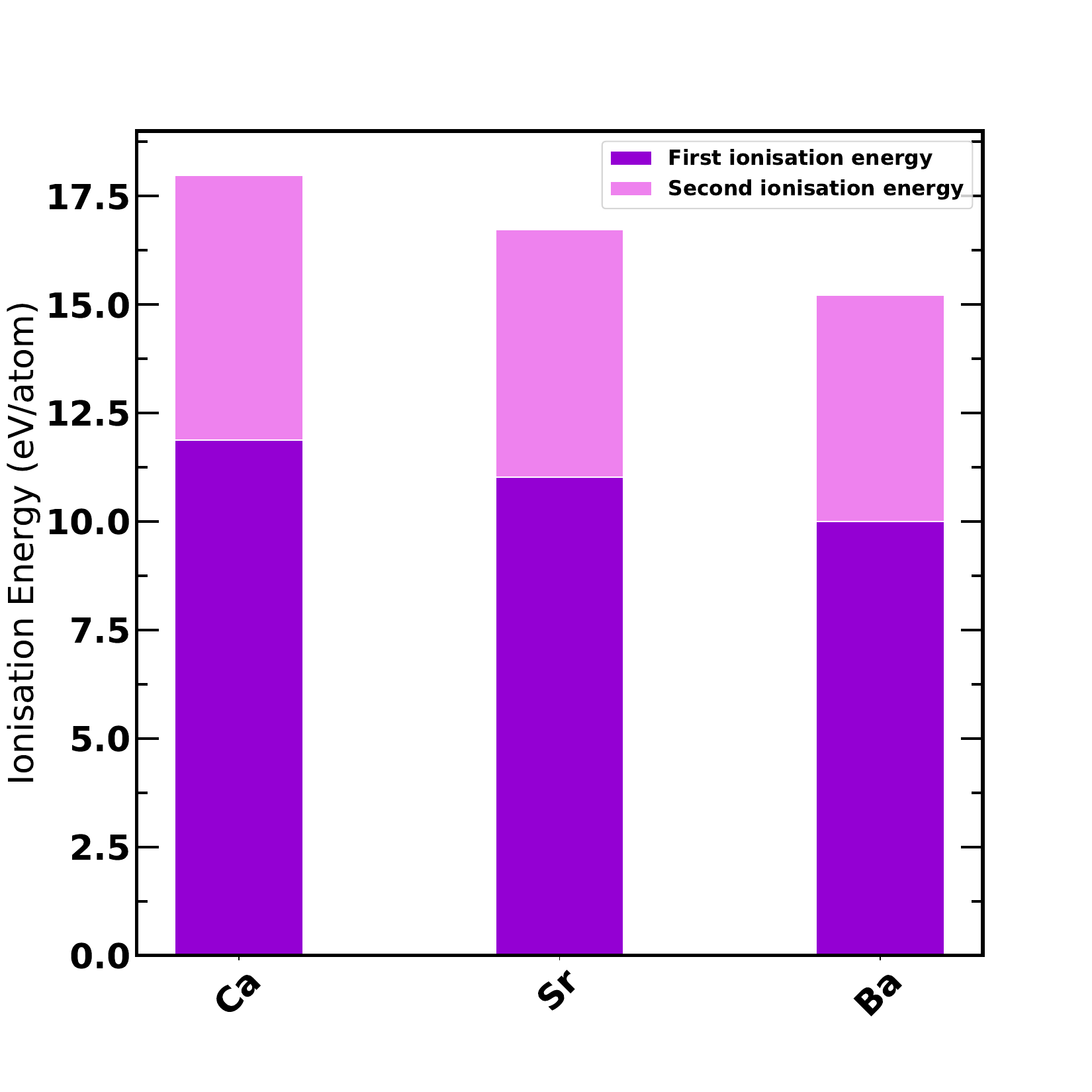}
    \caption[Ionisation energies]{First and second ionisation energies of Ca, Sr, and Ba. Values obtained from Ref.~\cite{CRC2016}}
    \label{fig:ionisation}
\end{figure}

The ionisation energy of an element is typically defined as the  amount of energy needed to remove the highest-energy electron from the orbital of the neutral atom when in the gas phase.
The second relates to the energy required to do the same for a charged ion (with total charge $+|e|$, where $|e|$ is the magnitude of charge of one electron).

When a surface is formed bonds are broken, and the ideal electronic charge distribution that is usually found in the bulk is disturbed. In order for a surface to remain stable, charge must be redistributed in such a way that the surface remains energetically favourable. Redistributing electrons incurs an energetic cost. We suggest that when the energetic cost of redistributing charge is cheaper, the energetic cost of forming the surface is also cheaper, i.e. the surface formation energy for that surface is lower.
The trend of decreasing first and second ionisation energies with increasing period~\cite{CRC2016}, as shown in \figref{fig:ionisation}, closely mirrors the reduction in surface formation energy observed for the \ao{} surface of oxide perovskites across the same period in this study.

\section{Methods}
\label{sec:methods}

\subsection{Slab structure generation}
\label{sec:slab_generation}

\begin{figure}[ht!]
    \centering
    \subfloat[Cubic/tetragonal perovskite cells]{\includegraphics[width=0.45\linewidth]{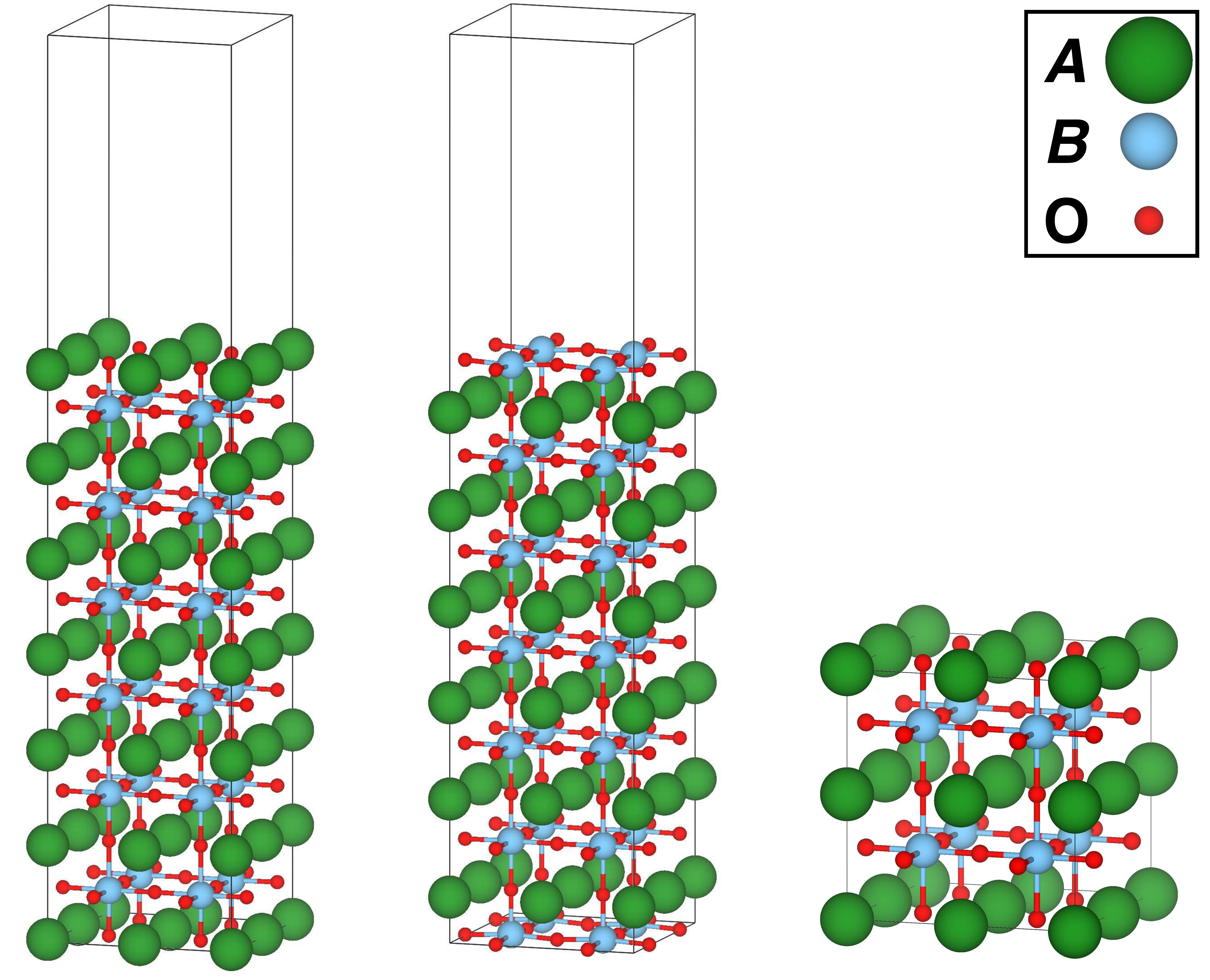}\label{fig:structures:cubic}}%
    \hspace{2em}%
    \subfloat[Orthorhombic perovskite cells]{\includegraphics[width=0.45\linewidth]{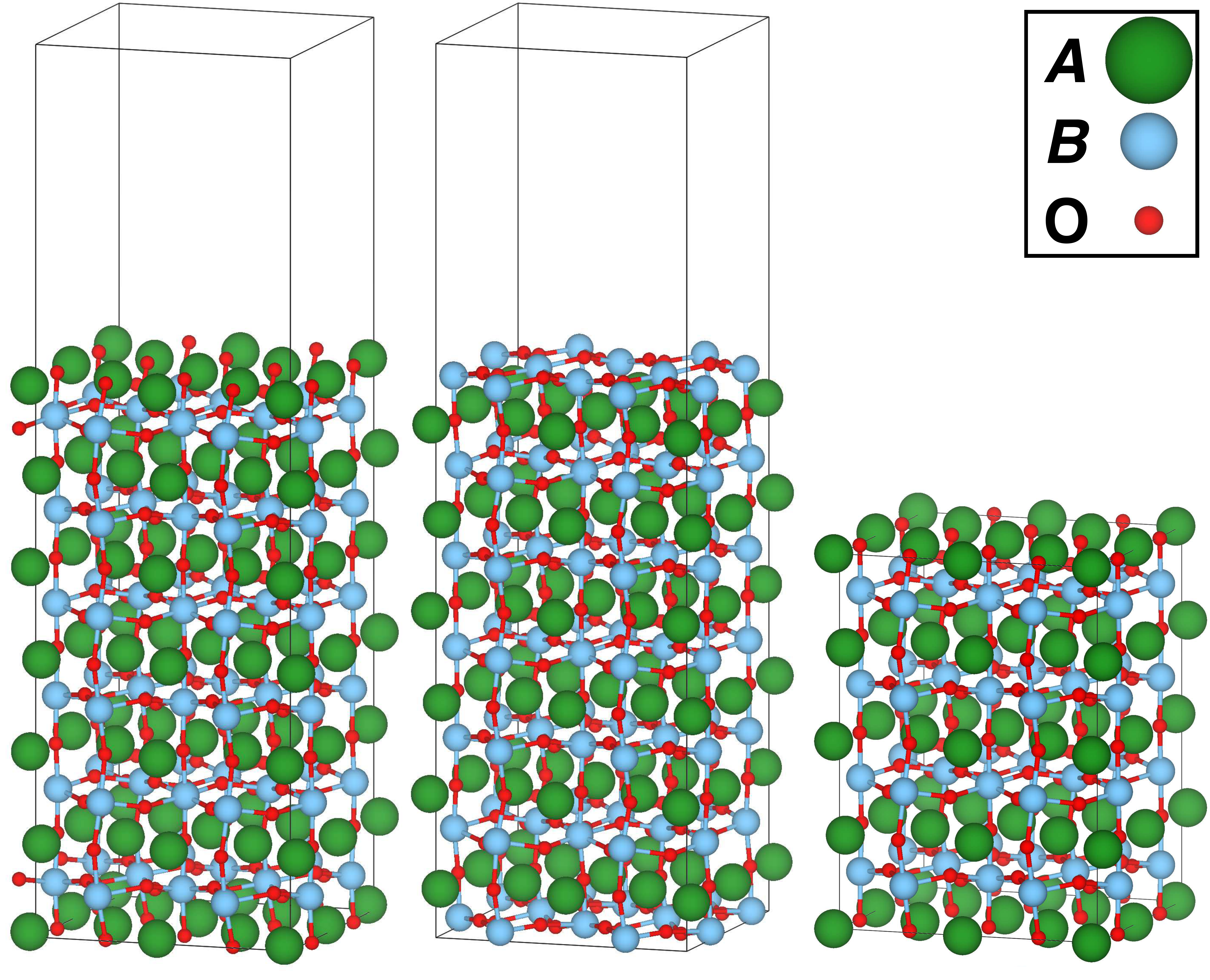}\label{fig:structures:ortho}}%
    \caption[Schematics of bulk and slab supercells]{
        Ball-and-stick representation of surface and bulk supercells used for vacancy calculations in \protect\subref{fig:structures:cubic} cubic/tetragonal and \protect\subref{fig:structures:ortho} orthorhombic oxide perovskites.
        The surface models include \ao{}- (left) and \bo{}-terminated (middle) undefected $2\times2$ slab supercells, each composed of six repeating units along the surface normal, plus an additional layer of atoms to reflect the chosen termination.
        For cubic/tetragonal (orthorhombic) phases, this results in 8 (32) \abo{} units for the bulk and 24 (48) \abo{} units for the slabs, plus an additional 4 (8) units of the corresponding surface termination stoichiometry.
        To minimise polar effects, both slab surfaces are constructed with equivalent stoichiometry.
        The bulk model (right) shows an undefected $2\times2\times2$ supercell used for bulk vacancy calculations.
        Figure generated using VESTA~\cite{Momma2011}.
    }
    \label{fig:structures}
\end{figure}

In this work, supercells are always generated using an even number of unit cell extensions; this decision is made due to symmetry constraints of odd-numbered unit cells, particularly those enforced upon tilting structures (like seen in the orthorhombic phase).
Supercells generated with odd-numbered lattice vector extensions (e.g. $3\times1\times1$), cannot, due to the supercell symmetry enforced on the perovskite structure, support octahedral tilting. As such, even numbered supercells were the only viable option and are those explored in this work.

For surfaces, the $2 \times 2 $ ($\mathbf{a} \times \mathbf{b}$) supercell extension is performed for cubic, tetragonal and orthorhombic phases (see \figref{fig:structures}).
This means that, due to the large cell size of orthorhombic unit cells, the orthorhombic slabs contain double the number of chemical units (\abo{}) than the cubic and tetragonal phases.

Whilst larger even supercells, such as a pseudo-orthorhombic form, could be taken for the cubic/tetragonal cells for both bulk and slabs, the focus of this study lies in the difference between vacancies in the bulk and in slabs (as a function of distance from the surface).
Therefore, as long as the same in-surface plane concentration is maintained, then the results between the bulk and slabs will be directly comparable.
As a note of comparison, for \bso{}, the vacancies in a 32 \abo{} unit supercell (effectively a $2\sqrt{2}\times2\sqrt{2}\times4$ expansion) for Ba, Sn, and O exhibit energy differences of $-0.10$, $0.09$, and $-0.52$~\si{\electronvolt}/vac when compared to the $2\times2\times2$ supercell (where negative denotes more favourable in the larger supercell).

\subsection{Formation energy of bulk oxide perovskites}
\label{sec:formation}

As formation energy is merely a comparative value, it strongly depends on the reference materials used to compare against.
The two methods we consider for the perovskites are comparison against 1) the constituent elements in their perfect bulks and 2) the constituent binary oxides.
These two forms of formation energy are

\begin{equation}
    E_\textrm{form,\abo{}} = E_\textrm{\abo{}}^\textrm{bulk} - E_{A}^\textrm{bulk}     - E_{B}^\textrm{bulk} - 3E_\textrm{O}^\textrm{bulk},
    \label{eq:formation:bulk}
\end{equation}

\noindent
and
\begin{equation}
    E_\textrm{form,\abo{}} = E_\textrm{\abo{}}^\textrm{bulk} - E_\textrm{\ao{}}^\textrm{bulk} - E_\textrm{\bo{}}^\textrm{bulk}.
    \label{eq:formation:oxide}
\end{equation}

\noindent
Here, $E$ corresponds to a density functional theory (DFT) calculated total energy. $E_\textrm{\abo{}}^\textrm{bulk}$ is the energy of the bulk perovskite, $E_\textrm{X}^\textrm{bulk}$ is the energy of element X in bulk form, and $E_\textrm{\ao{}}^\textrm{bulk}$ ($E_\textrm{\bo{}}^\textrm{bulk}$) is the energy of the binary oxide \ao{} (\bo{}) in bulk. The form of formation energy seen in \eqref{eq:formation:oxide} is presented in \tabref{tab:bulk} for the nine oxide perovskites studied in the main article. The definition of formation energy used here is sometimes referred to as free energy.

\subsection{Surface Gibbs free energy}
\label{sec:gibbs}

The above equation does not take into consideration the chemical environment in which the surface is formed. To account for this surface Gibbs free energy must be used. The method used to assess surface Gibbs free energy is somewhat more demanding and is outlined below. A fully detailed description of the method can be found in reference~\cite{Heifets2007}.

The surface Gibbs free energy per unit cell for an $i$-terminated slab, $\Omega^{i}$, is defined as
\begin{equation}
    \Omega^{i}=\frac{1}{2}\left[G_\textrm{slab}^{i}-N_{A}^{i} \mu_{A}-N_{B}^{i} \mu_{B}-N_\textrm{O}^{i} \mu_\textrm{O}\right],
\label{eq:gibbs:surface}
\end{equation}

\noindent
where $G_\textrm{slab}^{i}$ is the Gibbs free energy of the slab, $N_{A}^{i}$, $N_{B}^{i}$, $N_\textrm{O}^{i}$ are the number of $A$, $B$, and O atoms in the slab respectively, and $\mu_{A}$, $\mu_{B}$, $\mu_\textrm{O}$ are the chemical potentials of the $A$, $B$, and O species respectively. The chemical potential, $\mu_\textrm{\abo{}}$, is taken as the sum of the chemical potentials of its constituents:

\begin{equation}
    \mu_\textrm{\abo{}}=\mu_{A}+\mu_{B} + 3 \mu_\textrm{O}.
\label{eq:gibbs:chempot}
\end{equation}

\noindent
By assuming the surface is stable, we introduce the requirement that the bulk must be in equilibrium with the surface. The chemical potential is set to the bulk crystal Gibbs free energy,

\begin{equation}
    \mu_\textrm{\abo{}}=g_\textrm{\abo{}}^\textrm{bulk}.
\label{eq:gibbs:mu:o}
\end{equation}

\noindent
With these definitions, we can now simply \eqref{eq:gibbs:surface} to

\begin{equation}
\Omega^{i}=\frac{1}{2}\left[G_\textrm{slab}^{i}-N_{B}^{i}g_\textrm{\abo{}}^\textrm{bulk}\right]-\Gamma_{{B},{A}}^{i}\mu_{A}-\Gamma_{{B},\textrm{O}}^{i} \mu_\textrm{O},
\label{eq:gibbs:surface:reduced:1}
\end{equation}

\noindent
where $\Gamma_{{B},\textrm{X}}^{i}$ is the number of excess atoms of species X, in $i$-terminated slab, with respect to the number of atoms of species $B$. This $\Gamma$ factor can be expressed as 

\begin{equation}
    \Gamma_{{B}, X}^{i}=\frac{1}{2}\left(N_{X}^{i}-N_{B}^{i} \frac{N_{x}^\textrm{bulk}}{N_{B}^\textrm{bulk}}\right).
\label{eq:gibbs:gamma}
\end{equation}

We want to explore the chemical environments in which $A$ and $B$ are more favourable in the perovskite slab than their individual bulks. To negate the precipitation of the individual bulks on the surface, the element chemical potentials must be less than or equal to the Gibbs free energy of their bulk,

\begin{equation}
    \mu_{A}\leqslant g_{A}^\textrm{bulk},
    \label{eq:gibbs:mu:a}
\end{equation}

\noindent
and

\begin{equation}
    \mu_{B}\leqslant g_{B}^\textrm{bulk}.
    \label{eq:gibbs:mu:b}
\end{equation}

\noindent
The same is also true for the formation of oxides,

\begin{equation}
    \mu_{A}+\mu_\textrm{O} \leqslant g_\textrm{\ao{}}^\textrm{bulk},
    \label{eq:gibbs:mu:ao}
\end{equation}

\noindent
and

\begin{equation}
    \mu_{B} +2 \mu_\textrm{O} \leqslant g_\textrm{\bo{}}^\textrm{bulk}.
    \label{eq:gibss:mu:bo}
\end{equation}

Gibbs free energies are approximated using the DFT total energy calculations. A detailed explanation has been discussed by Heifets \etal{} in reference~\cite{Heifets2007}.

\begin{equation}
    g_{j}^\textrm{bulk} \approx E_{j}^\textrm{bulk}.
    \label{eq:gibbs:g:bulk}
\end{equation}

A change in the chemical potential is given by:

\begin{equation}
    \Delta \mu_{A}=\mu_{A}-g_{A}^\textrm{bulk} \approx \mu_{A}-E_{A}^\textrm{bulk},
    \label{eq:gibbs:mu:delta:a}
\end{equation}

\begin{equation}
    \Delta \mu_{B}=\mu_{B}-g_{B}^\textrm{bulk} \approx \mu_{B}-E_{B}^\textrm{bulk},
    \label{eq:gibbs:mu:delta:b}
\end{equation}
 
\begin{equation}
    \Delta \mu_\textrm{O} =\mu_\textrm{O}-\frac{1}{2} E_{\textrm{O}_{2}}.
    \label{eq:gibbs:mu:delta:o}
\end{equation}

\noindent
Substituting these deviations into \eqref{eq:gibbs:surface:reduced:1} gives:

 \begin{equation}
 \begin{split}
     \\ & \Omega^{i}=\frac{1}{2}\left[G_{\text{slab}}^{i}-N_{B}^{i}g_\textrm{\abo{}}^{\text{bulk}}\right]  
     \\ & -\Gamma_{{B},{A}}^{i}\mu_{A}+\Gamma_{{B},{A}}^{i}g_{A}^{\text{bulk}}-\Gamma_{{B},{A}}^{i}g_{A}^{\text{bulk}}
     \\ & -\Gamma_{{B},\textrm{O}}^{i}\mu_\textrm{O}+\Gamma_{{B},\textrm{O}}^{i} \frac{1}{2} E_{\textrm{O}_{2}}-\Gamma_{{B},\textrm{O}}^{i} \frac{1}{2} E_{\textrm{O}_{2}}.
 \end{split}
 \label{eq:gibbs:surface:reduced:2}
 \end{equation}

\noindent
The Gibbs free energy can then be rewritten as

\begin{equation}
    \Omega^{i}=\phi^{i}-\Gamma_{{B}, {A}}^{i} \Delta \mu_{A}-\Gamma_{{B}, \textrm{O}}^{i} \Delta \mu_\textrm{O},
    \label{eq:gibbs:surface:reduced:3}
\end{equation}

\noindent
where

 \begin{equation}
 \begin{aligned}
 \phi^{i} &=\frac{1}{2}\left[G_{\text {slab }}^{i}-N_{B}^{i} g_\textrm{\abo{}}^\textrm{bulk}\right]-\Gamma_{{B}, {A}}^{i} g_{A}^\textrm{bulk}-\frac{1}{2} \Gamma_{{B}, \textrm{O}}^{i} E_{\textrm{O}_{2}} \\
 & \approx \frac{1}{2}\left[E_\textrm{slab}^{i}-N_{B}^{i} E_\textrm{\abo{}}^\textrm{bulk}\right]-\Gamma_{{B}, {A}}^{i} E_{A}^\textrm{bulk}-\frac{1}{2} \Gamma_{{B}, \textrm{O}}^{i} E_{\textrm{O}_{2}}.
 \end{aligned}
 \label{eq:gibbs:phi}
 \end{equation}

If the surface Gibbs free energy is negative then energy is released during surface formation and the surface will form spontaneously~\cite{Heifets2007}. The boundary condition for surface formation is therefore:

\begin{equation}
    \Omega^{i}\left(\Delta \mu_{A}, \Delta \mu_\textrm{O}\right)=0.
\label{eq:gibbs:omega}
\end{equation}

Under conditions in which the chemical potentials are static, and at 0 K, \eqref{eq:gibbs:surface:reduced:3} can be used independently to assess surface stability. Under a varying chemical potential surface stability will also vary. Under varying chemical potentials the boundary between two regions of stability for two surfaces $i$ and $j$ is given by the solution to:

\begin{equation}
    \Omega^{i}\left(\Delta \mu_{A}, \Delta \mu_\textrm{O}\right)=\Omega^{j}\left(\Delta \mu_{A}, \Delta \mu_\textrm{O}\right).
\label{eq:gibbs:omega:2}
\end{equation}

The solutions for different surfaces differ by a constant factor:

\begin{equation}
    \Delta \mu_{A}+\Delta \mu_\textrm{O}=g(i, j) .
\label{eq:gibbs:mu:delta:ao}
\end{equation}

Boundary conditions for phase diagrams of (001) surfaces are produced using Equations \ref{eq:gibbs:omega} and \ref{eq:gibbs:omega:2}.

It should be noted that the surface Gibbs free energy can be modified further to include temperature and entropy terms, in addition to oxygen correction terms for deviation from experiment  (as seen in Ref.~\cite{Heifets2007}). These can be accounted for by applying a further shift to the chemical potential values (namely oxygen), where the shift is dependent on the desired temperature and pressure environments. However, such additional factors have not been considered here.

\subsection{Dielectric values of slabs}
\label{sec:dielectric}

As mentioned in the main article, vacancies introduce the possibility of charged states occurring in the material to compensate for the defect. The introduced excess charge into the material creates its own electric field. Methods~\cite{Freysoldt2009,Suo2020} exist to account for the energy introduced by this compensatory charge to give a more accurate value of charged point defect formation energy. However, these methods rely on obtaining an accurate value of the material permittivity, often obtained from experiment. Accurate experimental values of dielectric constant for some of the systems considered here are not well known.  Experimentally, obtaining these values requires having a high quality sample free from defects or grains, it is therefore possible that these values will not be appropriate for computational study of an ideal crystal, or vice versa. Computational methods employed to calculate permittivity are often plagued with many issues, ranging from difficulty of applying continuous electric fields to periodic systems, to the number of different electric field setups that need to be sampled, to more accurately capturing the excited states, to the greater accuracy required to precisely sample the subsequent electronic and atomic displacements (i.e. the polarisation). These issues culminate in drastic computational cost. As such, accurately obtaining the full dielectric tensor for bulk systems is demanding, and systems of greater complexity only become more challenging.

\begin{figure*}[ht!]
    \centering
    \subfloat[\ao{}-terminated slab]{\includegraphics[width=0.45\linewidth]{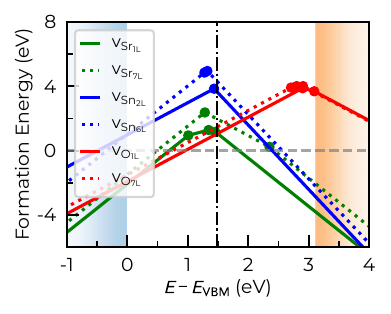}\label{fig:charge_state:diel:AO}}%
    \hspace{1em}%
    \subfloat[\bo{}-terminated slab]{\includegraphics[width=0.45\linewidth]{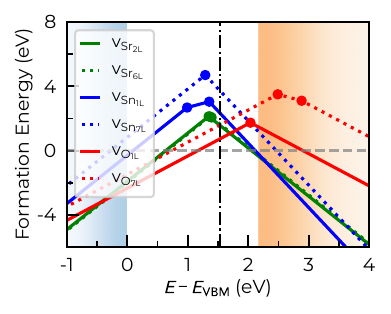}\label{fig:charge_state:diel:BO2}}%
    \caption[Dielectric effects on charged defects formation energies]{
        Formation energies of $A$-, $B$- and, O-site charged vacancies under the O-rich limit in \sso{} \protect\subref{fig:charge_state:weston:bulk} bulk, and \protect\subref{fig:charge_state:diel:AO} \ao{}- and \protect\subref{fig:charge_state:diel:BO2} \bo{}-terminated slabs.
        The relative dielectric permittivity parallel to the surface normal vector has been set to 10.
        Line gradients correspond to the charge state (in $|$\si{\elementarycharge}$|$) and points of discontinuity identify transition levels.
        The valence (conduction) band region is shaded in blue (orange) and the vertical dash-dotted line is the equilibrium Fermi level.
        Corrections to the charge state energies are calculated using \texttt{doped}~\cite{Kavanagh2024DopedPythonToolkit}, which employs band edge shifting~\cite{Broberg2023HighThroughputCalculations} and image charge~\cite{Kumagai2014ElectrostaticBasedFinite} corrections.
        Slab charge corrections~\cite{Freysoldt2009,Freysoldt2018FirstPrinciplesCalculations} are accounted for in subfigures \protect\subref{fig:charge_state:weston:AO} and \protect\subref{fig:charge_state:weston:BO2} using \texttt{qdef2d}~\cite{Tan2019ChargedDefectsFramework} and \texttt{sxdefectalign2d}~\cite{Freysoldt2009,Freysoldt2018FirstPrinciplesCalculations}.
    }
    \label{fig:charge_state:diel}
\end{figure*}

Whilst the method employed to correct for charges in slabs requires the 
For the bulk dielectric tensor fo \sso{}, the values have been taken from the Materials Project~\cite{Jain2013}.
For the \sso{} slab the bulk values have also been used; it is known that dielectric permittivity decreases for decreasing slab thickness~\cite{Zaccone2024ExplainingThicknessDependent}.
To mimic a large grain, the bulk value has been taken.
However, to check reliability under different dielectric values, the corrections have also been reapplied using a dielectric permittivity of 10 parallel to the surface normal vector, with the results presented in \figref{fig:charge_state:diel} for the same O-rich conditions used in the main article.
The same qualitative results as in the main paper are identified, including surface charge states being lower than in the slab-centre.

\subsection{Vacancy concentrations}
\label{sec:concentration}

Defect concentrations used in this work closely align with previous work investigating oxide perovskites~\cite{Wexler2021,Weston2018}.
As defect concentration decreases (i.e. the crystal approaches perfect bulk), vacancy formation energy is also expected to decrease, tending towards the value for the isolated vacancy~\cite{Freysoldt2022LimitationsEmpiricalSupercell}.
In higher concentrations, the vacancy formation energy will be artificially increased due to interactions with periodic images.
So far, correction schemes exist that account for the effects of introduced charge and ferroelectric distortion~\cite{Kim2020} within the material.
In agreement with others~\cite{Freysoldt2022LimitationsEmpiricalSupercell},  we note that, to date, there does not exist to account for defect periodic image interactions.
Regardless, results at these concentrations are still useful for prediction of qualitative results, i.e. trends.

\subsection{Competing phases for charged vacancies}
\label{sec:competing_phases}

\begin{table}[h]
    \centering
    \begin{tabular}{P{6em}P{6em}P{6em}}
        \hline
        Formula & $k$-mesh & $E_\mathrm{form}$ (eV/fu) \\\hline
        \sso{}    & 7$\times$7$\times$5 & -11.282 \\
        \o{}      & 1$\times$1$\times$1 & 0.000 \\
        Sn        & 5$\times$5$\times$7 & 0.000 \\
        SnO       & 5$\times$5$\times$5 & -2.686 \\
        SnO$_{2}$ & 5$\times$5$\times$5 & -5.096 \\
        Sr        & 5$\times$5$\times$5 & 0.000 \\
        Sr$_{2}$SnO$_{4}$  & 7$\times$7$\times$4 & -16.919 \\
        SrO       & 5$\times$5$\times$5 & -5.473 \\\hline
    \end{tabular}
    \caption[\sso{} competing phase energetics]{
        Formation energies per formula unit ($E_\mathrm{form}$) of \sso{} and all competing phases, with Monhorst-Pack $\Gamma$-centred $k$-grids~\cite{Monkhorst1976} used in calculations.
        Only the lowest energy polymorphs are included.
    }
    \label{tab:competing_phases:formation_energies}
\end{table}

The formation energies of the competing phases considered for vacancies in \sso{} are presented in \tabref{tab:competing_phases:formation_energies}.

\subsection{Machine learned potentials}
\label{sec:machine_learned_potentials}

With the rise of foundation models, there is significant interest in defining their domains of applicability and identifying where further training is necessary.
Many machine-learned interatomic potentials are trained primarily on bulk datasets, such as the Materials Project~\cite{Jain2013}, making them well-suited to modelling perfect crystals but less reliable for more complex systems -- such as surfaces and interfaces -- not well represented in the training data.

To investigate this, we assess the performance of the MACE-MPA-0 potential~\cite{batatia2023FoundationModelAtomistic}, which has shown promising results for interfaces between single-element materials~\cite{Taylor2025RAFFLEActiveLearning}.
As a foundation model, some level of fine-tuning is expected to be needed for systems that diverge from the training distribution; however, none has been applied here

To enable fair comparison, we analyse formation energies rather than absolute energies, which can vary due to inconsistent reference states.
Formation energies are defined relative to appropriate baselines: from binary oxides for bulk perovskites, surface formation energy for slabs, and vacancy formation energy under O-rich conditions for both bulk and slabs.
Slab vacancy formation energies are computed relative to their bulk counterparts, as discussed in the main text.
An alternative for the slab vacancy formation energy is presented (bulk-ref), using the following formulae:

\begin{align}
   E_{\mathrm{form,A}} &= E_{\mathrm{vac}} - E_{\mathrm{surf}} + E_{\mathrm{AO}} - \frac{1}{2} E_{\mathrm{O_2}}, \\
   E_{\mathrm{form,B}} &= E_{\mathrm{vac}} - E_{\mathrm{surf}} + E_{\mathrm{bulk}} - E_{\mathrm{AO}} - E_{\mathrm{O_2}}, \quad\mathrm{and} \\
   E_{\mathrm{form,O}} &= E_{\mathrm{vac}} - E_{\mathrm{surf}} + \frac{1}{2} E_{\mathrm{O_2}}.
\end{align}

The mean absolute error (MAE) and root mean squared error (RMSE) for each model and system type are summarised in \tabref{tab:foundation_errors}.
Notably, all models show reasonable surface formation energies, but struggle significantly with vacancy energies, suggesting deficiencies in modelling localised defect chemistry—even though surfaces, conceptually, are extended defects.

\begin{table}[h]
\centering
\begin{tabular}{lccccccr}
\toprule
\multirow{2}{*}{\textbf{System}} & \multicolumn{2}{c}{\textbf{MACE-MPA-0}} & \multicolumn{2}{c}{\textbf{MACE-MP-0}} & \multicolumn{2}{c}{\textbf{CHGNet}} & \multirow{2}{*}{\textbf{Units}} \\
\cmidrule(lr){2-3} \cmidrule(lr){4-5} \cmidrule(lr){6-7}
 & MAE & RMSE & MAE & RMSE & MAE & RMSE & \\
\midrule
Bulk perovskites          & 0.0146 & 0.0189 & 0.0258 & 0.0329 & 0.0953 & 0.1201 & \si{\electronvolt/fu} \\
Surfaces                  & 0.0049 & 0.0063 & 0.0027 & 0.0031 & 0.0111 & 0.0125 & \si{\electronvolt/\angstrom\squared} \\
Bulk vacancies            & 1.3997 & 1.8798 & 1.7694 & 2.2013 & 2.3574 & 2.9577 & \si{\electronvolt/vac} \\
Slab vacancies            & 0.5582 & 0.7599 & 0.6178 & 0.9070 & 0.6615 & 0.9191 & \si{\electronvolt/vac} \\
Slab vacancies (bulk-ref) & 1.2770 & 1.5881 & 1.4524 & 1.6785 & 1.7360 & 2.2354 & \si{\electronvolt/vac} \\
\bottomrule
\end{tabular}
\caption[Comparison of energetics between DFT and foundation models]{
Mean absolute error (MAE) and root mean squared error (RMSE) for formation energies of oxide perovskite bulks and surfaces with and without vacancies predicted by a set of foundation models (MPACE-MPA-0~\cite{batatia2023FoundationModelAtomistic}, MACE-MP-0~\cite{batatia2023FoundationModelAtomistic}, and CHGNet~\cite{Deng2023CHGNETPretrainedUniversalNeural}).
The errors are given with respect to the energies of formation calculated from GGA-PBE~\cite{Perdew1996} DFT energetics (where structures have been relaxed using GGA-PBE and are fed into the foundation models without any additional structural relaxation).
All errors are in units of the final column.
}
\label{tab:foundation_errors}
\end{table}

These results underscore the challenges of using current foundation models in chemically complex environments.
Although surface energies are reasonably captured, vacancy-related energies are not, despite their conceptual similarity—both involve under-coordinated atoms.
This highlights the need for either expanded training datasets or targeted fine-tuning to improve defect modelling fidelity.

\section{Results}
\label{sec:results}

\subsection{Bulk properties}
\label{sec:bulk}

\begin{figure}[ht!]
    \centering
    \includegraphics[scale=0.6]{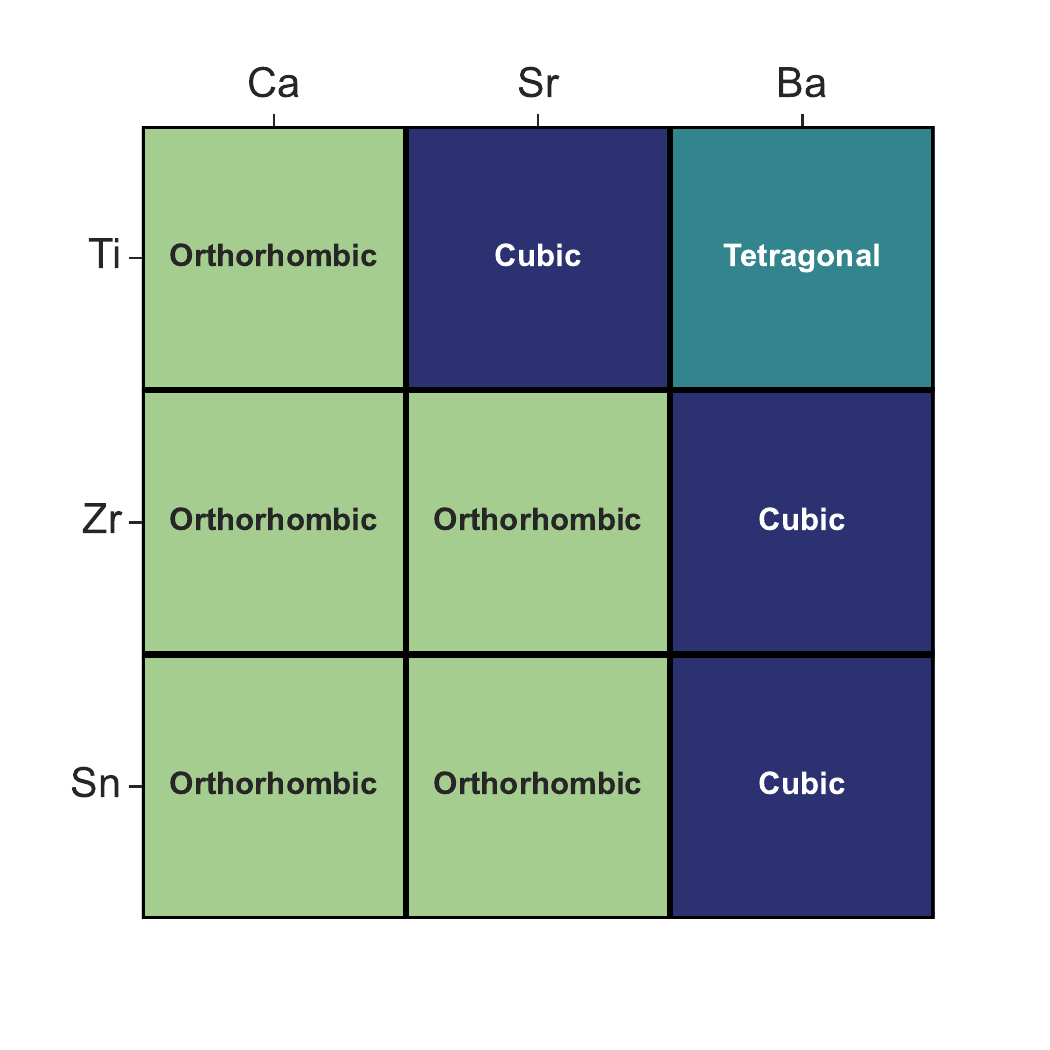}
    \caption[Room temperature phase of oxide perovskite]{Heat map detailing the phases of the nine oxide perovskites at room temperature.}
    \label{fig:phasemap}
\end{figure}

The oxide perovskites considered in the main article are studied in one of three possible phases -- their most favourable phase at room temperature~\cite{Fischer1993,Levin2003,Honorio2018}. The chosen phases are presented in \figref{fig:phasemap}, except \sto{}, which was taken as tetragonal. Furthermore, the theoretical lattice constants, formation energies, and band gaps of the perovskites are made available in \tabref{tab:bulk}. Note, whilst \sto{} is known to be cubic at room temperature, thin films and nanoparticles are often found to be tetragonal~\cite{Yuan2012}. In our study, slabs of cubic \sto{} were found to relax to the tetragonal phase due to the freedom in the $c$-axis, imposed by the presence of the vacuum gap along that axis. As such, we study \sto{} in the tetragonal phase in this work. However, studies of vacancies in bulk cubic \sto{} and in thinner slabs of cubic \sto{} (4 unit cells thick) showed comparable results, and the same trends remained.

\begin{table}[ht]
    \centering
    \begin{tabular}{cP{4em}P{4em}P{4em}P{6em}P{6em}P{6em}}
    \hline
        \multirow{2}{*}{Perovskite} & \multicolumn{3}{c}{Lattice constants (\si{\angstrom{}})} & \multicolumn{2}{c}{Energies} & Volume\\\cmidrule(lr){2-4}\cmidrule(lr){5-6}
        & $a$ & $b$ & $c$ & $E_\textrm{form}$~(\si{\electronvolt}/fu) & $E_g~(\si{\electronvolt})$ & change (\%)\\\hline
        \cto{} & 5.403 & 5.499 & 7.690 & -0.687 & 2.382 & 0.51 \\
        \sto{} & 3.938 & 3.938 & 3.939 & -1.120 & 1.809 & 0.28 \\
        \bto{} & 4.031 & 4.031 & 4.033 & -1.263 & 1.703 & 0.66 \\\hline
        \czo{} & 5.613 & 5.806 & 8.074 & -0.364 & 3.999 & 0.22 \\
        \szo{} & 5.826 & 5.880 & 8.264 & -0.762 & 3.717 & 0.11 \\
        \bzo{} & 4.234 & 4.234 & 4.234 & -1.169 & 3.124 & 0.04 \\\hline
        \cso{} & 5.579 & 5.752 & 7.993 & -0.397 & 2.333 & 0.19 \\
        \sso{} & 5.778 & 5.812 & 8.185 & -0.712 & 1.735 & 0.23 \\
        \bso{} & 4.184 & 4.184 & 4.184 & -0.982 & 0.346 & 1.29 \\\hline
    \end{tabular}
    \caption[Energetic and structural properties of oxide perovskites]{
        Theoretical lattice constants, formation energy, and band gap for the nine oxide perovskites, obtained using the GGA-PBE functional.
        All lattice constant angles ($\alpha$, $\beta$, and $\gamma$) are $90$\si{\degree}, as is the case for cubic, tetragonal, and orthorhombic phases.
        The volume change relates to the average (over $A$-, $B$-, and O-site vacancies) volume expansion of one chemical unit of the perovskite caused by introducing a vacancy into a $2\times2\times2$ supercell of said perovskite, whilst a couple of $A$-site vacancies are found to result in a volume compression, the averages over the three sites exhibit only expansion.}
    \label{tab:bulk}
\end{table}

\subsection{Bulk vacancies}
\label{sec:bulk:vacs}

For orthorhombic phase oxide perovskites, there exist two symmetrically inequivalent oxygen sites -- the $4c$ and $8d$ Wyckoff positions (whilst the tetragonal phase also exhibits two symmetrically inequivalent sites, the difference is so small that energetic differences in taking vacancies in those sites is well below the accuracy of the methods used).
In  $2\times2\times2$ supercell (i.e. 32 \abo{} units), the most energetically favourable Wyckoff position for an oxygen vacancy in \cto{}, \szo{}, and \cso{} is the $4c$ position; in the same supercell for \czo{} and \sso{}, the most favourable Wyckoff site is $8d$.
The energy difference between sites (i.e. $E_{4c} - E_{8d}$) is as follows:
\cto{} $ = -0.0030$~\si{\electronvolt}/vac,
\czo{} $ = 0.0003$~\si{\electronvolt}/vac,
\szo{} $ = -0.0075$~\si{\electronvolt}/vac,
\cso{} $ = -0.0773$~\si{\electronvolt}/vac, and
\sso{} $ = 0.0047$~\si{\electronvolt}/vac.
These energy differences are well within the expected error margin of DFT calculations~\cite{Lejaeghere2016ReproducibilityDFT} (e.g. due to the choice of functional or comparison to experiment).

\begin{figure}[ht!]
    \centering
    \subfloat[From elemental references]{\includegraphics[width=0.45\linewidth]{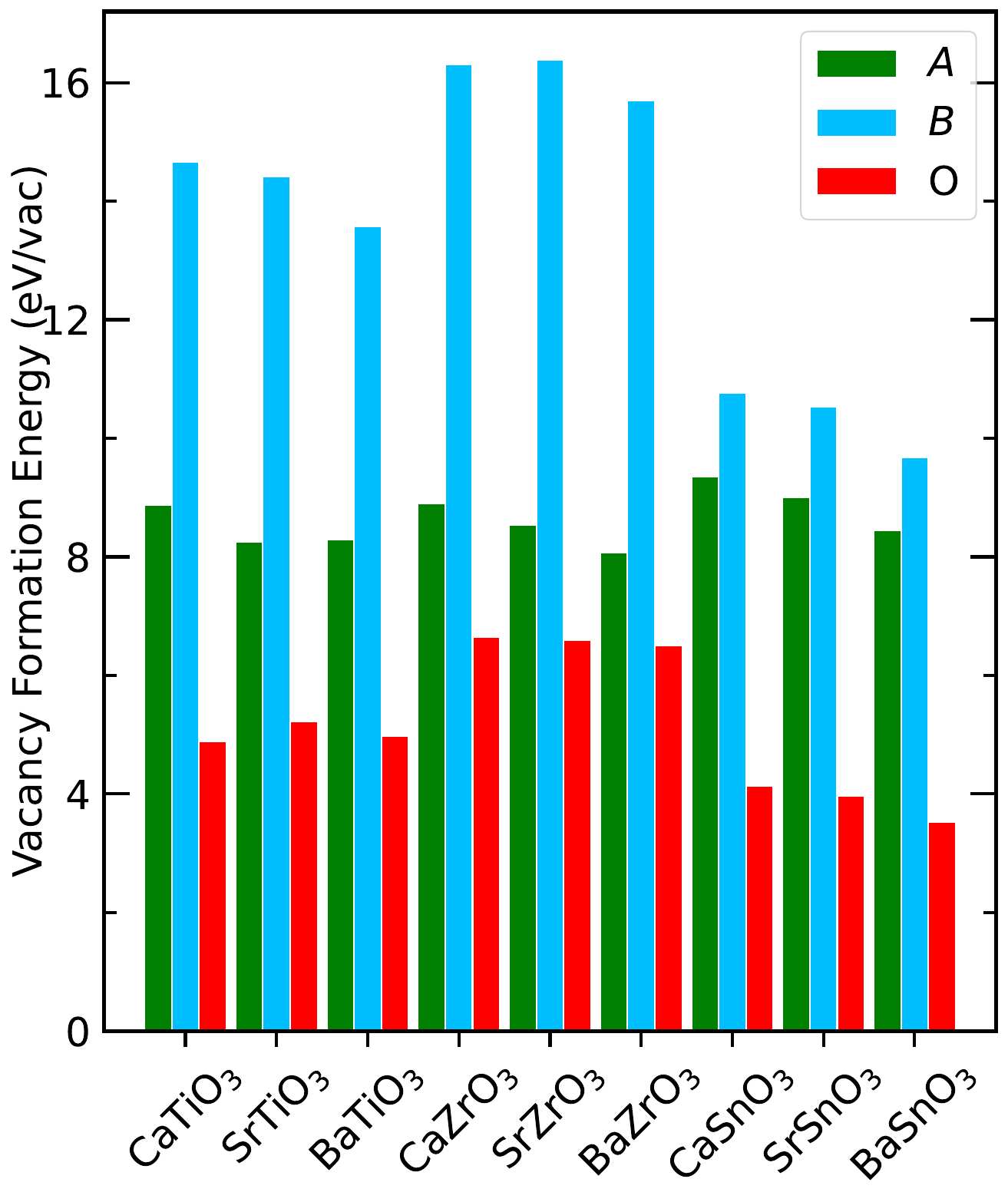}\label{fig:vac:bulk:from_elements}}%
    \hspace{1.5em}%
    \subfloat[From oxides]{\includegraphics[width=0.45\linewidth]{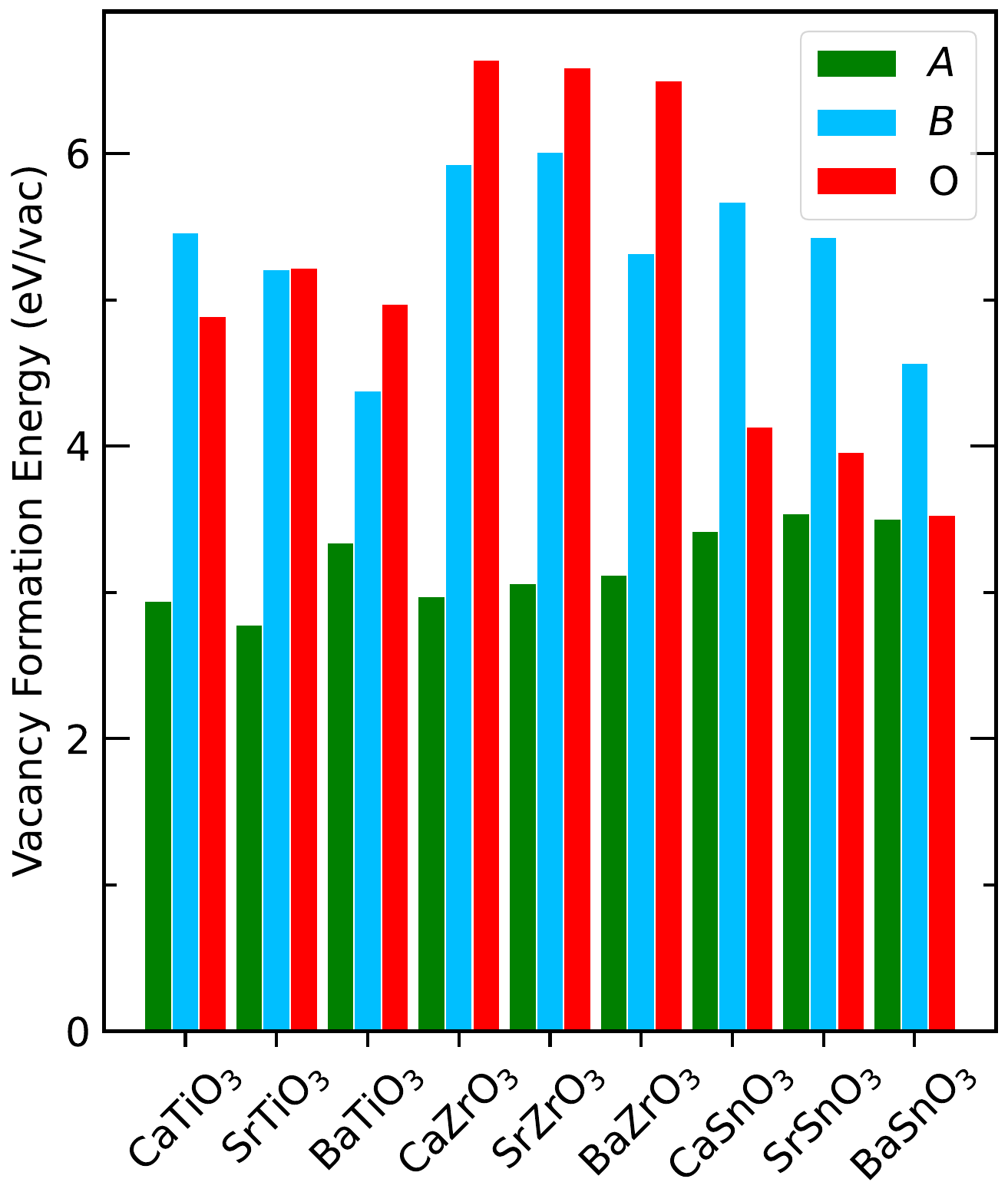}\label{fig:vac:bulk:from_oxides}}%
    \caption[Bulk vacancy formation energies]{
    The formation energy of $A$-, $B$- and O-site vacancies in various oxide perovskite (\abo{}) bulk crystals. 
    The formation energies in \protect\subref{fig:vac:bulk:from_elements} are given with respect to the $A$ and $B$ bulk metals and the \o{} molecule.
    The formation energies in \protect\subref{fig:vac:bulk:from_oxides} are given with respect to the \ao{} and \bo{} oxides and the \o{} molecule.
    }
    \label{fig:vac:bulk:alt}
\end{figure}

\figref{fig:vac:bulk:alt} shows the formation energy of $A$-, $B$-, and O-site vacancies within the perovskites.
These figures differ from that seen in the main article in how they calculate the formation energies.
For \figref{fig:vac:bulk:from_elements}, the formation energies are given with respect to the bulk metals ($A$ and $B$), whilst \figref{fig:vac:bulk:from_oxides} gives them with respect to the \ao{} and \bo{} oxides.
The main article figure gives them with respect to the O-rich environment (defined as the \abo{}-\ao{}-\o{} limit).
For the two subfigures of \figref{fig:vac:bulk:alt}, the O-site vacancies are the same, as an \o{} molecule is used as reference in both.

\begin{table}[ht]
\begin{tabular}{P{5em}c} 
  \hline
  Oxide & $E_\textrm{form}$ (\si{\electronvolt}/fu) \\\hline
  CaO & -5.931\\
  SrO & -5.473\\
  BaO & -4.945\\\hline
  TiO$_2$ & -9.202\\
  ZrO$_2$ & -10.371\\
  SnO$_2$ & -5.096\\\hline\hline
  TiO & -5.149 \\
  ZrO & -4.581 \\
  SnO & -2.686 \\\hline
  Ca$_{2}$TiO$_{4}$ & -21.512 \\
  Sr$_{2}$TiO$_{4}$ & -21.466 \\
  Ba$_{2}$TiO$_{4}$ & -21.141 \\\hline
  Ca$_{2}$ZrO$_{4}$ & -21.602 \\
  Sr$_{2}$ZrO$_{4}$ & -21.867 \\
  Ba$_{2}$ZrO$_{4}$ & -21.624\\\hline
  Ca$_{2}$SnO$_{4}$ & -17.624 \\
  Sr$_{2}$SnO$_{4}$ & -16.919 \\
  Ba$_{2}$SnO$_{4}$ & -16.489 \\\hline
\end{tabular}
  \caption[Formation energies of binary oxides and oxide perovskites]{
  Formation energy of the considered oxides $A$=Ca, Sr, and Ba, and $B$=Ti, Zr, and Sn.
  The formation energy is given with respect to the bulk $A$ or $B$ metal, and the \o{} molecule, and are presented as per formula unit.
  As expected, all formation energies of oxides are negative, meaning that they are more favourable than the bulk metals and \o{} gas for equal concentrations.
  }
\label{tab:form:oxide}
\end{table}

\tabref{tab:form:oxide} highlights the formation energy of the \ao{} and \bo{} binary oxides considered in this work. It can be seen that all the oxides here have negative formation energy, meaning that they are more favourable than their bulk metals and \o{} gas.

The averaged percentage volume distortions for the perovskites with a vacancy within a $2\times2\times2$ supercell are presented in \tabref{tab:bulk} (averaged over the $A$-, $B$-, and O- site vacancies).
In these bulk vacancy calculations, the cells are allowed to relax to mimic the $c$-axis degree of freedom available to vacancies in slab structures due to the $14$~\si{\angstrom} vacuum gap present in the latter.
It can be seen that introducing a vacancy induces a volume expansion of around $0.4$~\% (with respect to the volume of the undefected bulk unit).
Note that, for the charged defect calculations performed for \sso{}, fixed-cell relaxations are performed (fixed to the undefected bulk lattice constants) and used as this is required by the charge correction methods presented in literature~\cite{Kumagai2014ElectrostaticBasedFinite}.

\subsection{Charged vacancies compared to literature}
\label{sec:charge_states:lit_comparison}

\begin{figure*}[ht]
    \centering
    \subfloat[bulk]{\includegraphics[width=0.31\linewidth]{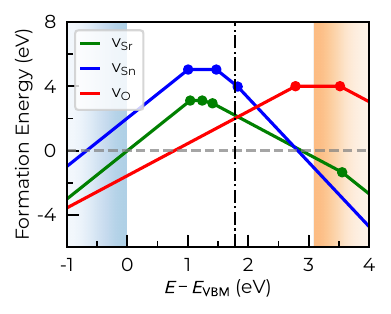}\label{fig:charge_state:weston:bulk}}%
    \hspace{1em}%
    \subfloat[\ao{}-terminated slab]{\includegraphics[width=0.31\linewidth]{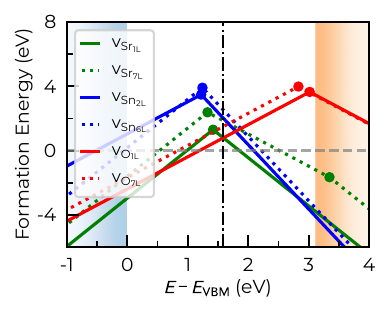}\label{fig:charge_state:weston:AO}}%
    \hspace{1em}%
    \subfloat[\bo{}-terminated slab]{\includegraphics[width=0.31\linewidth]{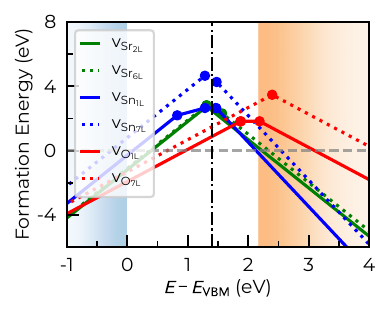}\label{fig:charge_state:weston:BO2}}%
    \caption[Direct literature comparison of charged vacancies in \sso{}]{
        Formation energies of $A$-, $B$- and, O-site charged vacancies under O-rich conditions in \sso{} \protect\subref{fig:charge_state:weston:bulk} bulk, and \protect\subref{fig:charge_state:weston:AO} \ao{}- and \protect\subref{fig:charge_state:weston:BO2} \bo{}-terminated slabs.
        Line gradients correspond to the charge state (in $|$\si{\elementarycharge}$|$) and points of discontinuity identify transition levels.
        The valence (conduction) band region is shaded in blue (orange) and the vertical dash-dotted line is the equilibrium Fermi level.
        The oxygen rich environment as defined in reference~\cite{Weston2018} is used, i.e. $\mu_{\mathrm{Sr}} = -5.91$~\si{\electronvolt}, $\mu_{\mathrm{Sn}} = -5.55$~\si{\electronvolt}, and $\mu_{\mathrm{O}} = 0.00$~\si{\electronvolt}, with respect to the elemental references.
        Corrections to the charge state energies are calculated using \texttt{doped}~\cite{Kavanagh2024DopedPythonToolkit}, which employs band edge shifting~\cite{Broberg2023HighThroughputCalculations} and image charge~\cite{Kumagai2014ElectrostaticBasedFinite} corrections.
        Slab charge corrections~\cite{Freysoldt2009,Freysoldt2018FirstPrinciplesCalculations} are accounted for in subfigures \protect\subref{fig:charge_state:weston:AO} and \protect\subref{fig:charge_state:weston:BO2} using \texttt{qdef2d}~\cite{Tan2019ChargedDefectsFramework} and \texttt{sxdefectalign2d}~\cite{Freysoldt2009,Freysoldt2018FirstPrinciplesCalculations}.
    }
    \label{fig:charge_state:weston}
\end{figure*}

The accuracy of the band edge shift method for correcting charged vacancy formation energies in \sso{} from GGA-PBE to HSE06 results is assessed.
Previous studies have reported charge states for Sr, Sn, and O vacancies~\cite{Weston2018,Lin2024TraceImpurityMatters} under O-rich conditions, though with differing chemical potentials.
Reference~\cite{Weston2018} employs $\mu_{\mathrm{Sr}} = -5.91$~\si{\electronvolt}, $\mu_{\mathrm{Sn}} = -5.55$~\si{\electronvolt}, and $\mu_{\mathrm{O}} = 0.00$~\si{\electronvolt}, relative to elemental references.
To enable direct comparison, charged state figures from the main article are recalculated using these values and presented in \figref{fig:charge_state:weston}.

Weston~\etal{} report the V$_{\mathrm{O}}$ (+2/0) transition at $2.85$~\si{\electronvolt} above the valence band maximum (VBM) and V$_{\mathrm{O}}^{+2}$ crossing zero formation energy at $0.55$~\si{\electronvolt}.
The V$_{\mathrm{Sr}}^{-2}$ and V$_{\mathrm{Sn}}^{-4}$ charge states cross zero formation energy at $2.82$ and $2.94$~\si{\electronvolt}, respectively.
V$_{\mathrm{O}}^{0}$ shows a formation energy of $4.59$~\si{\electronvolt} 

\begin{figure*}[ht!]
    \centering
    \includegraphics[width=0.31\linewidth]{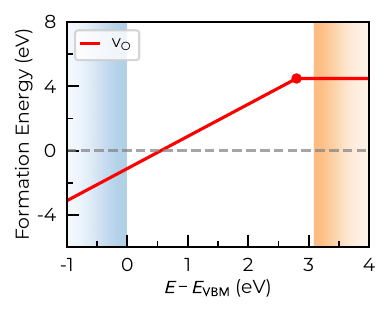}%
    \caption[Bulk \sso{} charged oxygen vacancy energies from HSE06]{
        Formation energies the O-site charge neutral and $+2$ charged vacancy under O-rich conditions in bulk \sso{}, calculated using the one-shot HSE06 method.
        Line gradients correspond to the charge state (in $|$\si{\elementarycharge}$|$) and points of discontinuity identify transition levels.
        The valence (conduction) band region is shaded in blue (orange) and the vertical dash-dotted line is the equilibrium Fermi level.
        The oxygen rich environment as defined in reference~\cite{Weston2018} is used, i.e. $\mu_{\mathrm{Sr}} = -5.91$~\si{\electronvolt}, $\mu_{\mathrm{Sn}} = -5.55$~\si{\electronvolt}, and $\mu_{\mathrm{O}} = 0.00$~\si{\electronvolt}, with respect to the elemental references.
        Corrections to the charge state energies are calculated using \texttt{doped}~\cite{Kavanagh2024DopedPythonToolkit} to employ the image charge~\cite{Kumagai2014ElectrostaticBasedFinite} correction.
    }
    \label{fig:charge_state:weston:bulk:Ovac_only}
\end{figure*}

Using the same chemical potentials, the present calculations yield a V$_{\mathrm{O}}$ (+2/0) transition at $2.784$~\si{\electronvolt} above the VBM and V$_{\mathrm{O}}^{+2}$ crossing zero at $0.80$~\si{\electronvolt}.
The V$_{\mathrm{Sr}}^{-2}$ and V$_{\mathrm{Sn}}^{-4}$ crossing points shift to $2.89$ and $2.82$~\si{\electronvolt}, respectively.
V$_{\mathrm{O}}^{0}$ has a formation energy of $4.01$~\si{\electronvolt} (for fixed-cell vacancy relaxation, whereas the value presented in \tabref{tab:vacancies:o} is for cell-free vacancy relaxation).
The two discrepancies between this work and literature are the $0.5$~\si{\electronvolt} downward shift in oxygen chemical potential and V$_{\mathrm{Sr}}^{-2}$ and V$_{\mathrm{Sn}}^{-4}$ crossing above rather than below zero formation energy.
As seen in \figref{fig:charge_state:weston:bulk:Ovac_only}, the chemical potential error is corrected for when doing all calculations using HSE06 (returning a charge neutral oxygen vacancy formation energy of $4.49$~\si{\electronvolt}), so it can be said that this difference in chemical potential arises from a difference in choice of functional (where one-shot hybrid approach has been taken, in which all structures have been relaxed using GGA-PBE and then a single HSE06 calculation is performed on each of the systems to correct the energetics).
Due to the number of and computational cost associated with vacancy calculations in slabs, the authors choose to perform the study using the GGA-PBE functional, corrections of $-0.07$, $-0.12$, and $0.5$~\si{\electronvolt} could be applied to all Sr, Sn, and O vacancy formation energies, respectively.

The band edge shift method demonstrates good agreement with literature values, confirming its applicability to \sso{} and supporting its use in qualitative analysis.
The \ao{}- and \bo{}-terminated slabs are also recalculated using the chemical potentials from~\cite{Weston2018} and presented in \figref{fig:charge_state:weston}.

\subsection{Surface formation energy}
\label{sec:form:surf}

\begin{table}[ht]
\begin{tabular}{P{8em}P{3em}P{3em}P{5em}P{3em}P{3em}P{5em}} 
  \hline
  \multirow{3}{*}{Perovskite} &\multicolumn{6}{c}{Surface Energy (\si{\electronvolt}/su)} \\\cmidrule(lr){2-7}
  & \multicolumn{3}{c}{Literature} & \multicolumn{3}{c}{Our work}   \\\cmidrule(lr){2-4}\cmidrule(lr){5-7}
  & \ao{} & \bo{} & $\Delta$ & \ao{} & \bo{} & $\Delta$ \\
  \hline
  CaTiO$_3$~\cite{Eglitis2009} & 0.83 & 0.74 &  0.09   &   0.98 & 0.90 &  0.08 \\
  SrTiO$_3$~\cite{Eglitis2009} & 1.15 & 1.23 & -0.08   &   0.96 & 0.97 & -0.01 \\
  BaTiO$_3$~\cite{Eglitis2009} & 1.19 & 1.07 &  0.12   &   0.98 & 0.84 &  0.14 \\
  CaZrO$_3$~\cite{Eglitis2019} & 0.87 & 1.33 & -0.46   &   1.09 & 1.10 & -0.01 \\
  BaZrO$_3$~\cite{Eglitis2009} & 1.30 & 1.31 & -0.01   &   1.11 & 1.06 &  0.05 \\
  BaSnO$_3$~\cite{Slassi2017} & 1.50 & 1.48 &  0.02   &   1.02 & 1.08 & -0.06 \\
  PbTiO$_3$~\cite{Eglitis2009} & 1.15 & 1.23 &  -0.08  &   -- & -- & --\\
  KNbO$_3$~\cite{Slassi2017}  & 1.21 & 0.75 &  0.46   &   -- & -- & -- \\\hline
  NaTaO$_3$*~\cite{Zhao2019}  & 0.10 & 0.07 &  0.03   &   -- & -- & -- \\  
  KTaO$_3$*~\cite{Zhao2019}   & 0.09 & 0.07 &  0.02   &   -- & -- & -- \\  
\hline
\end{tabular}
  \caption[Direct literature comparison of perovskite surface formation energies]{Surface formation energies of various oxide perovskites presented in literature. $\Delta$ is the difference in formation energy between the two surfaces $\Delta = E_\textrm{form,\ao{}} - E_\textrm{form,\bo{}}$. The surface energies for all compounds except those denoted with an asterisks (*) are calculated using the formalism outlined in reference~\cite{Eglitis2007}, with units of \si{\electronvolt}/su, where u is one surface formula unit. Those marked with an asterisks -- NaTaO$_3$ and KTaO$_3$ -- are calculated using the method detailed in these works, and, as such, the formation energy is given in units of \si{\electronvolt/\angstrom\squared}.}
\label{tab:form:lit}
\end{table}

\tabref{tab:form:lit} shows surface energies for a variety of \abo{} perovskites. Aside from NaTaO$_3$ and KTaO$_3$, which use the formalism employed in this work, the surface formation energies are are obtained using the equations outlined by Eglitis \etal{}~\cite{Eglitis2007}. We start with the energy cost associated with cleaving a bulk along the (001) plane into a pair of \ao{} and \bo{} slabs,

\begin{equation}
    E_\textrm{surf}^\textrm{unr} = \frac{1}{4}\left[ E_\textrm{slab}^\textrm{unr}(\textrm{\ao{}}) + E_\textrm{slab}^\textrm{unr}(\textrm{\bo{}}) - N E_\textrm{bulk} \right],
    \label{eq:altform:1}
\end{equation}

\noindent
where $X$=\ao{} or \bo{} highlights the termination, $E_\textrm{slab}^\textrm{unr}(X)$ is the energy of the unrelaxed (initially cleaved) energy of the $X$-terminated slab. $N$ is the total number of formula units in the \ao{} and \bo{} slabs. We divide by $4$ because there exist four surfaces in this total system (two surfaces on each slab), to provide us with an energy per (001) surface termination. The energy here relates to the energy of cleaving, but does not account for the energy of relaxation of those surfaces due to the cleaved bonds. To account for this relaxation energy, we determine the difference in energy between the unrelaxed and relaxed structure of slab $X$,

\begin{equation}
    \Delta E_\textrm{surf}^\textrm{rel}(X) = \frac{1}{2}\left[ E_\textrm{slab}(X) - E_\textrm{slab}^\textrm{unr}(X)\right],
    \label{eq:altform:2}
\end{equation}

\noindent
where $E_\textrm{surf}(X)$ is the energy of the relaxed $X$-terminated slab. With both of these, we can determine the energy of forming surface $X$ through cleaving as

\begin{equation}
    E_\textrm{surf}(X) = \frac{1}{U}\left[ E_\textrm{surf}^\textrm{unr} + \Delta E_\textrm{surf}^\textrm{rel}(X) \right].
    \label{eq:altform:3}
\end{equation}

\noindent
$U$ is the number of formula units on a surface.

This process involves determining the energy gained through relaxation of a particular surface after it has been cleaved. As such, this does not rely on chemical potential, as stoichiometry is maintained. However, this also means that the process cannot be as easily used for understanding of surface growth and, instead, is better suited for crystal cleaving.

The difference in surface formation energies $\Delta$ highlights the difference in surface energy of one termination over that of the other (where a positive value relates to a more favourable \bo{} surface, and a negative value for a more favourable \ao{} surface). Whilst there appears no general rule of favourability, it can be seen that the \bo{} is often more favourable than the \ao{} surface. A comparison of our results (employing the methodology outlined above) to those found in literature shows reasonable agreement of $\Delta$ value trends, with the largest discrepancies being for \czo{} and \bso{}, with deviations of $0.45$~\si{\electronvolt}/su and $0.08$~\si{\electronvolt}/su, respectively (with our results showing a preference for \ao{}, whilst the literature shows a preference for \bo{}). In the case of \czo{}, the literature value is obtained from a study exploring it in the cubic phase, whereas our results use the orthorhombic phase. For \bso{}, when performing the same analysis on smaller thickness slabs, we recover results closer to that of literature, with a \bso{} slab of four unit cells thickness resulting in a $\Delta$ of $0.01$, in good agreement with the $0.02$ reported by Slassi \etal{}~\cite{Slassi2017}.

Comparing the formation energies for NaTaO$_3$ and KTaO$_3$ to our results in the main article, we see a similar order of magnitude for the formation energy of \ao{} and \bo{} (001) surface terminations.

\subsection{Surface structures}
\label{sec:surf_struc}

\begin{figure*}[ht]
    \centering
    \subfloat[Relaxed \sto{} \ao{} slab]{\includegraphics[scale=0.16]{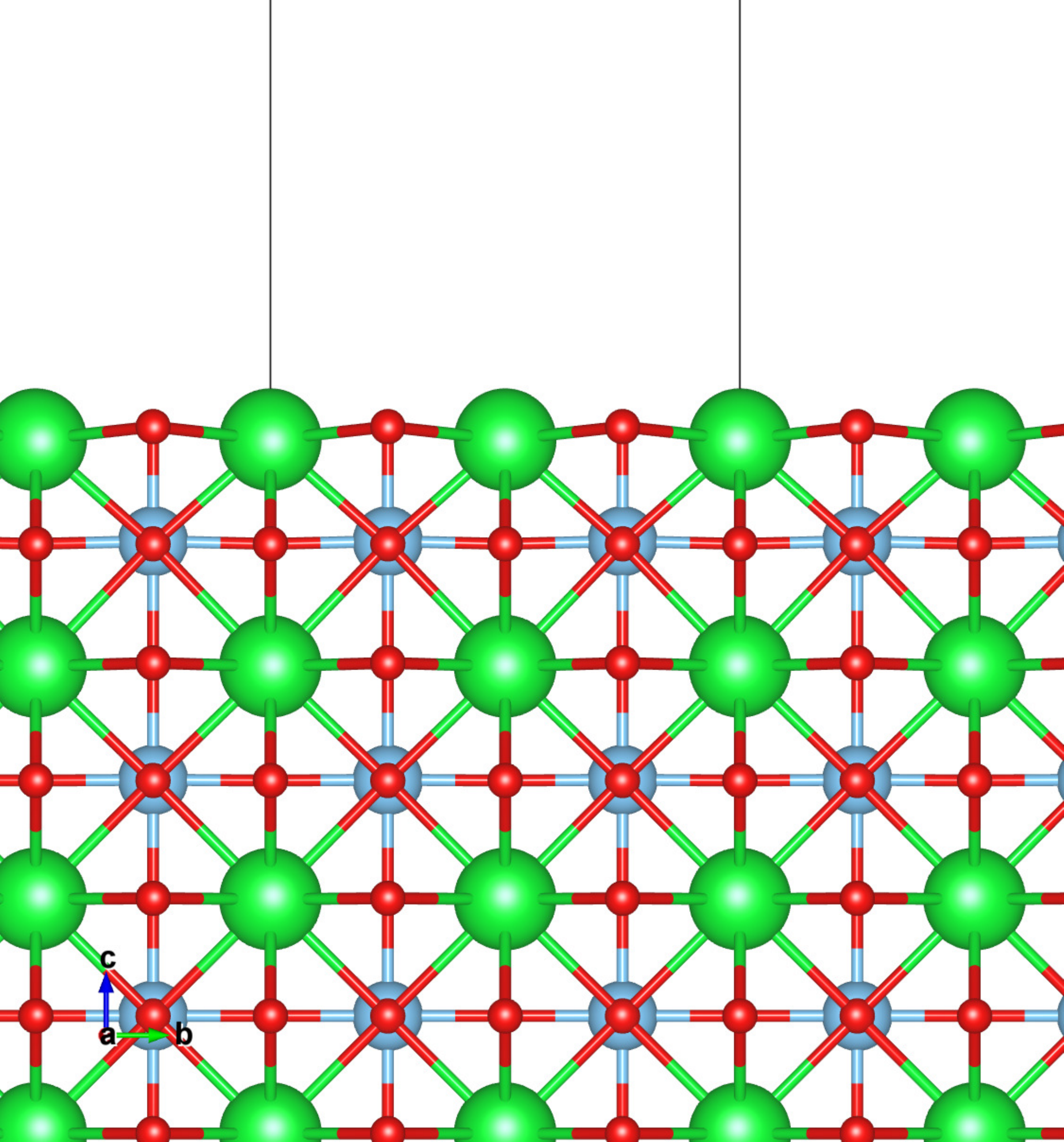}\label{fig:sto:ao}}%
    \hspace{2em}%
    \subfloat[Relaxed \sto{} \bo{} slab]{\includegraphics[scale=0.16]{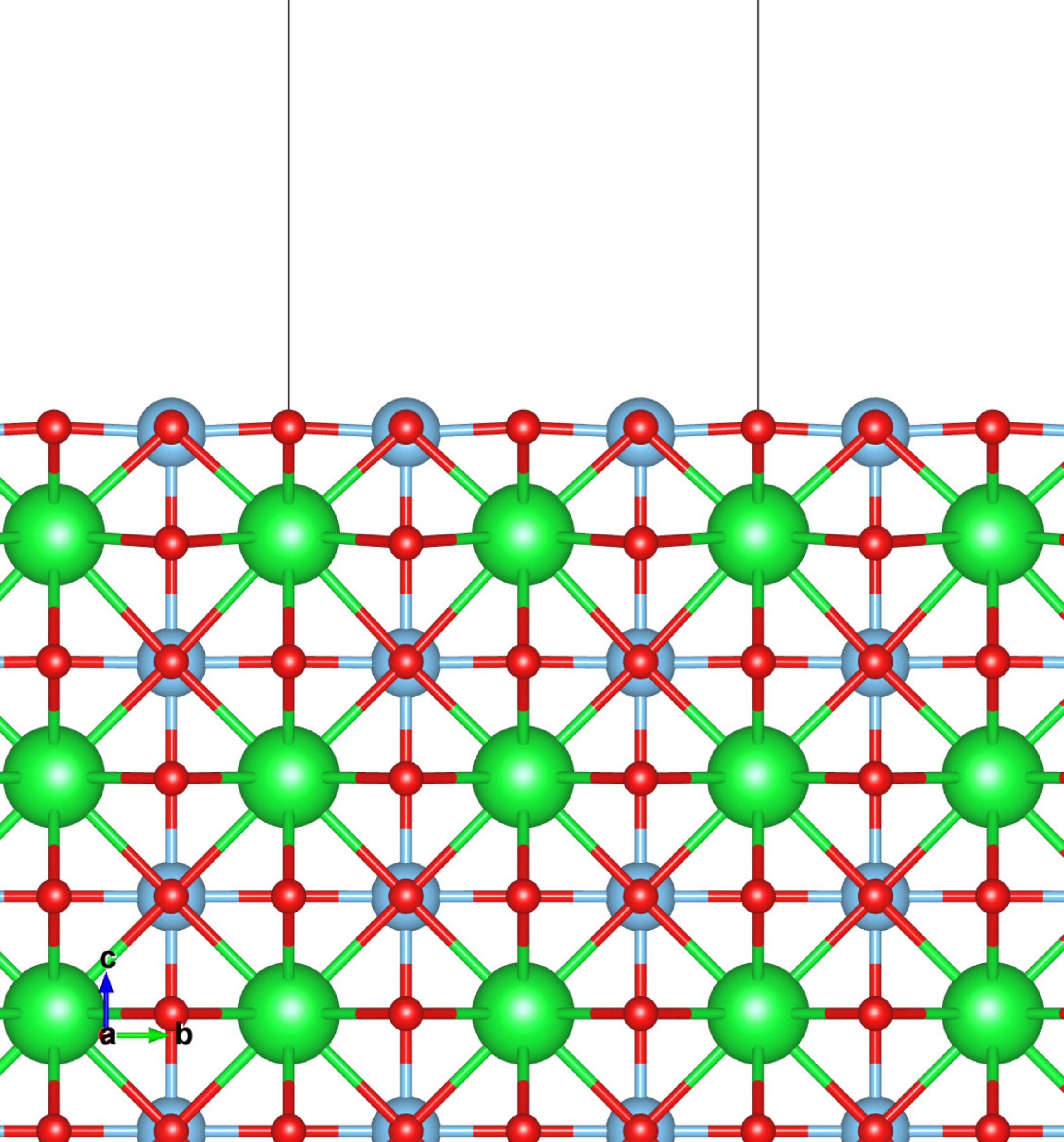}\label{fig:sto:bo2}}%
    \caption[Schematic of near-surface structural tilting]{
        Ball-and-stick representation of fully-relaxed tetragonal-phase \sto{} slabs.
        Tilting of the O--A--O and O--B--O bonds are present near the surface, pointing out of plane.
        Figure generated using VESTA~\cite{Momma2011}.
    }
    \label{fig:sto}
\end{figure*}

The tilting of bonds near the surface of undefected \sto{} slabs are presented in \figref{fig:sto}. This is provided as an example of the bond tilting near the surface.

\subsection{Chemical potential plots}
\label{sec:chem}

\begin{figure*}[ht]
    \centering
    \subfloat{\includegraphics[scale=0.22]{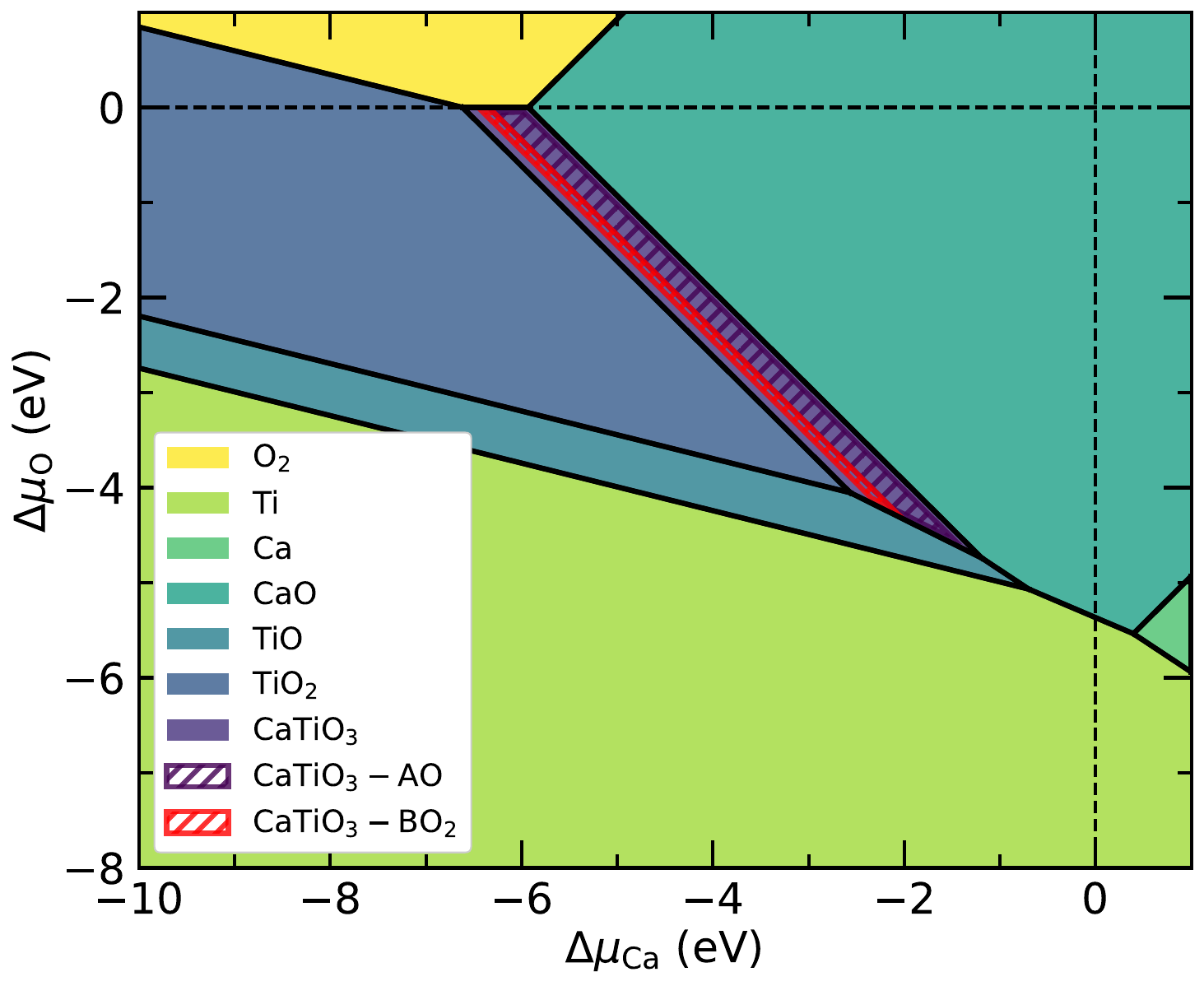}\label{fig:chem:cto}}%
    \hspace{0em}%
    \subfloat{\includegraphics[scale=0.22]{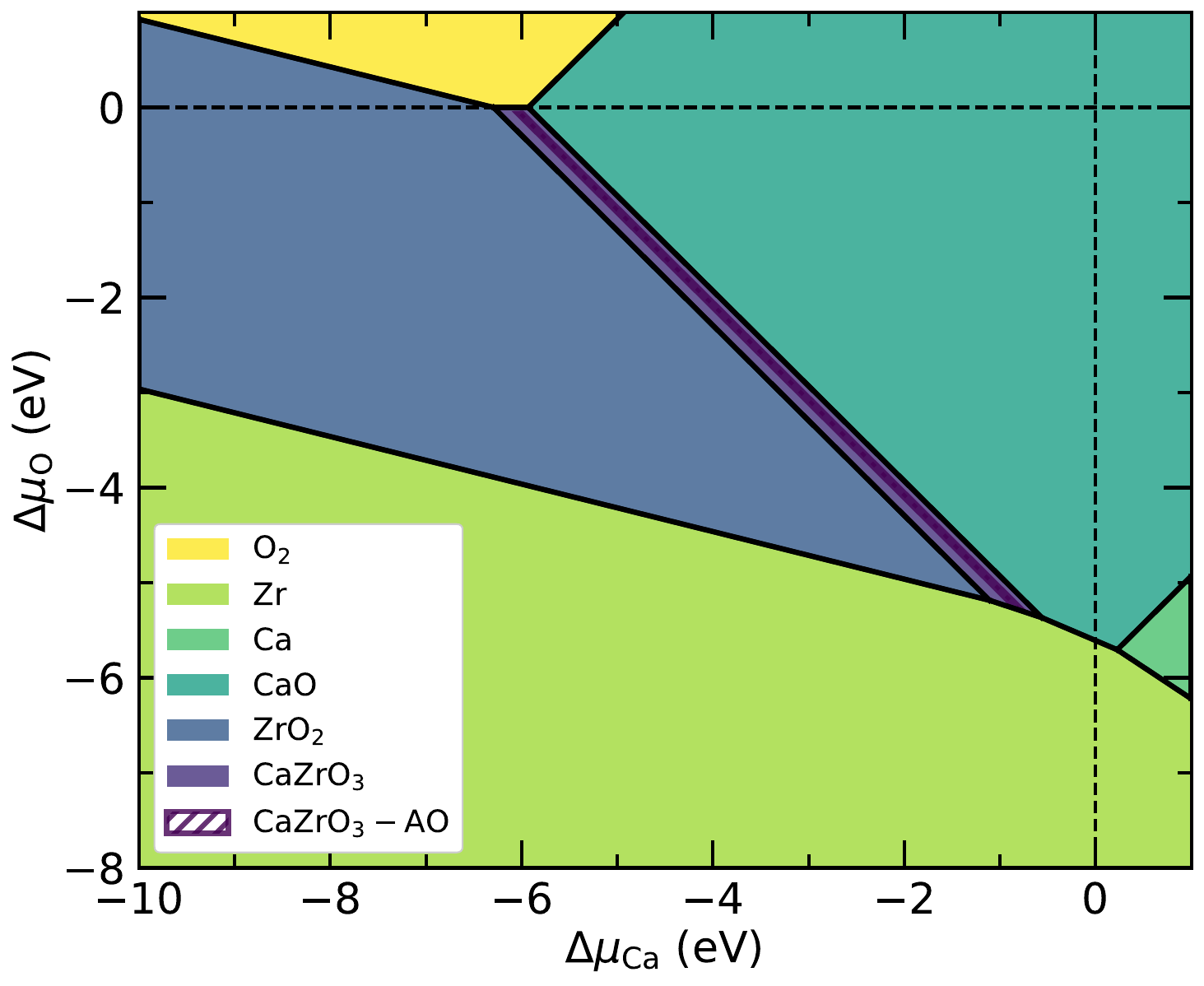}\label{fig:chem:czo}}%
    \hspace{0em}%
    \subfloat{\includegraphics[scale=0.22]{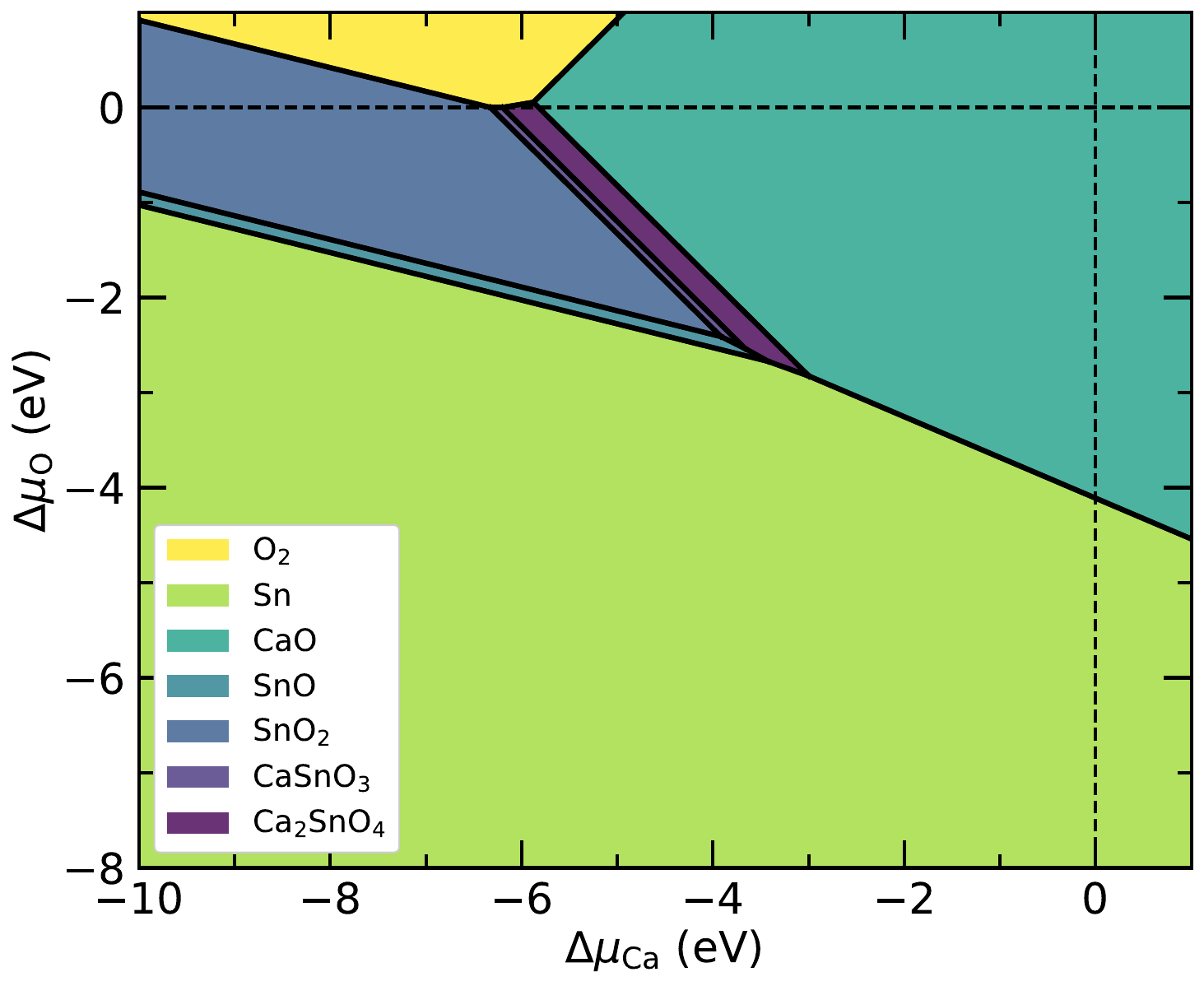}\label{fig:chem:cso}}\\
    \subfloat{\includegraphics[scale=0.22]{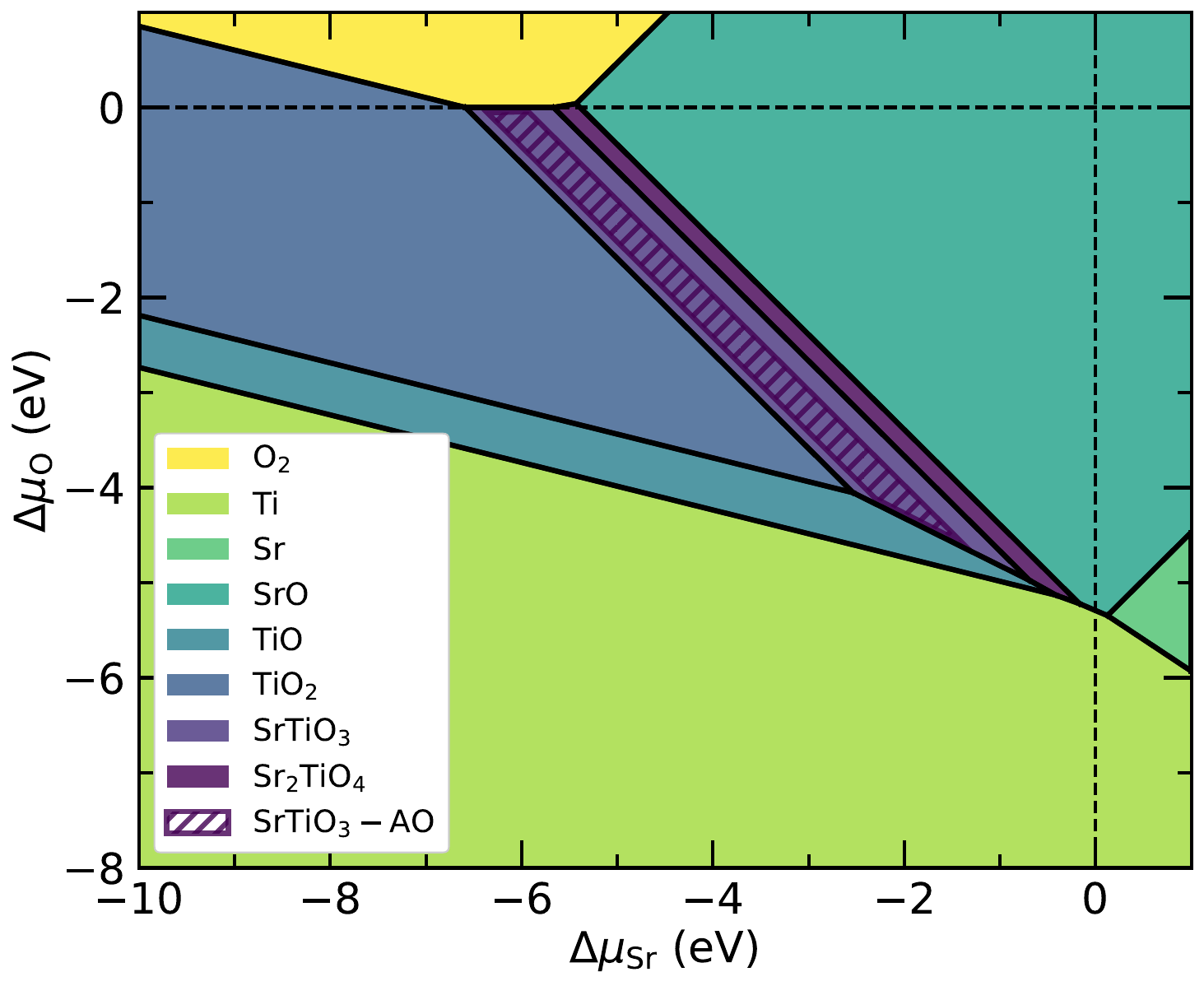}\label{fig:chem:sto}}%
    \hspace{0em}%
    \subfloat{\includegraphics[scale=0.22]{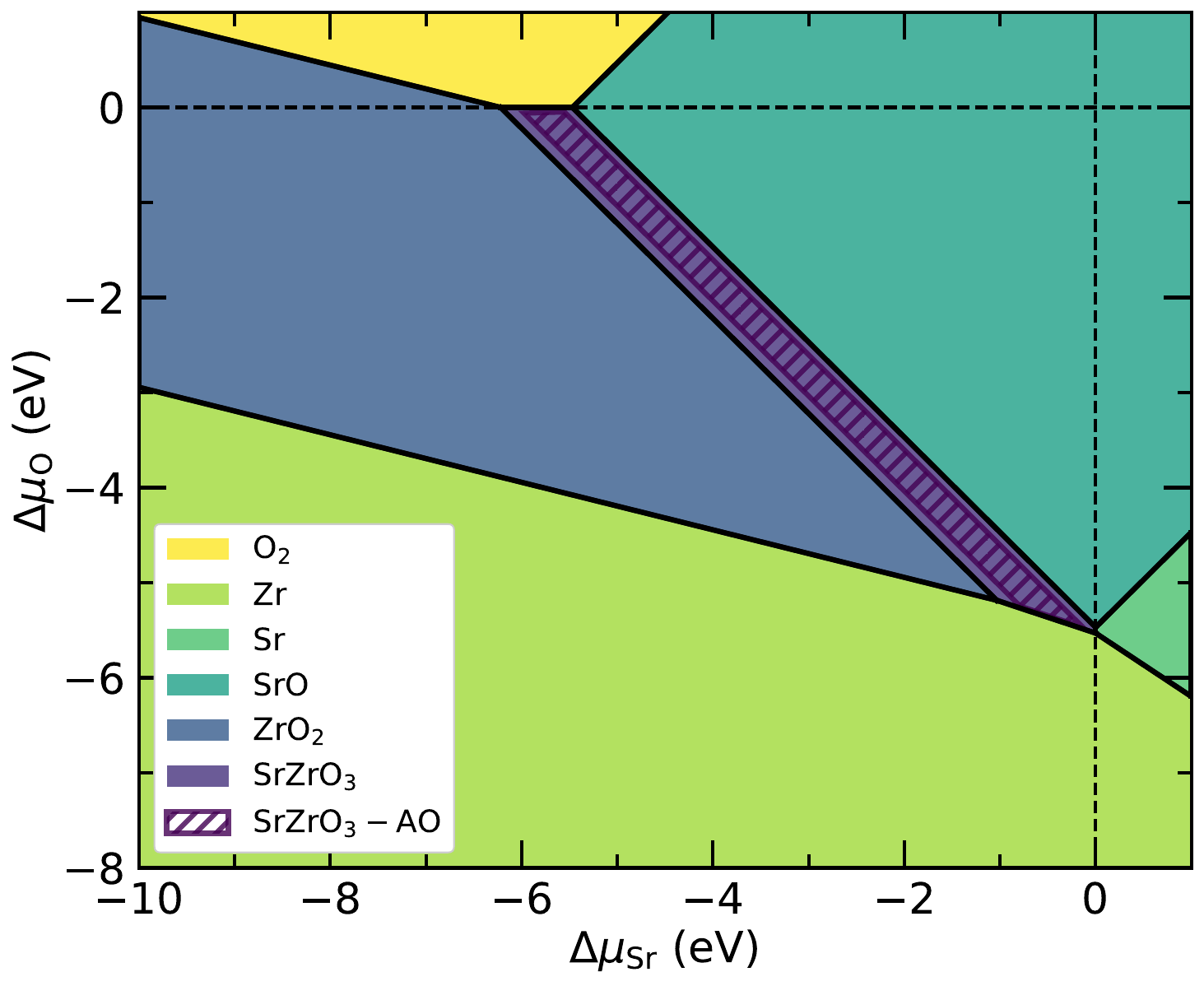}\label{fig:chem:szo}}%
    \hspace{0em}%
    \subfloat{\includegraphics[scale=0.22]{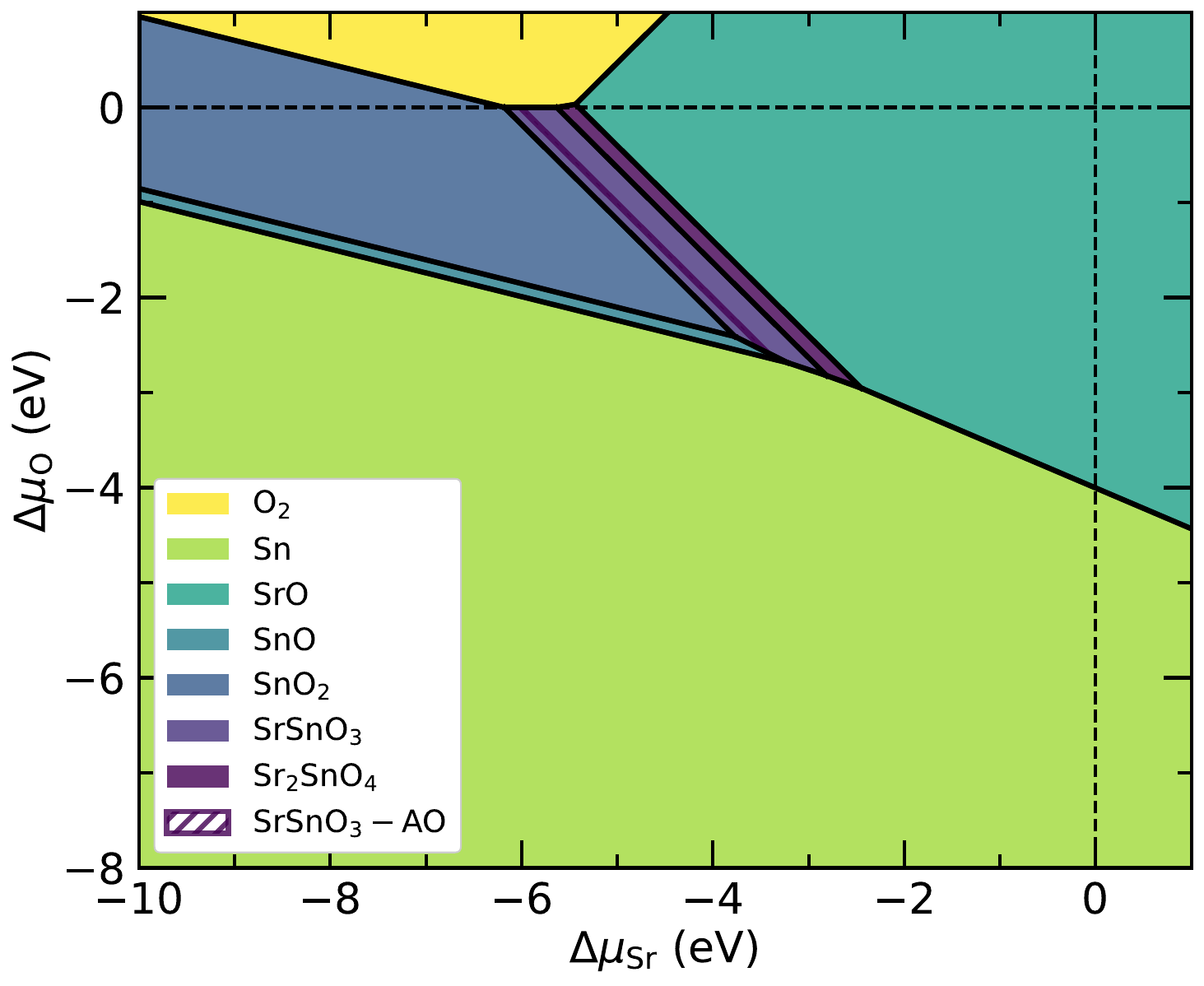}\label{fig:chem:sso}}\\
    \hspace{0em}%
    \subfloat{\includegraphics[scale=0.22]{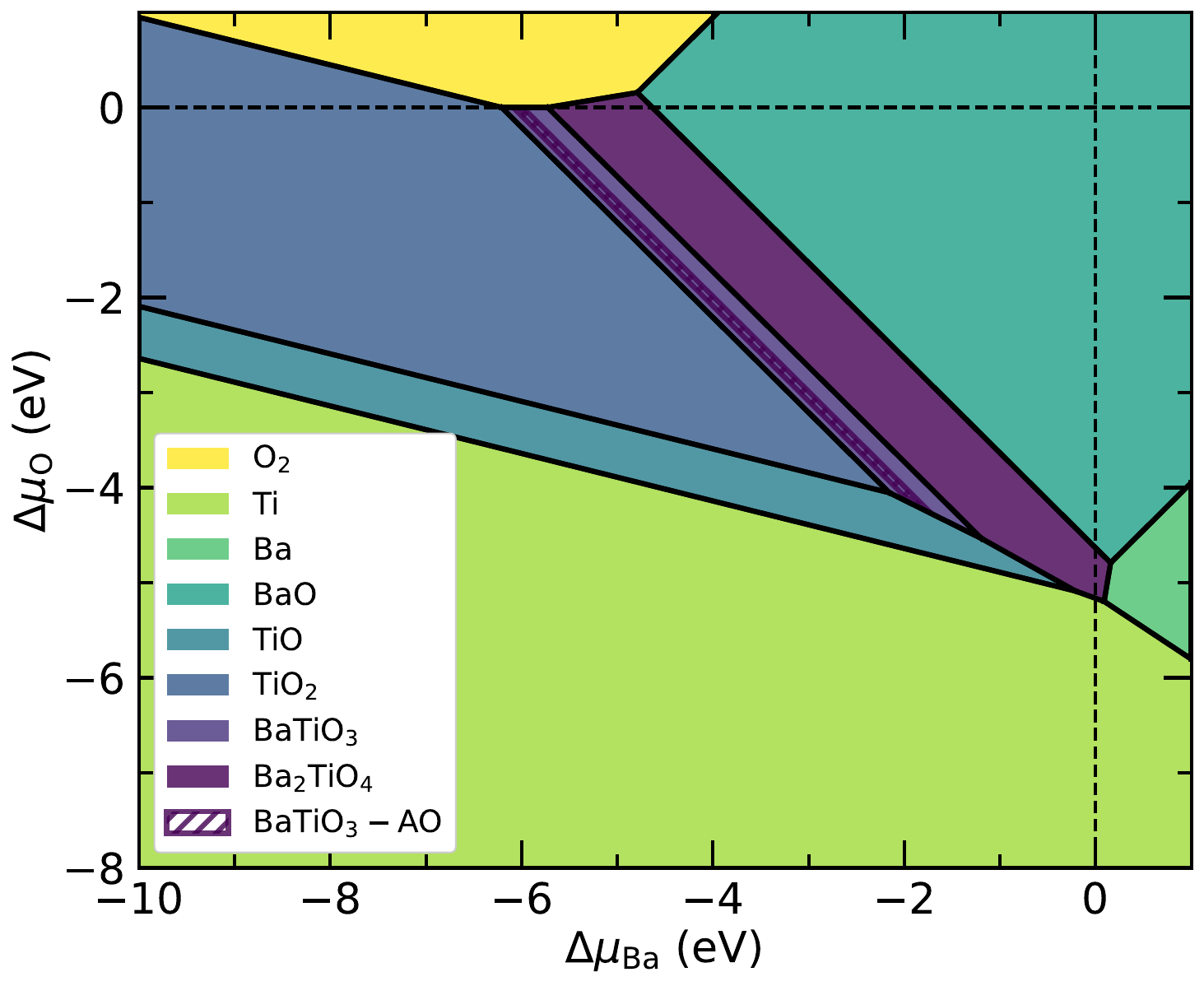}\label{fig:chem:bto}}%
    \hspace{0em}%
    \subfloat{\includegraphics[scale=0.22]{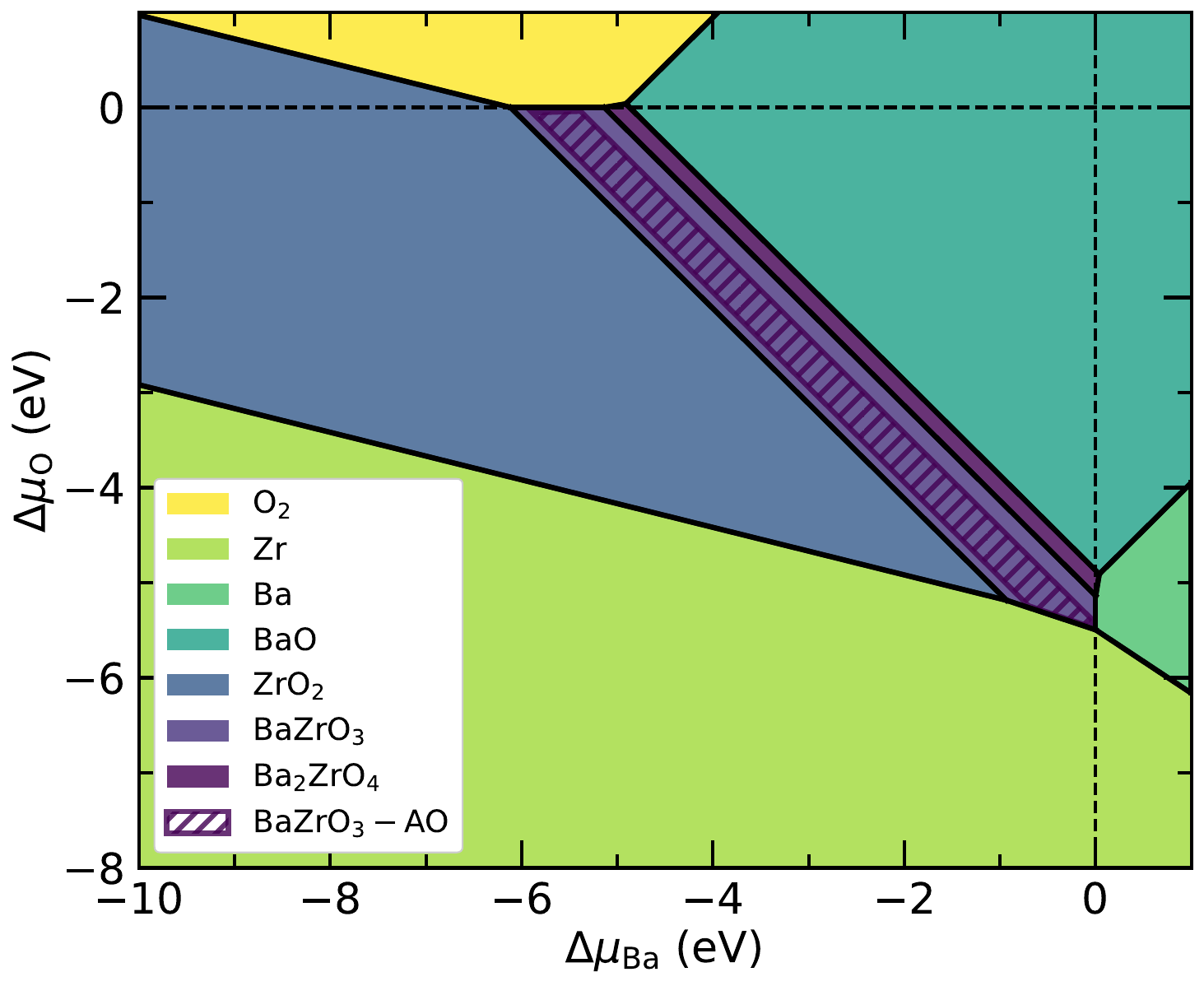}\label{fig:chem:bzo}}%
    \hspace{0em}%
    \subfloat{\includegraphics[scale=0.22]{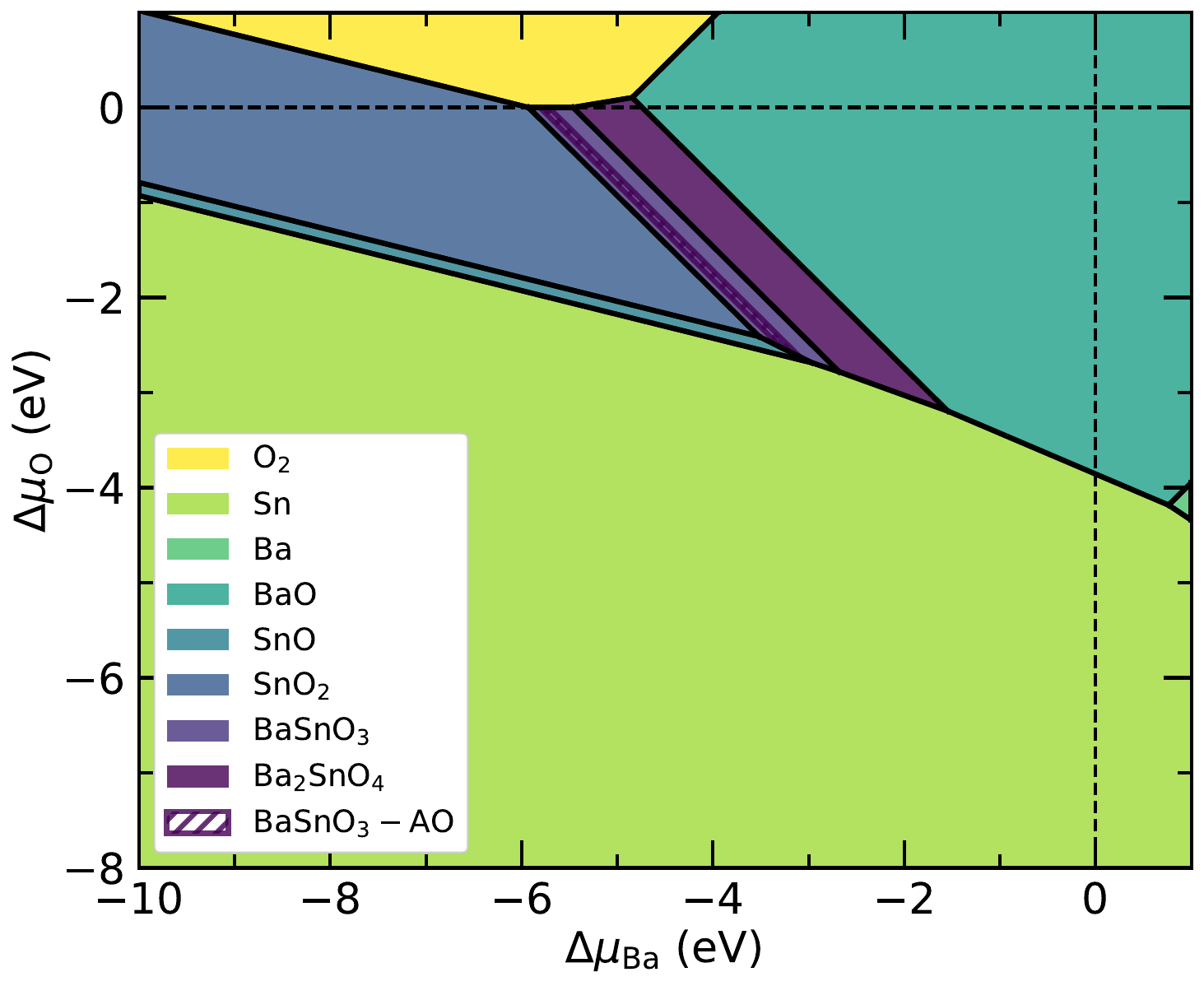}\label{fig:chem:bso}}%
    \caption[Chemical potential plots for oxide perovskites]{
        Chemical potential plots for the nine oxide perovskites studied in this work.
        The graphs are produced following the method outlined in \cite{Heifets2007}.
        The values of chemical potential for $A$ and O are varied for each, with $B$ depending on both of those based on the chemical formula of \abo{}.
        The shaded regions correspond to the region of stability for the labelled material, whilst the hashed region corresponds to the region of stability for the labelled surface.   Note, the surfaces are never more favourable than the undefected bulks, so we highlight the surfaces for where they would form if the perfect crystal did not exist.
        For plots where the \abo{}--\ao{} and/or \abo{}--\bo{} regions are not present, they are not stable when compared to the other materials considered.
    }
    \label{fig:chem}
\end{figure*}

In \figref{fig:chem}, we present the chemical potential plots for all nine of the oxide perovskites, and their surfaces. As the surfaces cost energy to form -- it must cost energy to cleave a bulk to form its surfaces, otherwise, that bulk would not be energetically stable -- they would not possess stable regions in these chemical potential plots. As such, to highlight the possibility of these surfaces, we present them as hatched regions in the chemical potential plots. For most of the structures presented here, we see that the \ao{} surface is the only one to appear to be energetically stable against the composite binary oxides (where the energy of the bulk perovskite has been removed as it is more stable than its surfaces within the region of its stability). We find that the \ao{} surface is favourable for most of the stability region of the perovskite.

We find two caveats to this general rule of \ao{} being the only stable surface, and those are for \cto{} and \cso{}. For \cto{}, we find that the \bo{} surface has a region of stability, but is far smaller than that the region dominated by the \ao{} surface, so the \ao{} will be preferable for most chemical environments. Secondly, we find that, for \cso{}, neither of the (001) surfaces are energetically stable when compared to the constituent binary oxide compounds. This is interesting, as it highlights the preference of the \ao{} surface for most, if not all, chemical environments. This is of particular interest for systems such as \sto{}, which have shown both surfaces experimentally~\cite{Sokolovi2021}.

\subsection{Vacancy energy comparison}
\label{sec:vacancy_energies}

\begin{table}[ht]
    \begin{tabular}{ccccccc} 
      \hline
      \multirow{3}{*}{Perovskite} & \multicolumn{6}{c}{Vacancy formation energy (\si{\electronvolt})}\\\cmidrule(lr){2-7}
      & \multicolumn{3}{c}{\ao{}} & \multicolumn{3}{c}{\bo{}} \\\cmidrule(lr){2-4}\cmidrule(lr){5-7}
      & bulk & surface & total & bulk & surface & total \\
      \hline
       \cto{}  & 2.26 & -1.62 & 0.64      & 5.46 & -2.13 & 3.33 \\
       \sto{}  & 1.66 & -1.68 & -0.02     & 5.21 & -1.71 & 3.50 \\
       \bto{}  & 2.08 & -2.35 & -0.27     & 4.38 & -2.68 & 1.70 \\\hline
       \czo{}  & 2.61 & -1.60 & 1.01      & 5.93 & -1.62 & 4.31 \\
       \szo{}  & 2.30 & -1.56 & 0.74      & 6.01 & -1.82 & 4.19 \\
       \bzo{}  & 1.96 & -1.10 & 0.85      & 5.32 & -1.19 & 4.13 \\\hline
       \cso{}  & 3.02 & -1.62 & 1.40      & 5.67 & -1.96 & 3.71 \\
       \sso{}  & 2.82 & -1.51 & 1.32      & 5.43 & -2.33 & 3.10 \\
       \bso{}  & 2.52 & -1.08 & 1.44      & 4.57 & -1.22 & 3.35 \\\hline\hline
      Averages & 2.36 & -1.57 & 0.79      & 5.33 & -1.85 & 3.48 \\
    \hline
      
    \end{tabular}
    \caption[$A$- and $B$-site vacancy formation energies]{Table of vacancy formation energies for $A$- and $B$-site vacancies in bulk and at their respective surface (\ao{} for $A$-site surface vacancies and \bo{} for $B$-site surface vacancies). As surface vacancy energies are given with respect to the bulk, the column designated \textit{total} refers to the sum of bulk and surface energies, retrieving surface vacancy energies relating to their corresponding binary oxide (i.e. \ao{} or \bo{}).}
    \label{tab:vacancies:a+b}
\end{table}

The vacancy formation energies for $A$- and $B$-site bulk and surface vacancies are detailed in \tabref{tab:vacancies:a+b}, given with respect to their corresponding binary oxides (\ao{} and \bo{}).
Average bulk $A$- and $B$-site vacancy formation energies are $3.91$ and $4.50$~\si{\electronvolt}, respectively.
The average surface $A$- and $B$-site vacancies have energy differences from their bulk values of $-1.57$ and $-1.85$~\si{\electronvolt}, respectively.
As such, the average total formation energy of the $A$- and $B$-site surface vacancies are $1.62$ and $2.65$~\si{\electronvolt}, respectively.
From this, we can see that the $A$-site vacancies are more likely to form at the \ao{} surface than the $B$-site vacancies at the \bo{} surface.
This should lead to more rapid degradation of the \ao{} surface.

\begin{table}[ht]
    \begin{tabular}{cccccc} 
      \hline
      \multirow{3}{*}{Perovskite} & \multicolumn{5}{c}{Vacancy formation energy (\si{\electronvolt})}\\\cmidrule(lr){2-6}
      & \multirow{2}{*}{bulk} & \multicolumn{2}{c}{\ao{}} & \multicolumn{2}{c}{\bo{}} \\\cmidrule(lr){3-4}\cmidrule(lr){5-6}
      & & surface & total & surface & total \\
      \hline
       \cto{}  & 4.89 &     0.00 & 4.89     & -1.39 & 3.50 \\
       \sto{}  & 5.22 &    -0.73 & 4.50     & -1.55 & 3.68 \\
       \bto{}  & 4.97 &    -0.92 & 4.04     & -2.24 & 2.73 \\\hline
       \czo{}  & 6.64 &    -0.92 & 5.72     & -0.06 & 6.58 \\
       \szo{}  & 6.59 &    -1.21 & 5.38     & -0.16 & 6.43 \\
       \bzo{}  & 6.50 &    -1.04 & 5.47     & -0.17 & 6.33 \\\hline
       \cso{}  & 4.13 &    -0.12 & 4.01     & -1.94 & 2.19 \\
       \sso{}  & 3.96 &    -0.07 & 3.89     & -2.14 & 1.82 \\
       \bso{}  & 3.53 &     0.11 & 3.64     & -1.83 & 1.69 \\\hline\hline
      Averages & 5.16 &    -0.55 & 4.61     & -1.28 & 3.88 \\
    \hline
      
    \end{tabular}
    \caption[O-site vacancy formation energies]{Table of vacancy formation energies for O-site vacancies in bulk and at the \ao{} and \bo{} surfaces. As surface vacancy energies are given with respect to the bulk, the column designated \textit{total} refers to the sum of bulk and surface energies, retrieving surface vacancy energies relating to \o{} gas.}
    \label{tab:vacancies:o}
\end{table}

Additionally, the formation energies of O-site vacancies are compared in \tabref{tab:vacancies:o}.
We see that, for stannates and titanates, the \bo{} surface is easier to vacate, whereas, for the zirconates, the \ao{} surface is more energetically favourable to vacate.
However, considering both $A$- (B-) and O-site vacancies on the \ao{} (\bo{}) surface simultaneously is challenging and cannot be considered by simply summing the relevant \textit{isolated} vacancy energies.
One thing to note though, is that, with there being more ions on the \bo{} surface, it will likely require more energy to fully deplete than the \ao{} surface.

\subsection{Vacancy-related structural distortions}
\label{sec:angles}

Whilst the atomic structure of slabs are distorted due to the missing bonds at the surface, further distortions can be brought on through the inclusion of vacancies. In \tabref{tab:angles}, we highlight the tilting induced by the deepest vacancies in the slabs (where the deepest vacancies are expected to exhibit a vacancy formation energy equivalent to that of the vacancy in the bulk, i.e. where the vacancy should have no interaction with the surface or vacuum). By presenting the difference between the unvacated and vacated slabs, we can try to isolate the amount of tilting induced by the slab from that of the vacancy inclusion. Here, we use the difference between the unvacated and vacated slab for the O--B--O bond angle far from the vacancy. These angles are not rotationally-dependent, so there is no information here regarding the direction a tilt may be associated with. Studying the difference angle caused by these deep-level vacancies should show how the atomic structure of the slab differs from the bulk. Such distortions of the atomic structure are suggested to be the cause of the difference in vacancy formation energy of deep-level vacancies versus vacancies in bulk. These tilts or distortions can either be linked to a cubic or tetragonal structure transitioning towards an orthorhombic phase, or for the perovskite structure becoming polarised.

For the most part, \ao{}-terminated perovskites are seen to be well behaved for their vacancy formation energies (seen in the main article), which agrees with the distortion angles. We see that \ao{}-terminated \sto{} and \bto{} exhibit significant angular distortion for O-site vacancies (in addition Ti-site vacancies in \ao-terminated \bto{}), which coincides with the difference in energy between the deep-level vacancies and bulk vacancies of around $1$~\si{\electronvolt} (termed \textit{deep-bulk difference}). There exist three other large angular distortions in \ao{}-terminated perovskites (\czo{} and \cso{} Ca-site vacancies, and \bzo{} O-site vacancies), however, these show much smaller deep-bulk differences (on the order of $0.1$~\si{\electronvolt}).

For vacated \bo{}-terminated perovskites, we see a general trend for the titanate and zirconate perovskites, in which the $A$- and $B$-site vacancies both show large deep-bulk differences (in excess of $1$~\si{\electronvolt}), which is seen to coincide with large angular distortions, ranging from $2.6$ to $8.4$~\si{\degree}. The only other large angular distortion in \bo{}-terminated perovskites is the O-site vacancy in \bso{}, which lines up with the roughly $1$~\si{\electronvolt} deep-bulk difference.

\begin{table*}[ht!]
\centering
    \begin{tabular}{ccccccccc} 
      \hline
      \multirow{3}{*}{Perovskite} & \multicolumn{4}{c}{\ao{}} & \multicolumn{4}{c}{\bo{}} \\\cmidrule(lr){2-5}\cmidrule(lr){6-9}
      & unvacated & $A$-vac & $B$-vac & O-vac & unvacated & $A$-vac & $B$-vac & O-vac \\\cmidrule(lr){2-2}\cmidrule(lr){3-5}\cmidrule(lr){6-6}\cmidrule(lr){7-9}
      & $\theta$ (\si{\degree}) & \multicolumn{3}{c}{$\Delta\theta$ (\si{\degree})} & $\theta$ (\si{\degree}) & \multicolumn{3}{c}{$\Delta\theta$ (\si{\degree})} \\
      \hline
       \cto{} & 179.1 & -0.4 & -0.9 & -1.3     & 178.3 & \textbf{-2.6} & \textbf{-5.4} & 0.2  \\
       \sto{} & 179.4 & -0.1 & -1.3 & \textbf{-3.4}     & 178.9 & \textbf{-6.0} & \textbf{-7.1} & -0.2 \\
       \bto{} & 179.6 & -0.4 & \textbf{-5.4} & \textbf{-4.7}     & 178.8 & \textbf{-6.2} & \textbf{-8.4} & -1.7 \\\hline
       \czo{} & 179.6 & \textbf{-3.6} & -0.4 & -0.1     & 179.2 & \textbf{-3.0} & \textbf{-5.8} & 0.1 \\
       \szo{} & 179.7 & -0.4 & -0.4 & -1.6     & 178.6 & \textbf{-2.6} & \textbf{-4.3} & 0.1 \\
       \bzo{} & 179.7 & -1.0 & 0.2 & \textbf{-2.4}    & 179.3 & \textbf{-4.6} & \textbf{-5.9} & 0.0 \\\hline
       \cso{} & 179.9 & \textbf{-2.5} & -1.2 & -0.3     & 179.3 & -0.1 & -0.5 & -0.8 \\
       \sso{} & 179.7 & -0.4 & -0.3 & -0.2     & 179.1 & -0.2 & -0.2 & -1.1 \\
       \bso{} & 179.8 & -1.3 & -0.3 & -0.9     & 179.4 & -0.5 & -0.7 & \textbf{-2.3} \\
      \hline
      
    \end{tabular}
    \caption[Vacancy-related distortions]{
        Average distortions of the underlying perovskite structure induced by inclusion of an $A$-, $B$-, or O-site vacancy in the layer closest to the centre of the slab.
        $\theta$ equates to the in-layer-plane O--B--O bond angle in the sub-surface \bo{} layer above the bottom surface in the undefected \ao{} or \bo{} slabs.
        $\Delta\theta$ is the change in the angle -- the tilting distortion -- of the O--B--O bond in the vacated slab from the unvacated one, where negative (positive) values relate to a smaller (larger) angle.
        Bolded numbers highlight large distortions, termed as angle decreases greater than 2~\si{\degree}.
    }
    \label{tab:angles}
\end{table*}

\subsection{Dopability}
\label{sec:dopability}

\begin{table}[]
    {\smaller
    \centering
    \begin{tabular}{P{6em}cP{6.5em}P{5em}cP{6.5em}P{5em}}
        \hline
         \multirow{2}{*}{\makecell{\\System\\}}     & \multicolumn{3}{c}{$p$-type} & \multicolumn{3}{c}{$n$-type} \\\cmidrule(lr){2-4}\cmidrule(lr){5-7} 
                    & Limit & \makecell{Compensating \\ defect} & \makecell{Dopability \\ limit} & Limit & \makecell{Compensating \\ defect} & \makecell{Dopability \\ limit}\\\hline
         Bulk       & \sso{}-SnO$_{2}$-\o{} & V$_{\mathrm{O}}^{+2}$               &  0.78 & Sr$_{2}$SnO$_{4}$-\sso{}-Sn  & V$_{\mathrm{O}}^{-2}$ & 4.11 \\
         \ao{} slab & \sso{}-SnO$_{2}$-\o{} & V$_{\mathrm{O}_{1\mathrm{L}}}^{+2}$ &  1.19 & Sr$_{2}$SnO$_{4}$-\sso{}-Sn  & V$_{\mathrm{Sr}_{1\mathrm{L}}}^{-3}$ & 2.89 \\
         \bo{} slab & \sso{}-SnO$_{2}$-\o{} & V$_{\mathrm{O}_{1\mathrm{L}}}^{+2}$ &  0.96 & \sso{}-SnO$_{2}$-\o{} & V$_{\mathrm{Sr}_{2\mathrm{L}}}^{-3}$ & 2.14 \\\hline
    \end{tabular}
    }
    \caption[Dopability limits in \sso{}]{
        Dopability limits of \sso{} bulk and \ao{}- and \bo{}-terminated \sso{} slabs.
        Energy is referenced to the valence band maximum.
        The $p$-type ($n$-type) dopability limit corresponds to the highest (lowest) Fermi energy at which a native defect reaches zero formation energy, indicating the spontaneous formation of compensating native defects beyond these limits.
    }
    \label{tab:dopability_limits}
\end{table}

\begin{table}[]
{\smaller
    \centering
    \begin{tabular}{P{6em}cP{6.5em}P{4em}cP{6.5em}P{4em}}
        \hline
         \multirow{2}{*}{\makecell{\\System\\}} & \multicolumn{3}{c}{$p$-type} & \multicolumn{3}{c}{$n$-type} \\\cmidrule(lr){2-4}\cmidrule(lr){5-7} 
                    & Limit & \makecell{Compensating \\ defect} & \makecell{Doping \\ window} & Limit & \makecell{Compensating \\ defect} & \makecell{Doping \\ window} \\\hline
         Bulk       & \sso{}-SnO$_{2}$-\o{}         & V$_{\mathrm{O}}^{+2}$                & -1.561 & SnO-\sso{}-Sn              & V$_{\mathrm{O}}^{-2}$ & 2.185 \\
         \ao{} slab & Sr$_{2}$SnO$_{4}$-\sso{}-\o{} & V$_{\mathrm{Sr}_{1\mathrm{L}}}^{+3}$ & -2.68 & Sr$_{2}$SnO$_{4}$-\sso{}-Sn & V$_{\mathrm{Sr}_{1\mathrm{L}}}^{-3}$ & -0.68  \\
         \bo{} slab & \sso{}-SnO$_{2}$-\o{} & V$_{\mathrm{O}_{1\mathrm{L}}}^{+2}$          & -1.93 & \sso{}-SnO$_{2}$-\o{} & V$_{\mathrm{Sr}_{1\mathrm{L}}}^{-3}$ & -0.09  \\\hline
    \end{tabular}
    }
    \caption[Doping windows in \sso{}]{
        Dopability of \sso{} bulk and \ao{}- and \bo{}-terminated \sso{} slabs.
        The doping windows are determined by the lowest-energy compensating defects at the band edges -- valence band maximum for $p$-type and conduction band minimum for $n$-type.
        These define the dominant compensating native defects (vacancies in this study) for $p$- and $n$-type doping in \sso{}.
    }
    \label{tab:doping_windows}
\end{table}

The dopability limits of \sso{} are explored through analysis of the native charged defects.
The dopability limits and doping windows are provided in Tables \ref{tab:dopability_limits} and \ref{tab:doping_windows}, respectively.
Native defect thermodynamics suggest that \sso{} is more easily $n$-type dopable than $p$-type dopable, due to the larger (more positive) doping window exhibited for the $n$-type compensating defects.
However, as these doping windows are negative for the surfaces, it is likely that doping the surfaces will result in surface degredation, as compensating native vacancies are energetically favourable (i.e. negative formation energies).

\subsection{Charged defects for different chemical environments}
\label{sec:charge_state:chempots}

\begin{table}[]
    \centering
    \begin{tabular}{cccc}
    \hline
        \multirow{2}{*}{Limit} & \multicolumn{3}{c}{Chemical potential (\si{\electronvolt})} \\\cmidrule(lr){2-4}
         & Sr & Sn & O \\\hline
         \sso{}-SnO$_{2}$-\o{}         & -6.1852 & -5.0964 &  0.0000 \\
         Sr$_{2}$SnO$_{4}$-\sso{}-Sn   & -2.8153 &  0.0000 & -2.8221 \\
         Sr$_{2}$SnO$_{4}$-\sso{}-\o{} & -5.6374 & -5.6442 &  0.0000 \\
         SnO-\sso{}-Sn                 & -3.2238 &  0.0000 & -2.6859 \\
         SnO-\sso{}-SnO$_{2}$          & -3.7747 & -0.2755 & -2.4105 \\\hline
    \end{tabular}
    \caption[Chemical potential limits for \sso{}]{
    Chemical potential limits for the five limits of the Sr-Sn-O environment, considering the phases bulk metals Sr and Sn, \o{} gas, binary SrO, SnO, and SnO$_{2}$, Sr$_{2}$SnO$_{4}$, and orthorhombic perovskite \sso{}.
    The chemical potential values are given with respect to the elemental reference phases, i.e. Sr and Sn bulk metals and \o{} gas.
    }
    \label{tab:chem_pots:sso}
\end{table}

\begin{figure*}[ht]
    \centering
    \subfloat[Sr$_{2}$SnO$_{4}$-\sso{}-Sn]{\includegraphics[width=0.45\linewidth]{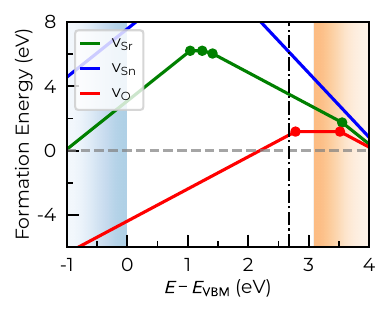}\label{fig:charge_state:chempots:bulk:Sr2SnO4-SrSnO3-Sn}}%
    \hspace{1em}%
    \subfloat[Sr$_{2}$SnO$_{4}$-\sso{}-\o{}]{\includegraphics[width=0.45\linewidth]{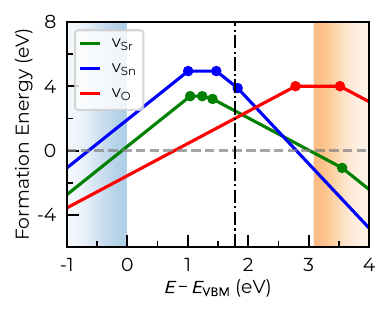}\label{fig:charge_state:chempots:bulk:Sr2SnO4-SrSnO3-O2}}%
    \hspace{1em}%
    \subfloat[SnO-\sso{}-Sn]{\includegraphics[width=0.45\linewidth]{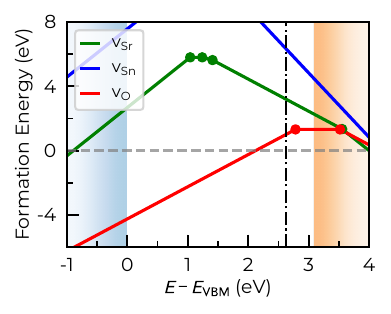}\label{fig:charge_state:chempots:bulk:SnO-SrSnO3-Sn}}%
    \hspace{1em}%
    \subfloat[SnO-\sso{}-SnO$_{2}$]{\includegraphics[width=0.45\linewidth]{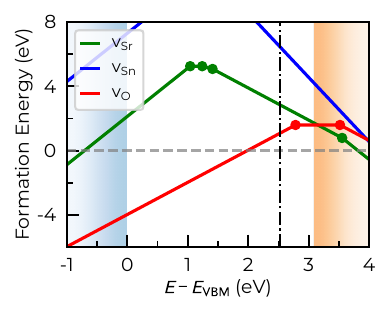}\label{fig:charge_state:chempots:bulk:SnO-SrSnO3-SnO2}}%
    \caption[Charged vacancies in bulk \sso{} under various chemical limits]{
        Formation energies of $A$-, $B$- and, O-site charged vacancies in bulk \sso{} under various chemical limits (see \tabref{tab:chem_pots:sso}).
        Line gradients correspond to the charge state (in $|$\si{\elementarycharge}$|$) and points of discontinuity identify transition levels.
        The valence (conduction) band region is shaded in blue (orange) and the vertical dash-dotted line is the equilibrium Fermi level.
        Corrections to the charge state energies are calculated using the following codes: \texttt{doped}~\cite{Kavanagh2024DopedPythonToolkit}, \texttt{qdef2d}~\cite{Tan2019ChargedDefectsFramework}, and \texttt{sxdefectalign2d}~\cite{Freysoldt2009,Freysoldt2018FirstPrinciplesCalculations}, which employ band edge shifting~\cite{Broberg2023HighThroughputCalculations}, image charge~\cite{Kumagai2014ElectrostaticBasedFinite}, and slab charge~\cite{Freysoldt2009,Freysoldt2018FirstPrinciplesCalculations} corrections.
    }
    \label{fig:charge_state:chempots:bulk}
\end{figure*}

The native charged vacancies discussed in the main article are recalculated under four additional chemical environment limits.
This study compares the formation of \sso{} and its vacancies to seven competing phases, with their respective formation energies shown in \tabref{tab:competing_phases:formation_energies}.
Five chemical potential limits are identified, being the environments in which \sso{} transitions between being stable and unstable to its competing phases.
These limits are presented in \tabref{tab:chem_pots:sso}.
\sso{}-\o{}-SrO is defined as the O-rich limit, whilst \sso{}-Sn-SrO is the Sr-rich limit.
The O-rich limit is presented in the main article, whilst the other four limits are presented in \figrefss{fig:charge_state:chempots:bulk}{fig:charge_state:chempots:AO}{fig:charge_state:chempots:BO2} for bulk \sso{}, and \ao{}- and \bo{}-terminated \sso{} slabs, respectively.

\begin{figure*}[ht]
    \centering
    \subfloat[Sr$_{2}$SnO$_{4}$-\sso{}-Sn]{\includegraphics[width=0.45\linewidth]{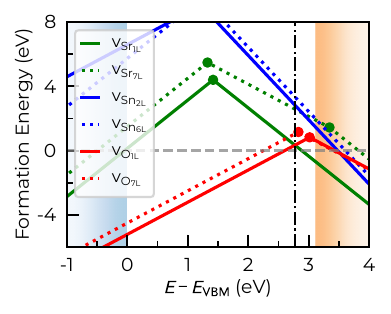}\label{fig:charge_state:chempots:AO:Sr2SnO4-SrSnO3-Sn}}%
    \hspace{1em}%
    \subfloat[Sr$_{2}$SnO$_{4}$-\sso{}-\o{}]{\includegraphics[width=0.45\linewidth]{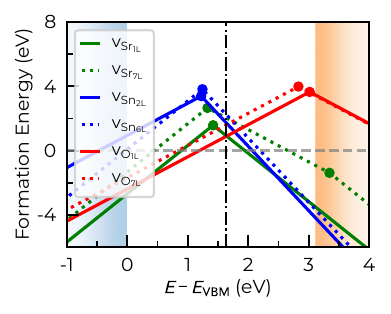}\label{fig:charge_state:chempots:AO:Sr2SnO4-SrSnO3-O2}}%
    \hspace{1em}%
    \subfloat[SnO-\sso{}-Sn]{\includegraphics[width=0.45\linewidth]{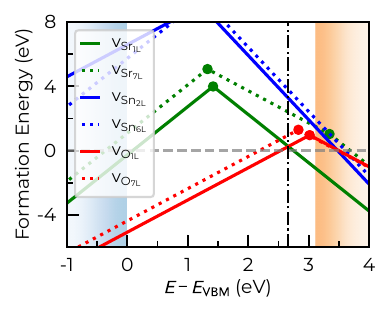}\label{fig:charge_state:chempots:AO:SnO-SrSnO3-Sn}}%
    \hspace{1em}%
    \subfloat[SnO-\sso{}-SnO$_{2}$]{\includegraphics[width=0.45\linewidth]{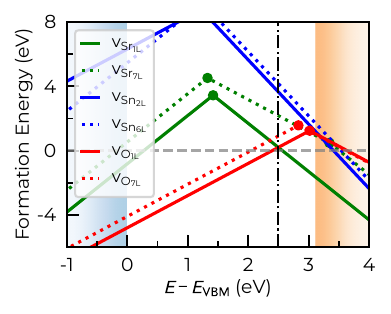}\label{fig:charge_state:chempots:AO:SnO-SrSnO3-SnO2}}%
    \caption[Charged vacancies in \ao{}-terminated \sso{} slabs under various chemical limits]{
        Formation energies of $A$-, $B$- and, O-site charged vacancies in the \sso{} \ao{}-terminated slab under various chemical limits (see \tabref{tab:chem_pots:sso}).
        Line gradients correspond to the charge state (in $|$\si{\elementarycharge}$|$) and points of discontinuity identify transition levels.
        The valence (conduction) band region is shaded in blue (orange) and the vertical dash-dotted line is the equilibrium Fermi level.
        Corrections to the charge state energies are calculated using the following codes: \texttt{doped}~\cite{Kavanagh2024DopedPythonToolkit}, \texttt{qdef2d}~\cite{Tan2019ChargedDefectsFramework}, and \texttt{sxdefectalign2d}~\cite{Freysoldt2009,Freysoldt2018FirstPrinciplesCalculations}, which employ band edge shifting~\cite{Broberg2023HighThroughputCalculations}, image charge~\cite{Kumagai2014ElectrostaticBasedFinite}, and slab charge~\cite{Freysoldt2009,Freysoldt2018FirstPrinciplesCalculations} corrections.
    }
    \label{fig:charge_state:chempots:AO}
\end{figure*}

\begin{figure*}[ht]
    \centering
    \subfloat[Sr$_{2}$SnO$_{4}$-\sso{}-Sn]{\includegraphics[width=0.45\linewidth]{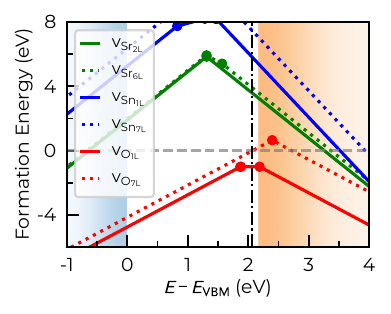}\label{fig:charge_state:chempots:BO2:Sr2SnO4-SrSnO3-Sn}}%
    \hspace{1em}%
    \subfloat[Sr$_{2}$SnO$_{4}$-\sso{}-\o{}]{\includegraphics[width=0.45\linewidth]{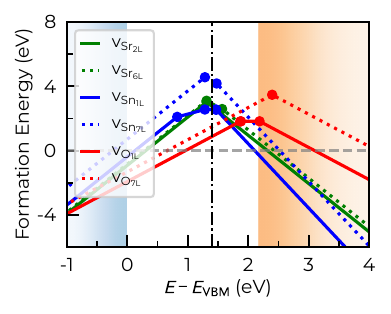}\label{fig:charge_state:chempots:BO2:Sr2SnO4-SrSnO3-O2}}%
    \hspace{1em}%
    \subfloat[SnO-\sso{}-Sn]{\includegraphics[width=0.45\linewidth]{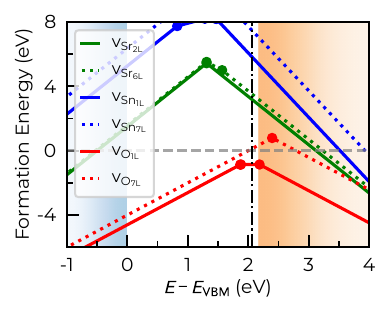}\label{fig:charge_state:chempots:BO2:SnO-SrSnO3-Sn}}%
    \hspace{1em}%
    \subfloat[SnO-\sso{}-SnO$_{2}$]{\includegraphics[width=0.45\linewidth]{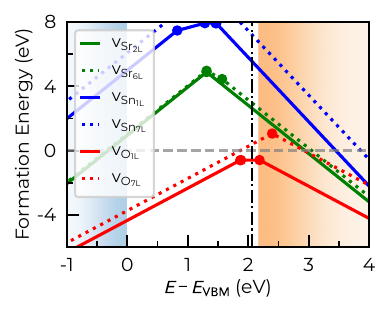}\label{fig:charge_state:chempots:BO2:SnO-SrSnO3-SnO2}}%
    \caption[Charged vacancies in \bo{}-terminated \sso{} slabs under various chemical limits]{
        Formation energies of $A$-, $B$- and, O-site charged vacancies in the \sso{} \bo{}-terminated slab under various chemical limits (see \tabref{tab:chem_pots:sso}).
        Line gradients correspond to the charge state (in $|$\si{\elementarycharge}$|$) and points of discontinuity identify transition levels.
        The valence (conduction) band region is shaded in blue (orange) and the vertical dash-dotted line is the equilibrium Fermi level.
        Corrections to the charge state energies are calculated using the following codes: \texttt{doped}~\cite{Kavanagh2024DopedPythonToolkit}, \texttt{qdef2d}~\cite{Tan2019ChargedDefectsFramework}, and \texttt{sxdefectalign2d}~\cite{Freysoldt2009,Freysoldt2018FirstPrinciplesCalculations}, which employ band edge shifting~\cite{Broberg2023HighThroughputCalculations}, image charge~\cite{Kumagai2014ElectrostaticBasedFinite}, and slab charge~\cite{Freysoldt2009,Freysoldt2018FirstPrinciplesCalculations} corrections.
    }
    \label{fig:charge_state:chempots:BO2}
\end{figure*}

\begin{table}[]
    \centering
    \begin{tabular}{cccc}
        \hline
         \multirow{2}{*}{Limit}           & \multicolumn{3}{c}{Pinned Fermi level (\si{\electronvolt})} \\\cmidrule(lr){2-4}
         & Bulk & \ao{} slab & \bo{}-slab\\\hline
         \sso{}-SnO$_{2}$-\o{}          & 1.78 & 1.53 & 1.40 \\
          Sr$_{2}$SnO$_{4}$-\sso{}-Sn   & 2.67 & 2.77 & 2.07 \\
          Sr$_{2}$SnO$_{4}$-\sso{}-\o{} & 1.78 & 1.64 & 1.40 \\
         SnO-\sso{}-Sn                  & 2.63 & 2.66 & 2.07 \\
         SnO-\sso{}-SnO$_{2}$           & 2.53 & 2.49 & 2.07 \\\hline
    \end{tabular}
    \caption[Pinned Fermi levels for \sso{} under various chemical limits]{
        Pinned Fermi level for \sso{} bulk and \ao{}- and \bo{}-terminated \sso{} slabs.
        The pinned Fermi levels are given with respect to the valence band maximum for the five limits of the Sr-Sn-O environment.
    }
    \label{tab:fermi_level}
\end{table}

The pinned Fermi level for \sso{} bulk and \ao{}- and \bo{}-terminated \sso{} slabs are presented in \tabref{tab:fermi_level} for the five chemical potential limits.

\subsection{Charged defect correction tables}
\label{sec:charge_state:correction_tables}

The formation energies and associated corrections applied to the DFT energies of charged native vacancies in \sso{} bulk and \ao{}- and \bo{}-terminated \sso slabs are provided in Tables \ref{tab:corrections:bulk}, \ref{tab:corrections:AO}, and \ref{tab:corrections:BO2}, respectively.

\begin{table}[ht]
\begin{tabular}{lr@{\hskip 1em}lllllll}
\hline
Defect            & $q$ & $\Delta E_{\mathrm{raw}}$ & $qE_{\mathrm{VBM}}$ & $qE_{\mathrm{F}}$ & $\Sigma\mu_{\mathrm{ref}}$ & $\Sigma\mu_{\mathrm{formal}}$ & $E_{\mathrm{corr}}$ & $E_{\mathrm{form}}$ \\\hline
V$_{\mathrm{Sr}}$ & $+3$ & 2.649   & 5.036    & 4.63     & -1.638   & -6.185   & -0.149   & 4.344   \\
V$_{\mathrm{Sr}}$ & $0$  & 10.662  & 0.0      & 0.0      & -1.638   & -6.185   & 0.0      & 2.838   \\
V$_{\mathrm{Sr}}$ & $-1$ & 13.41   & -1.679   & -1.543   & -1.638   & -6.185   & 0.171    & 2.535   \\
V$_{\mathrm{Sr}}$ & $-2$ & 16.224  & -3.358   & -3.087   & -1.638   & -6.185   & 0.447    & 2.402   \\
V$_{\mathrm{Sr}}$ & $-3$ & 21.179  & -5.036   & -4.63    & -1.638   & -6.185   & 0.725    & 4.414   \\\hline
V$_{\mathrm{Sn}}$ & $+3$ & 6.378   & 5.036    & 4.63     & -3.793   & -5.096   & -0.054   & 7.101   \\
V$_{\mathrm{Sn}}$ & $0$  & 14.384  & 0.0      & 0.0      & -3.793   & -5.096   & 0.0      & 5.494   \\
V$_{\mathrm{Sn}}$ & $-3$ & 22.962  & -5.036   & -4.63    & -3.793   & -5.096   & 0.879    & 5.285   \\
V$_{\mathrm{Sn}}$ & $-4$ & 25.933  & -6.715   & -6.174   & -3.793   & -5.096   & 1.414    & 5.567   \\\hline
V$_{\mathrm{O}}$  & $+2$ & -0.179  & 3.358    & 3.087    & -4.948   & 0.0      & 0.208    & 1.526   \\
V$_{\mathrm{O}}$  & $0$  & 8.954   & 0.0      & 0.0      & -4.948   & 0.0      & 0.0      & 4.006   \\
V$_{\mathrm{O}}$  & $-2$ & 19.259  & -3.358   & -3.087   & -4.948   & 0.0      & 0.092    & 7.958   \\\hline
\end{tabular}
    \caption[Charged vacancy formation energies in bulk \sso{}]{
        Formation energies and corrections for native charged vacancies in bulk \sso{}, as outputted by \texttt{doped}~\cite{Kavanagh2024DopedPythonToolkit}. 
        $\Delta E_{\mathrm{raw}}$ is the uncorrected DFT energy difference between pristine and defective slabs; 
        $qE_{\mathrm{VBM}}$ and $qE_{\mathrm{F}}$ are valence band and Fermi level contributions; 
        $\Sigma\mu_{\mathrm{ref}}$ and $\Sigma\mu_{\mathrm{formal}}$ are chemical potentials of the removed atom (reference and formal limit); 
        $E_{\mathrm{corr}}$ is the electrostatic charge correction; 
        $E_{\mathrm{form}}$ is the final corrected formation energy. 
        All energies are in \si{\electronvolt}, with $q$ in units of~|\si{\elementarycharge}|.
    }
    \label{tab:corrections:bulk}
\end{table}

\begin{table}[ht]
\begin{tabular}{lr@{\hskip 1em}lllllll}
\hline
Defect                          & $q$ & $\Delta E_{\mathrm{raw}}$ & $qE_{\mathrm{VBM}}$ & $qE_{\mathrm{F}}$ & $\Sigma\mu_{\mathrm{ref}}$ & $\Sigma\mu_{\mathrm{formal}}$ & $E_{\mathrm{corr}}$ & $E_{\mathrm{form}}$ \\\hline
V$_{\mathrm{Sr_\mathrm{1L}}}$ & $+3$  & 7.97    & -1.307  & 4.668   & -1.638   & -6.185   & -2.07    & 1.436   \\
V$_{\mathrm{Sr_\mathrm{1L}}}$ & $-3$  & 14.257  & 1.307   & -4.668  & -1.638   & -6.185   & -2.45    & 0.623   \\\hline
V$_{\mathrm{Sr_\mathrm{7L}}}$ & $+3$  & 9.458   & -1.307  & 4.668   & -1.638   & -6.185   & -2.203   & 2.792   \\
V$_{\mathrm{Sr_\mathrm{7L}}}$ & $-2$  & 12.231  & 0.871   & -3.112  & -1.638   & -6.185   & -0.515   & 1.652   \\
V$_{\mathrm{Sr_\mathrm{7L}}}$ & $-3$  & 15.162  & 1.307   & -4.668  & -1.638   & -6.185   & -0.536   & 3.441   \\\hline\hline
V$_{\mathrm{Sn_\mathrm{2L}}}$ & $+2$  & 12.238  & -0.871  & 3.112   & -3.793   & -5.096   & -0.985   & 4.603   \\
V$_{\mathrm{Sn_\mathrm{2L}}}$ & $-4$  & 19.41   & 1.743   & -6.224  & -3.793   & -5.096   & -3.434   & 2.605   \\\hline
V$_{\mathrm{Sn_\mathrm{6L}}}$ & $+3$  & 13.154  & -1.307  & 4.668   & -3.793   & -5.096   & -2.323   & 5.302   \\
V$_{\mathrm{Sn_\mathrm{6L}}}$ & $-4$  & 18.805  & 1.743   & -6.224  & -3.793   & -5.096   & -2.317   & 3.118   \\\hline\hline
V$_{\mathrm{O_\mathrm{1L}}}$  & $+2$  & 4.388   & -0.871  & 3.112   & -4.948   & 0.0      & -0.955   & 0.726   \\
V$_{\mathrm{O_\mathrm{1L}}}$  & $-2$  & 14.723  & 0.871   & -3.112  & -4.948   & 0.0      & -0.959   & 6.575   \\\hline
V$_{\mathrm{O_\mathrm{7L}}}$  & $+2$  & 4.74    & -0.871  & 3.112   & -4.948   & 0.0      & -0.601   & 1.432   \\
V$_{\mathrm{O_\mathrm{7L}}}$  & $-2$  & 14.843  & 0.871   & -3.112  & -4.948   & 0.0      & -1.12    & 6.534   \\\hline                 
\end{tabular}
    \caption[Charged vacancy formation energies in \ao{}-terminated \sso{} slab]{
        Formation energies and corrections for native charged vacancies in \ao{}-terminated \sso{} slabs, as outputted by \texttt{doped}~\cite{Kavanagh2024DopedPythonToolkit}. 
        $\Delta E_{\mathrm{raw}}$ is the uncorrected DFT energy difference between pristine and defective slabs; 
        $qE_{\mathrm{VBM}}$ and $qE_{\mathrm{F}}$ are valence band and Fermi level contributions; 
        $\Sigma\mu_{\mathrm{ref}}$ and $\Sigma\mu_{\mathrm{formal}}$ are chemical potentials of the removed atom (reference and formal limit); 
        $E_{\mathrm{corr}}$ is the electrostatic charge correction; 
        $E_{\mathrm{form}}$ is the final corrected formation energy. 
        All energies are in \si{\electronvolt}, with $q$ in units of~|\si{\elementarycharge}|.
    }
    \label{tab:corrections:AO}
\end{table}

\begin{table}[ht]
\begin{tabular}{lr@{\hskip 1em}lllllll}
\hline
Defect                        & $q$ & $\Delta E_{\mathrm{raw}}$ & $qE_{\mathrm{VBM}}$ & $qE_{\mathrm{F}}$ & $\Sigma\mu_{\mathrm{ref}}$ & $\Sigma\mu_{\mathrm{formal}}$ & $E_{\mathrm{corr}}$ & $E_{\mathrm{form}}$ \\\hline
V$_{\mathrm{Sr_\mathrm{2L}}}$ & +3  & 10.843  & -3.018   & 3.252    & -1.638   & -6.185   & -1.467   & 1.787   \\
V$_{\mathrm{Sr_\mathrm{2L}}}$ & -3  & 12.679  & 3.018    & -3.252   & -1.638   & -6.185   & -1.461   & 3.159   \\\hline
V$_{\mathrm{Sr_\mathrm{6L}}}$ & +3  & 11.173  & -3.018   & 3.252    & -1.638   & -6.185   & -1.716   & 1.867   \\
V$_{\mathrm{Sr_\mathrm{6L}}}$ & -2  & 11.162  & 2.012    & -2.168   & -1.638   & -6.185   & -0.173   & 3.009   \\
V$_{\mathrm{Sr_\mathrm{6L}}}$ & -3  & 12.626  & 3.018    & -3.252   & -1.638   & -6.185   & -1.074   & 3.494   \\\hline\hline
V$_{\mathrm{Sn_\mathrm{1L}}}$ & +3  & 12.825  & -3.018   & 3.252    & -3.793   & -5.096   & -0.751   & 3.418   \\
V$_{\mathrm{Sn_\mathrm{1L}}}$ & +1  & 12.118  & -1.006   & 1.084    & -3.793   & -5.096   & -0.402   & 2.904   \\
V$_{\mathrm{Sn_\mathrm{1L}}}$ & -1  & 12.35   & 1.006    & -1.084   & -3.793   & -5.096   & 0.1      & 3.482   \\
V$_{\mathrm{Sn_\mathrm{1L}}}$ & -4  & 15.903  & 4.024    & -4.336   & -3.793   & -5.096   & -2.052   & 4.649   \\\hline
V$_{\mathrm{Sn_\mathrm{7L}}}$ & +3  & 14.889  & -3.018   & 3.252    & -3.793   & -5.096   & -1.726   & 4.508   \\
V$_{\mathrm{Sn_\mathrm{7L}}}$ & -2  & 14.976  & 2.012    & -2.168   & -3.793   & -5.096   & -0.418   & 5.513   \\
V$_{\mathrm{Sn_\mathrm{7L}}}$ & -4  & 16.625  & 4.024    & -4.336   & -3.793   & -5.096   & -1.123   & 6.301   \\\hline\hline
V$_{\mathrm{O_\mathrm{1L}}}$  & +2  & 5.52    & -2.012   & 2.168    & -4.948   & 0.0      & -0.486   & 0.242   \\
V$_{\mathrm{O_\mathrm{1L}}}$  & +1  & 5.961   & -1.006   & 1.084    & -4.948   & 0.0      & -0.063   & 1.028   \\
V$_{\mathrm{O_\mathrm{1L}}}$  & -2  & 9.836   & 2.012    & -2.168   & -4.948   & 0.0      & -0.698   & 4.034   \\\hline
V$_{\mathrm{O_\mathrm{7L}}}$  & +2  & 5.935   & -2.012   & 2.168    & -4.948   & 0.0      & -0.297   & 0.847   \\
V$_{\mathrm{O_\mathrm{7L}}}$  & -2  & 11.898  & 2.012    & -2.168   & -4.948   & 0.0      & -0.695   & 6.099   \\\hline
\end{tabular}
    \caption[Charged vacancy formation energies in \bo{}-terminated \sso{} slab]{
        Formation energies and corrections for native charged vacancies in \bo{}-terminated \sso{} slabs, as outputted by \texttt{doped}~\cite{Kavanagh2024DopedPythonToolkit}. 
        $\Delta E_{\mathrm{raw}}$ is the uncorrected DFT energy difference between pristine and defective slabs; 
        $qE_{\mathrm{VBM}}$ and $qE_{\mathrm{F}}$ are valence band and Fermi level contributions; 
        $\Sigma\mu_{\mathrm{ref}}$ and $\Sigma\mu_{\mathrm{formal}}$ are chemical potentials of the removed atom (reference and formal limit); 
        $E_{\mathrm{corr}}$ is the electrostatic charge correction; 
        $E_{\mathrm{form}}$ is the final corrected formation energy. 
        All energies are in \si{\electronvolt}, with $q$ in units of~|\si{\elementarycharge}|.
    }
    \label{tab:corrections:BO2}
\end{table}

\clearpage
\section*{References}
\bibliography{main}